Ph.D. Thesis

# Non-Equilibrium Properties of Open Quantum Systems

Ihor Vakulchyk

Basic Science

UNIVERSITY OF SCIENCE AND TECHNOLOGY

December 2021

# Non-Equilibrium Properties of Open Quantum Systems

**Ihor Vakulchyk**

**A dissertation Submitted in Partial Fulfillment of Requirements for the Degree of Doctor of Philosophy**

**December 2021**

**UNIVERSITY OF SCIENCE AND TECHNOLOGY**
**Major of: Basic Science**

**Supervisor: Sergej Flach**
**Co-supervisor: Alexei Andreanov**

# We hereby approve the Ph.D of Ihor Vakulchyk

**December 2021**

Jung-Wan Ryu sign

Chairman of Thesis Committee

Alexei Andreanov sign

Thesis Committee Member

SERGEY DENISOV sign

Thesis Committee Member

Peter TALKNER sign

Thesis Committee Member

Hee Chul Park sign

Thesis Committee Member

Ivan Savenko sign

Thesis Committee Member

Sergej Flach sign

Thesis Committee Member

**UNIVERSITY OF SCIENCE AND TECHNOLOGY**

## ACKNOWLEDGEMENTS

I want to thank my supervisor, Dr. Sergej Flach, for amazing supervision and both professional and personal support during my Ph.D. years. I also want to thank Dr. Sergey Denisov for collaboration and countless educational discussions.

I thank Dr. Carlo Danieli, Dr. Juzar Thinga, and, especially, Dr. Ivan Savenko for support with the thesis preparation.

**ABSTRACT**

We study two classes of open systems: discrete-time quantum walks (a type of Floquet-engineered discrete quantum map) and the Lindblad master equation (a general framework of dissipative quantum systems), focusing on the non-equilibrium properties of these systems. We study localization and delocalization phenomena, soliton-like excitations, and quasi-stationary properties of open quantum systems.

In discrete-time quantum walks we study both analytically and numerically Anderson localization induced by disorder in several physically relevant fields. In particular, we show the existence of a regime with a uniform localization length value for every eigenstate and give an analytical expression for it. Based on this, we study wave packet spreading due to mean-field-like nonlinearity in presence of a random field, and confirm previous conjectures on this dynamics numerically up to previously unachievable time scales. We further study the effect of nonlinearity in an all-bands-flat setup. In this setup, the particle transport is only carried by nonlinearity. We show the existence of a plethora of both stationary and moving soliton-like solutions and analyze their stability.

In the framework of the Lindblad master equation, we continue research of localization properties and study signatures of many-body localization transition in the presence of dissipation. We show that the inclusion of engineered dissipation allows for detection of the transition in both dynamical – growth of entanglement rate; and steady-state properties – statistics of local observables, average entanglement entropy, and spectral statistics. Moreover, we show that engineered dissipation can prevent dephasing from completely removing traces of the Hamiltonian properties in the steady-state. For this study, we develop an efficient parallel supercomputer implementation of the time-evolving block decimation algorithm, that is used for the evolution of many-body dissipative 1D systems. Furthermore, we study quasi-stationary states in Lindblad systems by generalizing a well-known classical approach for Markovian processes. We derive projected Lindblad equation and propose a generalized quantum trajectories algorithm for efficient simulation of systems with quasi-stationary properties.

# 초록

본연구에서우리는두종류의열린시스템에대해서연구한다. 첫째는일종의플로케엔지니어링에의한이산양자보행인이산시간양자보행이고, 두번째는린드블라드마스터방정식이며, 두시스템의비평형성에집중한다. 우리는열린양자시스템의국소화/비국소화현상, 유사-솔리톤들뜸, 그리고준평형성에대해서연구한다.

이산시간양자보행에서우리는해석적, 수치적으로무질서에의해유발되는앤더슨국소화에대해연구한다. 특히, 우리는국소화길이가모든고유상태들에대해동일해지게되는영역이존재함을증명하며, 이영역에서국소화길이를계산하는해석적식을제시한다. 이를토대로, 우리는무작위장의존재하에서유사평균장근사 (Mean Field Approximation) 비선형성에의한파동묶음의퍼짐을연구하며, 이와같은역학에서이전연구들에서제안된추측들을이전까지도달할수없었던시간스케일까지계산함을통해확인하였다. 또한우리는올밴즈플랫 (All Bands Flat) 시스템에서비선형성에의한효과를연구한다. 이러한시스템에서입자의이동현상은단지비선형성에의해이뤄진다. 또한이러한시스템에서움직이는유사솔리톤해들의존재성과이들의안정성을보인다.

이러한린드블라드마스터방정식의프레임워크하에서앤더슨국소화성질과에너지소산하에서의다체국소화전이의표식들에대해연구한다. 우리는조작된에너지소산이동적 (얽힘정도), 정적인상태 (국소적인관측가능량들의통계) 들의특성들을검출가능하게만듦을보인다. 더욱이, 이러한조작된에너지소산이정상상태의해밀토니안특성들의자취가완전히지워지는것으로발생하는 dephasing 을방지할수있음을보인다. 이를위해, 우리는효율적인 Time-Evolving Block Decimation algorithm 의병렬계산을구현하였으며, 이를이용하여다체에너지소산 1D 시스템의시간변화를계산하였다. 또한, 우리는마르코프과정 (Markov Process) 를분석하기위해사용되는널리알려진고전적인방법을일반화함으로써린드블라드시스템의준평형상태를연구한다. 우리는사영된린드블라드방정식을유도한후, 효율적인준평형특성에대한시뮬레이션을위해고안된일반화된양자궤적알고리즘을제안한다.

# Contents

# List of Figures

# Chapter 1

# Introduction

## 1.1 Open Systems

A considerable part of physics since its inception has been dedicated to studying isolated systems. This stems from the very nature of the modern scientific research in natural sciences – the inductive approach towards building and verifying theories. In physics, the most fundamental of natural sciences, the pinnacle of theories are exact fundamental laws of nature. To become fundamental, these laws have to be formulated for isolated systems in order to encapsulate the most complete description of a system possible. Given these laws as first principles, it is only natural that the to-go approach in physics is to attempt to reduce every problem to a model of an isolated system.

However, all real-life systems do interact with the outside world. While sometimes the intrusion of the environment is negligible, as it was manifested by glorious successes of, first, Newton's [1], and then Einstein's theories [2] when applied to the description of the Solar system, many other problems differ. One notable example is the theory of mechanical stability against resonances. Citing Mark Buchanan [3], *"The gradual acceptance of resonance as a mechanical phenomenon only took place as dramatic failures in bridges..."*. Further research of stability against external forces revolutionized mechanical engineer-ing. Another phenomenon omnipresent in mechanics is friction. Though initially it was understood differently, the phenomenon represents a typical example of dissipation, one of the cornerstones of statistical physics.

These two examples illustrate two ways a system can be "open"[1], i.e. not isolated – coherent and incoherent (or dissipative) coupling to external systems.

We call coupling to an external system **coherent** if it does not lead to continuous loss of information on a quantum state of the system we study. Coherent coupling is typically associated with either macroscopically controlled interactions with external degrees of freedom, or with coupling the reservoirs which do not posses properties of a thermodynamic bath. Examples of associated phenomena include coherent lasing [4], often described using Rabi model [5]; Floquet systems, where the effective Hamiltonian of the system at each point of time is controlled by an experimental setup (see, for example, Ref. [6, 7]); multiple applications of the measurement theory with controlled deterministic measurements, for example, the quantum measurement control theory [8], etc.

We call coupling to an external system **incoherent** or dissipative if it results in continuous (except if the system is in the steady state) loss of information due to either dissipation of macroscopic structure into microscopic degrees of freedom or inherently stochastic nature of external excitations. This type of coupling is ubiquitous, and thus, virtually every real physical system is an open one.

The concept of an external thermal bath or reservoir is crucial in thermodynamics, and while many results within the conventional thermodynamics can be achieved without addressing these issues, thermodynamics of open non-equilibrium systems is built entirely around external dissipation [9]. Other closely related examples include Brownian motion [10] and phenomena discribed by the Langevin [11] and Fokker-Planck [12] equations; non-Hermitian quantum physics [13]; quantum master equations [14], etc. While in the majority of cases dissipation is uncontrolled and can only be mitigated as undesirable and destructive, there exist examples of engineered dissipation [15] that can protect an open quantum

[1] As in some fields of physics the term "open system" is used to describe exclusively dissipative systems, to avoid confusion we stress that within this work we use the term "open system" is a broad scope, encapsulating all the systems that are in any type of contact with any other entities.

system from complete decoherence [16].

In this thesis, we focus on two examples of open quantum systems, one from each of the described above classes, namely Floquet quantum systems and dissipative quantum systems.

### 1.1.1 Floquet theory

Here, we provide a brief overview of the main results achieved utilizing the Floquet theory within quantum mechanics.

The Floquet theory, introduced by Gaston Floquet in 1883 [17] describes linear differential equations with periodic coefficients

$$\dot{\mathbf{x}}(t) = \mathbf{A}(t)\mathbf{x}(t), \tag{1.1}$$

where $\mathbf{A}(t+T) = \mathbf{A}(t)$ for a certain positive $T$. The Floquet theorem states that linearly independent solutions of the equation (1.1) arranged as a matrix $\mathbf{F}(t)$ (the so called fundamental matrix of (1.1)) can be expressed as

$$\mathbf{F}(t) = \mathbf{D}(t)e^{\mathbf{B}t}, \tag{1.2}$$

where $\mathbf{D}(t)$ is also periodic with a period $T$, and the matrix $\mathbf{B}$ is constant.

The theorem can be applied to quantum mechanics with a periodic Hamiltonian

$$H(t+T) = H(t). \tag{1.3}$$

The solution of the Schrödinger equation reads

$$|\Psi(t)\rangle = U(t)|\Psi(0)\rangle = \mathcal{T}e^{-i\int_0^t H(t)}|\Psi(0)\rangle, \tag{1.4}$$

where $\mathcal{T}$ is the time ordering operator (here and throughout the thesis $\hbar = 1$); $U(t)$ is the evolution operator with eigenstates $\{e^{-i\lambda_j}\}_j$ and corresponding eigenvectors $\{|\tilde{\psi}_j(t)\rangle\}_j$. Using the Floquet theorem we can show that independent solutions of the Schrödinger equation can be described bt the ansatz

$$|\tilde{\Psi}_j(t)\rangle = e^{-i\lambda_j t}|\tilde{\psi}_j(t)\rangle, \tag{1.5}$$

where $|\tilde{\psi}_j(t)\rangle = |\tilde{\psi}_j(t+T)\rangle$. $\lambda_j$ are called Floquet eigenvalues (or eigenfrequencies, or quasienergies). They are defined up to a shift $2\pi/T$. As follows from Eq. (1.2), it is always possible to define an effective constant Hamiltonian that fully reproduces Floquet evolution on stroboscopic times $nT$ as

$$H_{\text{eff}} = i \ln U(T), \tag{1.6}$$

though such Hamiltonian is not uniquely defined.

These results will be used further in sec. 2.1.2.

### 1.1.2 Dissipative quantum systems

The core approach to the analysis of dissipative quantum systems is to reduce the full description of the system of interest and its environment to an effective description of the system itself. In other words, given the Hilbert space $\mathcal{H}_{\text{S}} \otimes \mathcal{H}_{\text{E}}$, where indices S and E correspond to system and environment, and the total Hamiltonian,

$$H = H_{\text{S}} + H_{\text{E}} + H_{\text{I}}, \tag{1.7}$$

where the rhs summands correspond to the system, environment, and their interaction, our target is to achieve the description in terms of "S" degrees of freedom only. The motivation of such approach vary from the inability to track huge amount of environmental degrees of freedom, to the impossibility to even know the specifics of $H_{\text{E}}$ or $H_{\text{I}}$. Moreover, these specifics are often of no consequence for the actual dissipative role the environment plays in the system's dynamics. Thus, the route is, using certain approximations suitable for the particular case, to trace out the environmental variables and still achieve closed-form equations. The term "to trace out" originates from the reduced density matrix approach, where the system's density matrix reads $\varrho_{\text{S}}(t) = \text{tr}_{\text{E}}\, \varrho(t)$. Direct application of this approach leads to one of the master equations, the term used for density matrix formalism approach within dissipative quantum theory. There also exist other approaches, such as non-Hermitian Hamiltonian formalism, and the stochastic Hilbert space dynamics approach, however both of them are closely connected to master equations [14].

The most striking influence dissipation brings to a quantum system is the loss of unitarity of its evolution, or, in other words, breaking of the time-reversal symmetry. Unlike purely-oscillatory isolated quantum systems, dissipative systems always have some type of fixed point or limit cycle solutions. In the case of master equations, these solutions tend to be unique steady states (except for certain highly symmetric cases). The structure of a steady state depends heavily on the type of environment (or bath) to which the system is coupled. An important class of baths is thermal baths, i.e. such environments that are already in thermal equilibrium. The steady states of systems connected to thermal baths are, naturally, Gibbs distributions. Another typical class of environments is dephasing baths, the effect of which can be described as suppression of quantum coherence. These baths also typically lead the system to the equilibrium steady states – the maximally mixed states $\varrho_S \sim \mathbb{1}$. A more intriguing situation arises when a system cannot be described in equilibrium. This may happen because of either the emergence of non-equilibrium steady states or inability of the system to reach the steady state and get stuck in quasi-stationary/metastable dynamics.

## 1.2 Breakdown of Transport and Equilibration due to Localization

Like there are myriads of transport mechanisms, there also exist multiple phenomena that hinder transport in quantum systems. For example, a single particle in a random 1D potential is usually exponentially localized due to Anderson localization [18]. Multiple interacting particles in random potentials may follow the same fate due to many-body localization [19, 20]. Also, interacting particles can localize due to a number of dynamical and geometrical reasons in such setups as the Josephson junction chains [21], quasi-1D orbital compass models [22], and 2D lattice gauge models [23]. These constitute a class of phenomena called disorder-free many-body localization. Transport in electronic devices can be weakened due to strong repulsion between the carriers of charge – the phenomenon referred to as the Coulomb blockade [24], or by the formation of a large energy gap in Mott insulators [25]. Both single-particle [26] and many-body [27, 28] systems can be localized due to a macroscopic number of compact localized states related to flat bands.

All of these phenomena are being extensively studied, but most of the existing research is based on the isolated systems approaches. In this thesis, we study aspects of Anderson localization, many-body localization, nonlinear delocalization, and related effects in open systems.

### 1.2.1 Anderson localization

In 1958 Philip W. Anderson studied the transport of a single particle in disorder medium [18]. He was initially motivated by Feher's experiment where it was shown that spin diffusion among donor electrons in silicon is negligible. Anderson's goal was to construct a minimal model adequately simulating multiple possible sources of disorder and irregularity typical for most macroscopic materials. He considered a single particle on a tight-binding chain

$$H = \sum_i \epsilon_i c_i^\dagger c_i - \sum_{i,j} V_{ij} c_i^\dagger c_j + \text{h.c.}, \tag{1.8}$$

where $c_i$ are annihilation operators, and disorder is encoded in random site-dependent local field values $\epsilon_i$ that are drawn independently and uniformly from $[-W, W]$. We call $W$ the strength of disorder. Anderson considered hoppings that decay faster than $V/|i-j|^3$, and showed that bellow a certain finite value (that is of the order of $1$) of $V/W$, all the eigenstates of (1.8) are exponentially localized, so at $i \to \infty$

$$|\Psi_i| \sim e^{-|i|/L_{\text{loc}}}, \tag{1.9}$$

where $i$ is the spatial coordinate and $L_{\text{loc}}$ is the energy-dependent length. The effect Anderson uncovered took after his name and today it is called Anderson localization (AL). Several years after Mott (who would latter share the Nobel prize with Anderson) and Twose generalized the results. Citing their paper [29], "...we reach the conclusion that all states in a one-dimensional lattice may be localized. This may correspond to the theorem that in one dimension any potential hole, however small, leads to a bound state." Thus, it was established that all the eigenstates at any value of $W > 0$ are localized in 1D. Moreover, all the results are qualitatively the same if disorder is non-uniform, as long as it is not heavily-tailed and uncorrelated. The universality of AL in 1D, i.e. its appearance regardless of the weakness of disorder, shows that AL is an inherently quantum interference phenomenon that cannot be explained by classical trapping.

What this means for transport is that any evolving compact wave packet in such a system will, first, spread, but then halt and cease from escaping its localization volume. The width of the wave packet then becomes of the order of the localization length [30]. Thus AL can be diagnosed by measuring an initially localized wave packet. Let us consider a wave packet's second moment

$$m_2(t) = \sum_i (i - \sum_i i|\Psi_i(t)|^2)^2 |\Psi_i(t)|^2. \tag{1.10}$$

and track its evolution. In the case of AL, saturation will occur at finite time scales (local mixing time). But, in principle, $m_2$ can be misleading as a localization indicator, as it may reach macroscopic values if the wave packet is localized in several distant, but confined areas. One of the surgically precise quantifiers is the participation ratio,

$$p(t) = \left( \sum_i |\Psi_i(t)|^4 \right)^{-1}, \tag{1.11}$$

which shows the portion of the space covered by the wave packet. When averaged over time $\lim_{t\to\infty} 1/t \int_0^t \mathrm{d}t_1\ p(t_1)$ it can be considered an order parameter of the Anderson localized phase, and it stays finite (even in thermodynamic limit) while in this phase. This is strikingly different from translationally invariant systems, where the Bloch theorem guarantees extended states and unbound wave packets' spread.

As for directional transport, (1.9) guarantees that for a wave packet with the center of mass at $x$, at any time $t \leq \infty$ the mean displacement is confined, so $\sqrt{< x^2 >} < \mathcal{O}(L_{\mathrm{loc}})$, where averaging is done over time and disorder realizations. Thus, there is no diffusion (which implies $\sqrt{< x^2 >} \sim t$), and in a system large enough (with a length larger than $L_{\mathrm{loc}}$) a wave packet launched from one side of the system can not reach the other side. It also follows that conductivity in AL phase tends to zero. Indeed, through the Einstein's relation conductivity is proportional to the diffusion coefficient $\sigma \sim D$, and $D \sim \lim_{t\to\infty} \sqrt{x(t)^2}/t = 0$. One can also show that $\sigma = 0$ directly from the structure of the eigenstates (1.9) using Kubo formula [31].

From the point of view of spectral statistics of eigenenergies and the Bohigas-Giannoni-Schmit conjecture [32], AL systems behave as integrable systems. Indeed, while extended states are generally all coupled under a parametric variation, localized states are independent up to exponentially small corrections. Thus, there is no level repulsion, and level splitting statistics is close to Poissonian, approaching it at $W \to \infty$.

First compelling proves of AL in higher dimensions came from numerics. In Ref. [33], strong indication of localization in 2D was demonstrated. But the most influential results came later in the paper by the "gang of four" [34]. The authors developed a scaling theory of conductivity. Let us consider conductance as a function of the system size $N$. If the conductance is large, classical ohmic scaling is applicable and we write

$$g(N) \sim N^{d-2}, \tag{1.12}$$

where $d$ is dimensionality of the system. In the localized phase conductance is suppressed by localization, and in any dimension reads

$$g(N) \sim e^{-N/L_{\mathrm{loc}}}. \tag{1.13}$$

The authors considered the scaling function

$$\beta(g(N), N) = \frac{\mathrm{d}\ln(g(N))}{\mathrm{d}\ln(N)}, \tag{1.14}$$

and used the single parameter scaling assumption $\beta(g(N), N) = \beta(g(N))$, and thus $\beta(g)$ is assumed to be a continuous function. Using (1.12) and (1.13) one can show that

$$\beta(g) \sim \begin{cases} d-2, & g \to \infty, \\ \log(g), & g \to 0. \end{cases} \tag{1.15}$$

The resulting schematic behavior is depicted in Fig. 1.1. In the localized limit ($g \to 0$) $\beta$ always decreases as $\sim \log(g)$. In the opposite limit the asymptotic values depend on the dimensionality and are negative in 1D, zero in 2D, and positive in 3D. Thus, while in 1D and 2D $\beta$ never reaches positive values and thus extended states are not supported, in 3D (and all subsequent dimensions) this transition does occur. There exists a critical value of the disorder strength, exceeding which brings the system from conducting to AL phase. Moreover, in higher dimensions under the same disorder strength, generally, both localized and extended states coexist. Those states are separated at certain energy scales called the “mobility edges”.

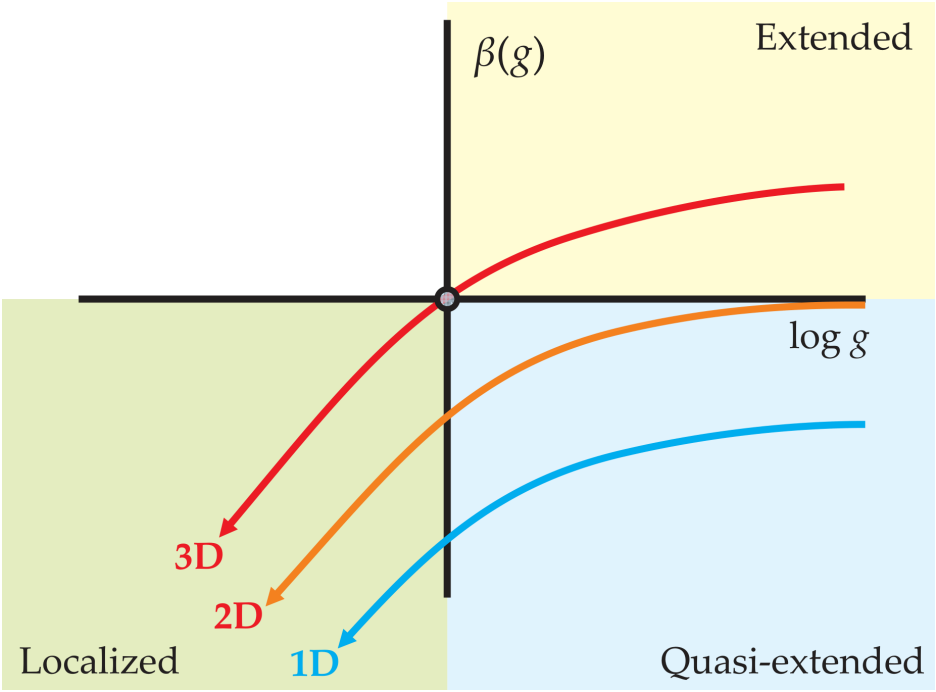

Figure 1.1: Modified Figure 4 from [35]. Schematic picture of the Anderson localization scaling function as a function of conductance for different dimensions. Note that the position of the vertical axis is not uniquely defined.

Let us comment on why both regimes are assumed to coexist in (1.15) while existence of transition itself is studied. Eq. (1.12) is correct even in localized phase as the maximal values of conductivity are achieved at the limit where $L_{\text{loc}} >> N$ and ohmic behavior is restored. This is shown as quasi-extended phase in Fig. 1.1. As for (1.13), the fact that localized phase always exists in any dimension for sufficiently large $W$ is rigidly proven [36]. We also note that this proof of the absence of extended states in 2D and existence in $d > 2$ is based entirely on the single parameter scaling assumption. So far, it was only proven to be true in systems with Cauchy disorder distribution (Lloyd model [37]) in Ref. [38]. Thus, the validity of the results in $d > 1$ still constitutes an open

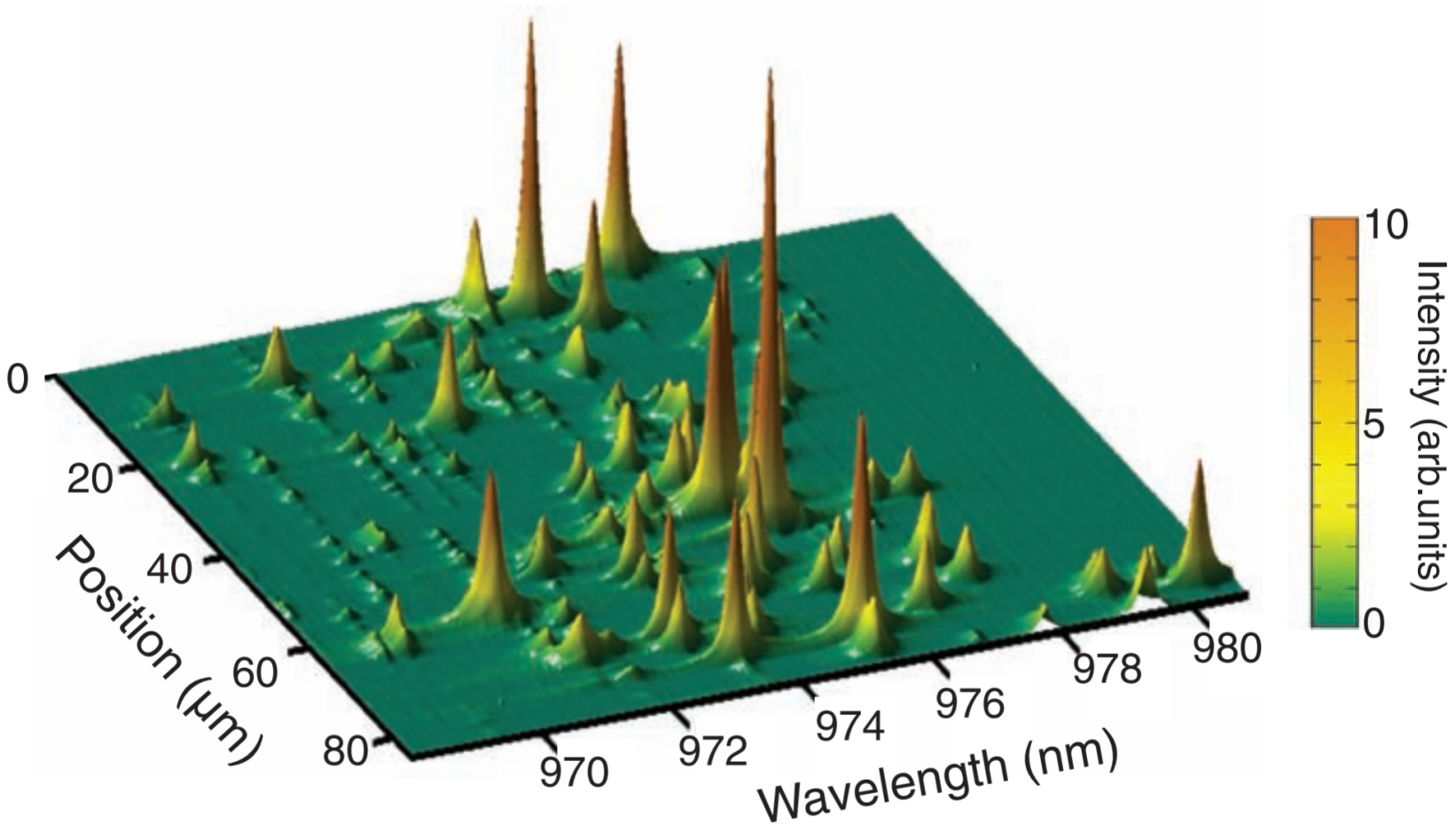

Figure 1.2: Figure 2 from [41]. Spectral signature of Anderson-localized modes in photonic crystal waveguides.

question [39, 40].

AL has a long history of experimental study. It has been experimentally observed in Bose-Einstein condensates of ultracold atomic gases loaded into optical potentials. The time evolution of a wave packet was used to prove and quantitatively characterize AL [42, 43]. Numerous further experimental studies of AL employed photonic crystal waveguides [41] (see Fig. 1.2), light [44, 45], microwaves [46], and ultrasound [47, 48].

### 1.2.2 Many-body localization

Many-body localization (MBL) is an extension of Anderson localization [18] into the realm of many-body quantum systems [19, 20]. The related research discusses localization transitions in presence of interaction. In addition to their fundamental importance for quantum thermodynamics and information, these systems are perspective from the technological point of view, e.g., most manufactured quantum computers are based on arrays of interacting superconducting qubits [49]. MBL systems are in the focus of the current research in theoretical and experimental quantum physics.

A conventional MBL model is an open-ended chain of $N$ (an even number) sites occupied by $N/2$ spinless fermions. The fermions are subjected to a random on-site potential $h_l$, $l = 1, \ldots, N$ and interact with each other when located on neighboring sites. The model Hamiltonian has the form,

$$H = -J \sum_{l=1}^{N} \left( c_l^{\dagger} c_{l+1} + c_{l+1}^{\dagger} c_l \right) + U \sum_{l=1}^{N} n_l n_{l+1} + \sum_{l=1}^{N} h_l n_l, \tag{1.16}$$

where $c_l^{\dagger}$ ($c_l$) creates (annihilates) a fermion at site $l$, and $n_l = c_l^{\dagger} c_l$ is the local particle number operator. Values $h_l$ are drawn from a uniform distribution on the interval $[-h, h]$. Note, that the same model is often used in terms of Heisenberg spin-one-half chains, which is given by performing Jordan-Wigner transformation

$$c_l = -\prod_{i}^{l} \sigma_i^{(z)} (\sigma_l^{(x)} - i\sigma_l^{(y)}), \tag{1.17}$$

and reads

$$H = -J \sum_{l=1}^{N} \left( \sigma_l^{(x)} \sigma_{l+1}^{(x)} + \sigma_l^{(y)} \sigma_{l+1}^{(y)} \right) + U \sum_{l=1}^{N} \sigma_l^{(z)} \sigma_{l+1}^{(z)} + \sum_{l=1}^{N} h_l \sigma_l^{(z)}. \tag{1.18}$$

The main statement of MBL theory in 1D is that there exists a critical value of disorder-to-interaction ratio $h_{\text{MBL}}$ (using normalization $J = U = 1$), at which the system undergoes a phase transition. Numerically, this value is estimated to be $h_{\text{MBL}} \backsimeq 3.6$ [50, 51].

There is a spectrum of definitions/quantifiers of this multi-faceted phenomenon developed to highlight the peculiar properties of MBL systems. The most straightforward property usually associated with localization phenomena is the absence of conductivity, or, to be more precise, its exponential suppression and eventual evanescence at the thermodynamic limit. The first works on MBL considered this transition in the context of varied temperature [19, 20]. Note that temperature here means average energy density, as thermodynamic temperature is ill-defined in 1D out-of-equilibrium systems. It was shown that there exists a critical

”temperature” below which conductivity disappears, and critical behavior reads

$$\ln \sigma(T) \sim -(T - T_{\text{cr}})^{-1/2}, \tag{1.19}$$

where $\sigma$ is conductivity, and $T$ is energy density. This result is strikingly different from the typical situation in crystal-like structures, as they support phonons, which, in thermodynamic limit, always support finite conductivity as for any energy difference between localized states, there can always be created a phonon with such energy. Thus, transport in MBL may be considered as a theory for materials with limited role of phonons.

More generally speaking, suppression of conductivity is one of the consequences of fail of the system to thermalize in the MBL phase. A simple indicator of this is behavior of the entanglement entropy. While in the thermal phase a typical eigenstate follows the volume law, meaning that it is proportional to the size of each subsystem, in localized phase a new, area, law emerges, so the entropy is proportional to the boundary between the subsystems [52, 53, 54]. This is a consequence of the locality, and thus inability of correlations propagate deep in the subsystems. Note that while MBL systems do, of course, follow the second law of thermodynamics, and thermalization out of the MBL regime does happen through tunneling, this process takes exponentially large times, as indicated, for example, by slow logarithmic growth of the entanglement entropy after an interaction quench [55, 56, 57, 58], so $S \sim \ln t$.

Slow spread of entanglement entropy can be seen as the sole reason of non-thermalization. As quantum evolution is unitary, no information about the initial conditions can be ever erased from the quantum state. Instead, this information gets erased from (quasi-)local physical observables. As entanglment spreads, non-local quantum correlations network makes it impossible to restore information on initial conditions without the application of global operations. This affects the system similarly to the action of external decoherence. In terms of the state, we say that a system thermalizes if

$$\lim_{t\to\infty} \langle \Psi(t)|\mathcal{O}|\Psi(t)\rangle = \operatorname{Tr} \mathcal{O} e^{-\beta H}/Z, \tag{1.20}$$

where $\beta$ is some positive number, $Z = \operatorname{Tr} e^{-\beta H}$, and $\mathcal{O}$ is a quasi-local observ-able, meaning that it occupies asymptotically small subspace of the whole operator space. This concept was used in first experimental confirmations of the existence of the MBL phase. A convenient observable for optical setups is the, so called, imbalance,

$$\mathcal{I} = \frac{\sum n_{\text{odd}} - n_{\text{even}}}{\sum n_{\text{odd}} + n_{\text{even}}}, \tag{1.21}$$

where $n_{\text{odd}}$, $n_{\text{even}}$ are particle numbers on odd/even sites. It is easy to see that in thermodynamic limit even and odd sites are identical as groups, and thus the expected thermal value is $\mathcal{I}_{\text{thermal}} = 0$. In Fig. 1.3 we see the result of the experimental measurement of the imbalance evolution for different values of interaction $\Delta$. If values are high enough (much higher than the critical value), the curves saturate at non-zero values.

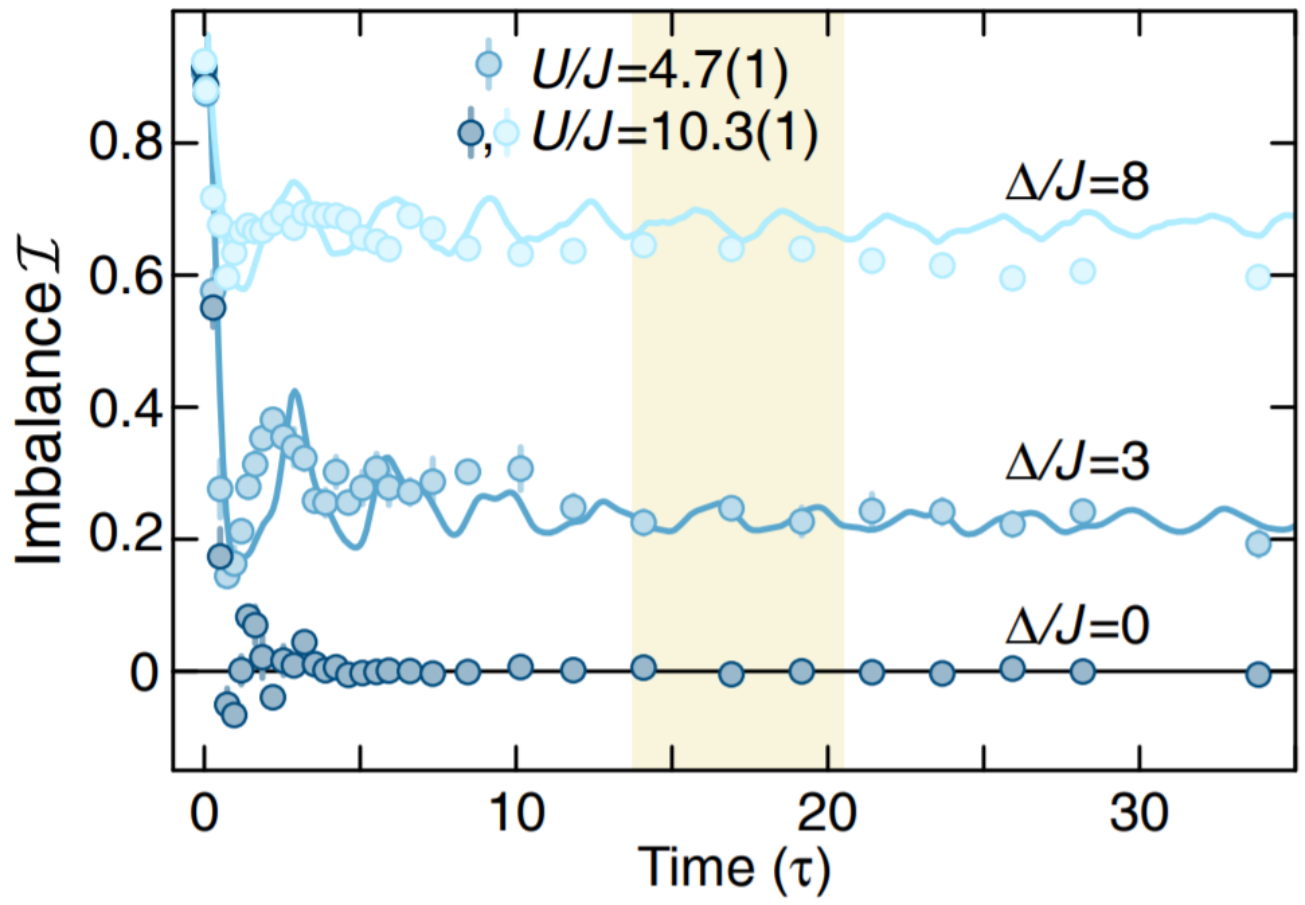

Figure 1.3: Figure from [7]. Imbalance (1.21) in a chain of interacting fermions as a function of time for different values of interaction strength. The initial conditions maximize the imbalance.

In terms of the Hamiltonian structure, a Hamiltonian can be called thermal if a typical (meaning each one up to an asymptotically diminishing portion) eigenstate is thermal, i.e. follows (1.20). This is called the eigenstate thermalization hypothesis (ETH), which is a sufficient condition for thermalization of a typical initial state. ETH fails in the MBL regime which was shown after an important bridge between MBL and integrability had been built while studying the existence of an extensive set of quasi-local integrals of motion (LIOM) [59]. It was proved that one can rewrite the Hamiltonian (1.16) in the form

$$H = \sum_{\alpha} h_{\alpha} I_{\alpha} + \sum_{\alpha,\beta} h_{\alpha,\beta} I_{\alpha} I_{\beta} + \sum_{\alpha,\beta,\gamma} h_{\alpha,\beta,\gamma} I_{\alpha} I_{\beta} I_{\gamma} + \dots, \tag{1.22}$$

where $h$ are some tensors, and $I_{\alpha}$ are the LIOMs. They are defined so they commute with each other, and the a priori (non-emergent unlike LIOMs) integrals of motion – the Hamiltonian and the total number of particles operator. They are shown to be quasi-local in the configuration space in the sense that their coefficients in the operator basis representation $I_{\alpha} = \sum_i c_i^{(\alpha)} \mathcal{O}_i$ drop exponentially with the distance between the support coordinates of $\mathcal{O}_i$. This representation still would not be local, if not for the additional condition that tensors $h$ of the second order and further also decay exponentially. The consequence of this structure for ETH is the proximity of eigenstates (formed as projectors on $I_{\alpha}$) to product states, which prevents possible thermalization. The existence of LIOMs is also reflected in spectral properties of MBL Hamiltonians [60, 61]. On the scale of a single (eigen)state of an MBL system, there is a class of related quantifiers and properties such as short-range correlations [50], and large fluctuations of local observables [62].

MBL has been observed in various experiments. One of the first manifestations of MBL was demonstrated in a system of fermions localized on quasiperiodic optical lattices [7]. Note, that while quasiperiodic disorder is technically not the same as truly random one, it is generally assumed not to cause qualitative difference in system dynamics. Moreover, experimental realization of truly random potentials is much more challenging than quasi-random ones, especially in cold-atoms experiments. MBL and its signatures were further confirmed in quantum simulators with programmable disorder [63], periodically driven systems [64], and 2D devices [65].

### 1.2.3 Nonlinear spreading in disordered systems

Nonlinearity can enter the stage of Anderson localized dynamics in several ways. One is interactions. While MBL that we discussed in section 1.2.2 is a purely linear quantum phenomenon, the modern research of it relies on relatively low particle density pictures. In the opposite limit, the mean-field approximation leads to effective nonlinear terms [66]. Another typical source of nonlinearity is the medium where particles move. Media without linear response are especially prevalent in optical systems, one example being Kerr-nonlinear wave-guide networks [67].

The question stands: does AL survive the introduction of nonlinearity, and if not, what is the resulting interplay? One may approach this question from two different perspectives – conductivity, or dynamical properties. Here we focus on the latter. In this context, the first fundamental question that should be answered is: What happens to an initially localized wave packet in a disordered nonlinear system? Naturally, the AL aspect prevents spreading far beyond the localization length scale, while nonlinearity transfers energy between linear modes, leading to spreading.

To make the discussion more concrete, but without substantial loss of generality, we consider the discrete nonlinear Schrödinger equation (DNLS), or the discrete Gross–Pitaevskii equation, which is one of the most utilized models in delocalization research. Consider a system on a chain with wave-function $\psi_n(t)$ governed by the evolution equation

$$i\dot{\psi}_n = \epsilon_n \psi_n - J\left(\psi_{n-1} - \psi_{n+1}\right) + g\left|\psi_n\right|^2 \psi_n, \qquad (1.23)$$

where $\epsilon_n$ is a random field, independently uniformly drawn from $[-W/2, W/2]$, $J$ is hopping rate, which defines the scale of time so we put $J = 1$, and $g$ is nonlinearity strength. Note that the sign of $g$ is irrelevant. Evolution (1.23) preserves both the total energy $H$, and the total norm $S = \sum_n |\psi_n|^2$. The values of g and $S$ are connected as evolution is defined by the product $gS$, so we put $S = 1$.

At the limit $g \to 0$ one recovers the Anderson model (1.8) with all the eigenstates being exponentially localized (1.9). In this case, an initially delta-localized wave packet consists of a number of eigenstates of the order of the localization volume (up to exponentially small corrections) and this volume confines possible expansion. Thus, after the local mixing time (that is usually of the order of $10^2$ for typical numerical experiments), the wave packet spread saturates at several localization lengths.

The addition of nonlinearity disturbs the linear pictures causing dynamical coupling and hybridization of linear modes that are also called normal modes. We can study this by transforming DNLS to the normal mode form. Let us expand the wave function in terms of linear Anderson modes (1.9), $\psi_n = \sum_\nu \phi_\nu \psi_{\nu,n}$. This leads to

$$i\dot{\phi}_\nu = E_\nu \phi_\nu + g \sum_{\nu_1,\nu_2,\nu_3} I_{\nu,\nu_1,\nu_2,\nu_3} \phi^*_{\nu_1} \phi_{\nu_2} \phi_{\nu_3}, \tag{1.24}$$

where $E_\nu$ is the linear energy level of $\phi_\nu$, and $I$ is the interaction integral that is given in terms of the linear normal mods as

$$I_{\nu,\nu_1,\nu_2,\nu_3} = \sum_n \psi_{\nu,n} \psi_{\nu_1,n} \psi_{\nu_2,n} \psi_{\nu_3,n}, \tag{1.25}$$

where we used the fact that in 1D all the eigenstates of a time-reversal symmetric short-range system can be gauged to be real-valued. The characteristic macroscopic time scales that will govern hybridization and, thus, wave packets spread are related to the following energy scales

- The total spectral width $\Delta$, which in this case is $s = 4 + W$ (as the eigenvalues of the linear model all lay in $[-W/2 - 2, W/2 + 2]$);

- Average spectral spacing of normal modes within the wave packet $\Delta E$. The number of these modes is proportional to the localization volume, which is of the same order as localization length [68]. Such states undergo the strongest level repulsion and thus are as distant as possible. Thus, $\Delta E \sim \Delta / L_{\text{loc}}$.

- Energy shift/renormalization due to non-linearity $\delta E$ can be estimated in the first order as proportional to nonlinearity strength $\delta E \sim g$.

Depending on the relation between these scales, three spreading regimes are possible. If $\delta E < \Delta E$ the nonlinear shift is smaller than the gaps between energy levels and thus is not sufficient to consistently couple normal modes, hence evolution is quasi-linear. If $\Delta E < \delta E < \Delta$, nonlinearity couples normal modes leading to resonances and excitation of previously uninvolved modes, and thus spreading. If $\Delta < \delta E$, nonlinearity can bring some states out of the linear spectrum, and as the support of the transport is purely linear, it leads to the formation of localized states intimately connected to discrete breathers [69]. This regime consists the described self-trapping and spreading similar to the previously described regime.

The self-trapping regime can be completely avoided by choosing moderate nonlinearity strength, such that $g < W + 4$. The quasi-linear regime starts with almost purely linear spreading up to the localization volume size within the mixing time scale. This may either, a) immediately bring a system to the nonlinear regime if $(4 + W) < gL_{\text{loc}}$, b) leave it in an intermediate quasi-linear regime up until rare resonances bring it to a nonlinear regime after a time scale that depends on $g$, or c) in terms of finite wave packets restore AL in a probabilistic manner [70, 71, 72]. In the nonlinear regime, coupling between normal modes leads to spreading. As the wave packet spreads, its average norm density decreases, which in turn lowers the local nonlinearity term which is proportional to $g/n$, where $n$ is the average number of sites occupied by the wave packet (which can be defined, for example, as the inverse participation ratio). Thus, as nonlinearity leads to spreading, and spreading reduces the nonlinear effect, the question stands: does spreading continue ad infinitum, or does it eventually halt?

Numerical results show universal subdiffusive spreading with the second moment $m_2 \sim t^{\alpha}$ with $\alpha \approx 1/3$ [74, 73], see Fig. 1.4. Though later, another intervening regime was discovered, where $\alpha \approx 1/2$ [75]. It occurs in-between the first near-linear stage, and the asymptotic $\alpha \approx 1/3$ regimes, though it can be observed for a long time by using strong disorder rate $W$, see Fig. 1.5.

Following [73, 69, 76], we demonstrate the origin of this behavior. Consider a wave packed that already has spread to a large size $\approx n$. The exterior of the wave packet of length $\sim L_{\text{loc}}$ is still not excited and thus does not support further spreading. There are two possible scenarios of what can happen to the mode in the exterior: it is either collectively heated by the wave packet, or if this can not happen, it will at some point come in resonance with a single triplet of modes in the packet and will be excited in this way. Let us consider the first case. The wave packet is chaotic and, thus, the phases can be considered random, incoherent, and possess a continuous spectrum [77]. Then, in the last term $\psi$ components are $\sim 1/\sqrt{n}$, and the summation and the interaction integral only depend on $L_{\text{loc}}$, but not on $g$ or $n$. Thus we find

$$i\dot{\phi}_\nu \approx E_\nu \phi_\nu + \frac{g}{n^{3/2}} f(t), \tag{1.26}$$

where $f(t)$ is a stochastic uncorrelated function with continuous spectrum, thus

$$< f(t) f(t') > \sim \delta(t - t'). \tag{1.27}$$

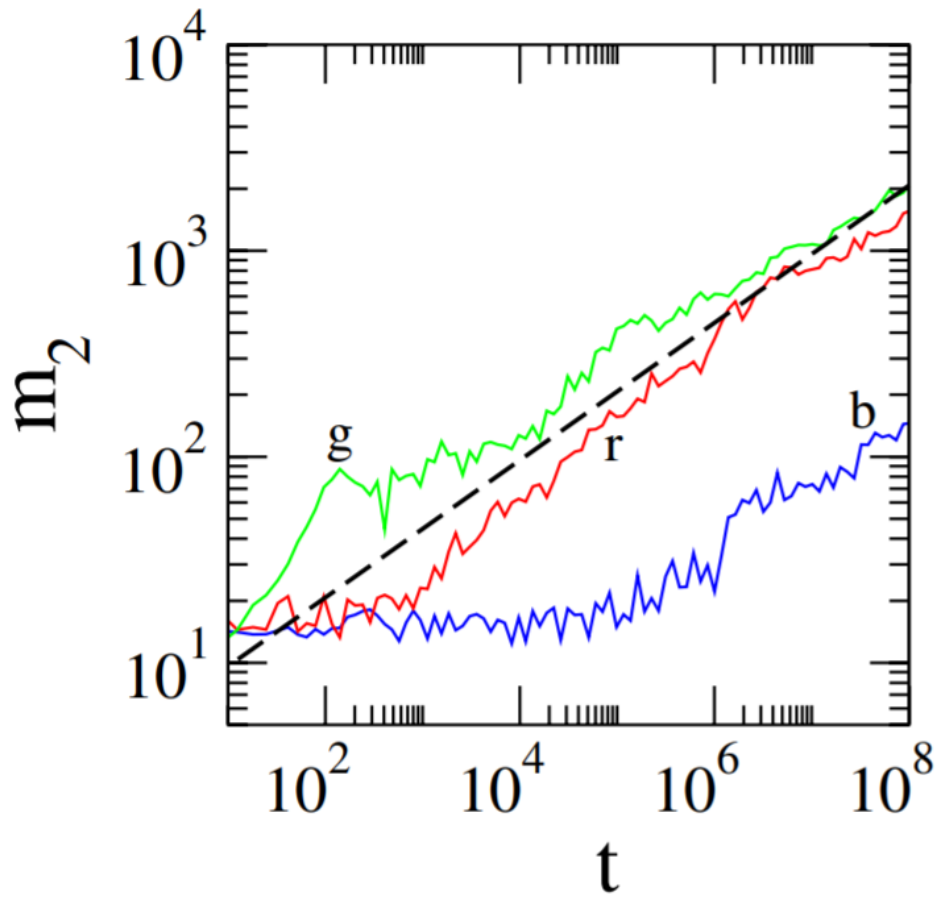

Figure 1.4: Modified figure from [73]. The second moment of a wave packet in DNLS against time for different parameter configurations. The dashed line corresponds to $m_2 \sim t^{1/3}$ spreading.

Integrating this equation one obtains

$$\phi_\nu(t) = e^{-iE_\nu t}(\phi_\nu(0) - i\frac{g}{n^{3/2}} \int_0^t e^{iE_\nu t'} f(t') \mathrm{d}t'). \tag{1.28}$$

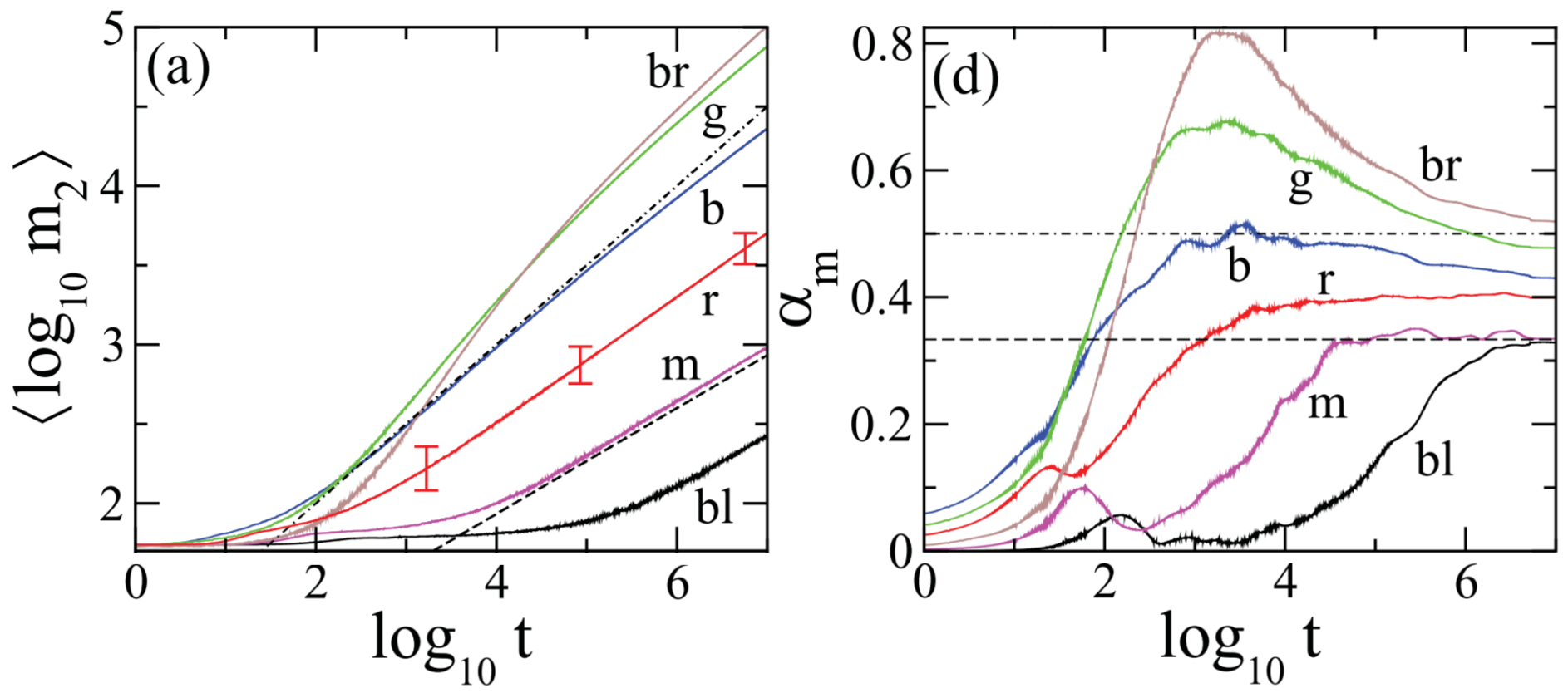

Figure 1.5: Modified figure from [75]. The second moment and the local subdiffusive exponent evolution for DNLS with different parameter configurations. Dashed lines correspond to $m_2 \sim t^{1/2}$ and $m_2 \sim t^{1/3}$ spreading.

As for the norm

$$|\phi_\nu(t)|^2 = |\phi_\nu(0)|^2 - 2i\frac{g}{n^{3/2}}\mathrm{Re}\phi_\nu(0)\int_0^t e^{iE_\nu t'}f(t')\mathrm{d}t' + \frac{g}{n^3}\int_0^t\int_0^t e^{iE_\nu(t'-t'')}f(t')f(t'')\mathrm{d}t'\mathrm{d}t''. \quad (1.29)$$

Integrating this using (1.27) and taking into account that $|\phi_\nu(0)|^2 \approx 0$, we find

$$|\phi_\nu(t)|^2 \sim g^2/n^3 t. \quad (1.30)$$

For the heated normal mode to reach the average norm of the wave packet $1/n$ it takes a typical time $T$ such that $1/n = g^2/n^3T$, thus $T = n^2/g^2$. This is the typical momentary diffusion time due to heating of a wave packet with length $n$, so we get $\frac{\mathrm{d}m_2}{\mathrm{d}t} \sim T \sim g^2/n^2 \sim g^2/m_2$, and solving this we get the subdiffusive law $m_2 \sim gt^{1/2}$.

This regime is called the strong chaos regime. However, as it has been mentioned above, it is an intermediate regime. When the packet spreads further, the assumption of fully chaotic behavior fails (though by introducing external dephasing, one recovers asymptotic strongly chaotic behavior). Strong chaos happens far away from any applicability of the perturbation theory. But in the opposite case, called the weak chaos, resonances happen in the vicinity of it. We use this fact to estimate the probability of a resonance. Consider the absolute value of the first-order perturbative correction to $\phi_\nu$

$$\begin{aligned} |\phi_\nu^{(1)}| &\approx gn^{-3/2}R^{-1}, \\ R &= \frac{I_{\nu,\nu_1,\nu_2,\nu_3}}{E_\nu + E_{\nu_1} - E_{\nu_2} - E_{\nu_3}}. \end{aligned} \tag{1.31}$$

The perturbation theory fails and a resonance happens when $|\phi_\nu^{(1)}| \approx |\phi_\nu^{(0)}| = n^{-1/2}$, thus when $R \sim g/n$. Values of $R$ are stochastic, and weakly mutually dependent, thus their distribution can be estimated as wide Poissonian distribution with probility density $\mathcal{P}(R) = \lambda e^{-\lambda R}$, $\lambda >> g/n$. Note, though, that this is probably the weakest part of this estimate. Thus the total probability of the resonance is $\int_0^{g/n} \lambda e^{-\lambda R} \mathrm{d}R \approx g/n$. Using this factor to modify (1.26), we find

$$i\dot{\phi}_\nu \approx E_\nu \phi_n u + \frac{g^2}{n^{5/2}} f(t), \tag{1.32}$$

and following the same procedure, we recover the weak chaos behavior $m_2 \sim gt^{1/3}$.

The validity of this estimate was confirmed with tests of its predictions for larger system dimensions [78], and different exponents of nonlinear terms which correspond to various $N$-body interactions [79]. Successful tests of systems with quasiperiodic (instead of random) potentials [80], and nonlinear versions of quantum kicked rotors [81, 82] yielded subdiffusion with $\alpha = 1/3$ as well, and revealed additional universality aspects of the observed process [76]. The largest times reached by these computations were $10^8 - 10^9$. In the case of one single disorder realization, a reported evolution for a Klein-Gordon chain reached time $10^{10}$ [83].

Recent relevant experimental results include granular chains [84], photonic waveguide lattices [45], light propagation in fiber arrays [85], and atomic Bose-Einstein condensates [86]. In particular, the latter case study the spatial extension of clouds of interacting $^{39}$K atoms revealed the destruction of AL through the onset of subdiffusion – an extremely slow process of wave packet spreading with its second moment $m_2 \sim t^\alpha$ with $\alpha < 1$.

Remarkably, a number of published arguments predict exactly the opposite outcome – that the spreading has to slow down [87, 70, 88, 89, 90], perhaps from a subdiffusive down to a logarithmic one [91]. These arguments use perturbation approaches which might (or might not) be of little help when chaotic dynamics hits. Various attempts to observe a slowdown were not successful but also limited by the computational horizons.

## 1.3 Outline

The thesis consists of two main parts, chapter 2 and 3. In chapter 2 we study a class of Flouqet quantum systems – discrete-time quantum walks (DTQW). In section 2.1 we introduce the model, explain its Floquet nature, motivate its benefits for research, and discuss its generic properties which will be relevant in what follows. In section 2.2 we study single-particle non-equilibrium regime of DTQW – Anderson localization. We provide a systematic and comprehensive study of localization length behavior for generalized unitary DTQW from both analytical and numerical points of view. In sections 2.3 and 2.4 we consider DTQW with a nonlinearity included as phenomenological mean-field many-body interaction. In section 2.3 we study nonlinear delocalization phenomenon with the model introduced in section 2.2. As the nonlinearity hybridizes the Anderson localized eigenstates, the ultimate fate of initially localized packets in disordered potentials is generally unclear. We numerically study this effect in DTQW. In section 2.4 we study super-localized soliton-like solutions formed in DTQW in the regime where only nonlinear transport is allowed. We study several types of such solutions and their stability.

Chapter 3 is dedicated to non-equilibrium properties of dissipative quantum systems. We start with an introduction of the Lindblad master equation formalism in section 3.1. We summarize the basic properties of this model, motivation for its use, and the reasons why the Lindblad equation is so widespread. In section 3.2 we describe a powerful numerical tool for attacking open many-body dissipative quantum systems in 1D – the time-evolving block decimation technique. We describe a highly parallel supercomputer implementation of this approach and benchmark its implementation against known from the literature results. With the help of this computational machinery, in section 3.3 we study the effects of MBL transition in dissipative systems with engineered dissipation. We show that both dynamical and non-equilibrium steady state features bear signatures of MBL. In section 3.4 we derive projected Lindblad master equation to study quasi stationary states and dynamics of dissipative quantum systems. Utilizing this equation we generalize the stochastic quantum trajectories approach to encompass numerical simulations of quasi-stationary evolution.

We finalize the thesis with final remarks in chapter 4.

# Chapter 2

# Localization and Delocalization in Discrete-Time Quantum Walks

## 2.1 Introduction

### 2.1.1 Discrete-time quantum walks

The notion of quantum walks (QW) can be intuitively introduced as a generalization of classical random walks to the quantum realm. A walker (now in a quantum state) may move on an arbitrary network, but instead of classical stochasticity, he is now governed by quantum interference and measurements. Two sub-classes of QW exist, continuous quantum walks represent scenario when a walker can continuously move without restrictions in time. Discrete-time quantum walks (DTQW) correspond to a situation where a walker can only move on discrete time steps, which leads to a unitary-map evolution. This thesis is dedicated to the latter. The origins and precedence of the concept of QW are currently being debated [92], and alongside this division, there exist several subfields of physics that heavily utilize QW.

One school of thought often follows the ideas described in the first paper that was completely dedicated to QW as we know them today, "Quantum random walks" by Aharonov, et al. [93]. This paper follows the previously introduced path of "quantization" of classical random walks. The walker becomes a wave function instead of a position (vector), and the evolution is governed by unitary maps that incorporate quantum "coins" in full analogy with the classical stochastic source. Interference leads to quantitatively different behavior with the most notable example being ballistic rather than diffusive transport. It was also shown that the mixing time of QW on a line is linear compared to quadratic of classical random walks; probability of absorption of QW with boundaries does not equal $1$ unlike that of the classical counterpart [94], and so on. Later, the connection between classical and (correlated) QW was shown to be even more profound [95].

From another hand, QW are also often traced back to Feynman's early proposal of a pre-qubit era quantum computer concept [96, 97, 98]. Related to this, QW are often studied in the context of quantum computing. QW might be used as high-level building blocks for quantum algorithms [99] that often provide significant speed-ups. The implementations include quantum NAND trees evaluation [100], search algorithms [101], graph traversal [102], etc. More generally, QW are shown to provide a complete basis for quantum computations [103, 104].

Another early mention of QW is by Godoy, et al. [105], where the term QW was introduced as a fusion of classical random walks and quantum effects to address a specific physical problem: one-dimensional tunneling diffusion. Later, QW became a staple model for studying various phenomena. This includes Dirac transport on lattice [106], chirality and bulk-boundary correspondence [107], topological phase transitions [108, 109, 110], localization and effect of disorder [111, 112, 113], etc.

There are several reasons to study physical phenomena' manifestations on QW. Some studies utilize them when the quantum generalization of classical random processes is required. Another important reason is a number of unique properties that arise from the discrete-in-time nature of the dynamics. One example is the independence of localization length on the position in the spectrum for the uncorrelated Anderson localization setup as we show in Sec. 2.2. For numerical purposes, discreetness in both time and space, and a relatively simple structure of the states provide a unique opportunity to reach previously unapproachable performance and simulation scales. We discuss this further in Sec. 2.1.4. DTQW were implemented in numerous experimental setups, including quantum optical systems [114], ion traps [115] and nuclear magnetic resonance systems [116].

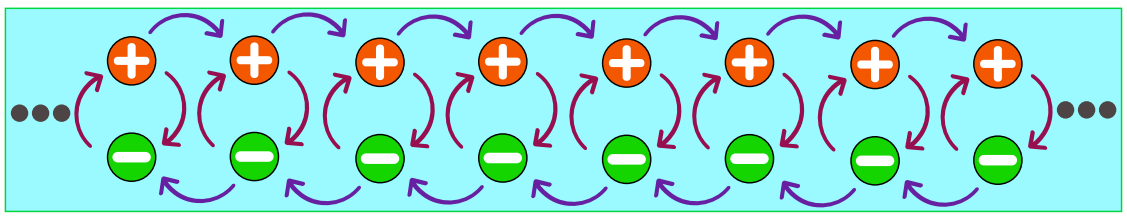

Figure 2.1: The schematics of the discrete-time quantum walk. The arrows indicate the directions of a single step transfer.

#### 2.1.1.1 Model

Consider a single quantum particle with an internal spin degree of freedom, hopping on a one-dimensional tight-binding lattice. The dynamics of the particle is characterized by a time- and lattice site-dependent *two-component wave function* $\hat{\psi}_n(t) = \{\psi_{+,n}, \psi_{-,n}\}$, which is defined at discrete times ($t$) and on lattice sites ($n$). Assume that it evolves under the influence of some periodic Floquet driving. Then, its evolution can be mapped onto a sequence of unitary maps. As a result, the components of the wave function transfer to the right or to the left, and *the quantum-mechanical amplitudes* of such hopping are determined by quantum coin operators acting independently on each site (Fig.2.1).

The DTQW has two ingredients – a quantum coin, and a shift (register) operation. Quantum coins are typically chosen mainly from single parameter (angle) operator distributions, including the well known case of the Hadamard coin [117, 118]. It has been shown [119] that generalized coins allow for improved control and optimization. Such a general coin can be implemented in an optical setup utilizing beam splitters [120], and some parameters may be controlled in other setups [121]. A single site coin operator $\hat{U}$ is a general unitary matrix of

rank 2:

$$\hat{U} = \begin{pmatrix} a & b \\ c & d \end{pmatrix} = e^{i\varphi} \begin{pmatrix} e^{i\varphi_1}\cos\theta & e^{i\varphi_2}\sin\theta \\ -e^{-i\varphi_2}\sin\theta & e^{-i\varphi_1}\cos\theta \end{pmatrix}. \tag{2.1}$$

A generic coin operator is completely determined by four angles $\varphi, \varphi_1, \varphi_2$ and $\theta$. As it will become evident below, they can be also related to a potential energy, external and internal synthetic flux, and a kinetic energy respectively. The coin operator can be implemented as an arbitrary two-level system subject to time-dependent perturbations of different durations. The coupling between the coin operators $\hat{U}_n$ and the quantum particle has a form, $\hat{S} = \sum_n |n\rangle\langle n| \otimes \hat{U}_n$, where the angles $\varphi, \varphi_1, \varphi_2$ and $\theta$ can vary from site to site. There exist several ways to implement a coin operator [122, 123, 124, 125].

The transfer operator is defined as

$$\hat{T}_\pm = \sum_n |n\rangle\langle n+1| \otimes |\mp\rangle\langle\mp| + |n\rangle\langle n-1| \otimes |\pm\rangle\langle\pm|, \tag{2.2}$$

and we will always use $T_+$. Thus, the discrete-time quantum walk is described as the sequence of successive $\hat{S}$ and $\hat{T}_+$ operators. The schematic of such dynamics is shown in Fig.2.1, and the equations read

$$\hat{\psi}_n(t+1) = \hat{M}_+\hat{\psi}_{n-1}(t) + \hat{M}_-\hat{\psi}_{n+1}(t), \tag{2.3}$$

where the matrices $\hat{M}_\pm$ for the translationally invariant case of identical quantum coins are written explicitly as

$$\hat{M}_+ = \begin{pmatrix} e^{i(\varphi_1+\varphi)}\cos\theta & e^{i(\varphi_2+\varphi)}\sin\theta \\ 0 & 0 \end{pmatrix}, \tag{2.4}$$

and

$$\hat{M}_- = \begin{pmatrix} 0 & 0 \\ -e^{i(\varphi-\varphi_2)}\sin\theta & e^{i(-\varphi_1+\varphi)}\cos\theta \end{pmatrix}. \tag{2.5}$$

The resulting unitary eigenvalue problem can be solved by finding the eigenvectors $\{\hat{\psi}_n\}$ with $\hat{\psi}_n(t+1) = e^{-i\omega}\hat{\psi}_n(t)$ and the eigenvalues $e^{-i\omega}$, where $\omega$ is the eigenfrequency of the discrete-time quantum walker.

#### 2.1.1.2 Diagonalization of ordered DTQW

In the absence of spatial disorder all coin operators are identical, and the unitary map equations (2.3) are invariant under discrete translations. The eigenvectors are then given by plane waves $\hat{\psi}_n = e^{ikn}\hat{\psi}(k)$, where $k$ is the wave vector, and $\hat{\psi}(k)$ is the two-component eigenvector in the Bloch basis (also called polarization vector). The quantum particle dynamics is then fully determined by the dispersion relation

$$\cos(\omega - \varphi) = \cos\theta \cos(k - \varphi_1) \ . \tag{2.6}$$

The spectrum consists of two bands. The polarization vectors are obtained as

$$\frac{\psi_{+,k}}{\psi_{-,k}} = e^{i(\varphi_2 - \varphi_1)} \frac{\cos\theta - e^{i([\omega(k)-\varphi]-[k-\varphi_1])}}{\sin\theta} \ . \tag{2.7}$$

It follows that $\theta$ is a kinetic energy parameter that controls the width of each band from its maximal value $\pi$ for $\theta = 0$ to a dispersionless (flat) band with width zero for $\theta = \pi/2$. The angle $\varphi$ corresponds to a potential energy term which renormalizes the frequency $\omega$. The angle $\varphi_1$ renormalizes the wave number $k$ similar to a flux threading a large one-dimensional chain with periodic boundary conditions. The angle $\varphi_2$ instead relates to an internal synthetic flux which impacts the phase shift between the two components of the polarization vector only.

For a generic value of $\theta$ the two bands have finite width and are gapped away from each other (e.g. blue lines, $\theta = \pi/4$ in Fig.2.2). For $\theta = 0$ the two bands turn into straight lines which cross, leading to a vanishing gap and a one-dimensional Dirac-like cone (black lines in Fig.2.2). Finally for $\theta = \pi/2$ the spectrum $\omega(k) = \varphi \pm \pi/2$ consists of *two flat bands* (Fig.2.2). This corresponds to macroscopic degeneracy. Linear combinations of Bloch eigenstates are easily shown to allow for *compact (2-sites) localized states* residing on a pair of neighbouring sites $m$ and $m+1$:

$$\hat{\psi}_n = \frac{1}{\sqrt{2}} \begin{pmatrix} 1 \\ 0 \end{pmatrix} \delta_{n,m} + \frac{1}{\sqrt{2}} \begin{pmatrix} 0 \\ ie^{-i\varphi_2} \end{pmatrix} \delta_{n,m+1} \ . \tag{2.8}$$

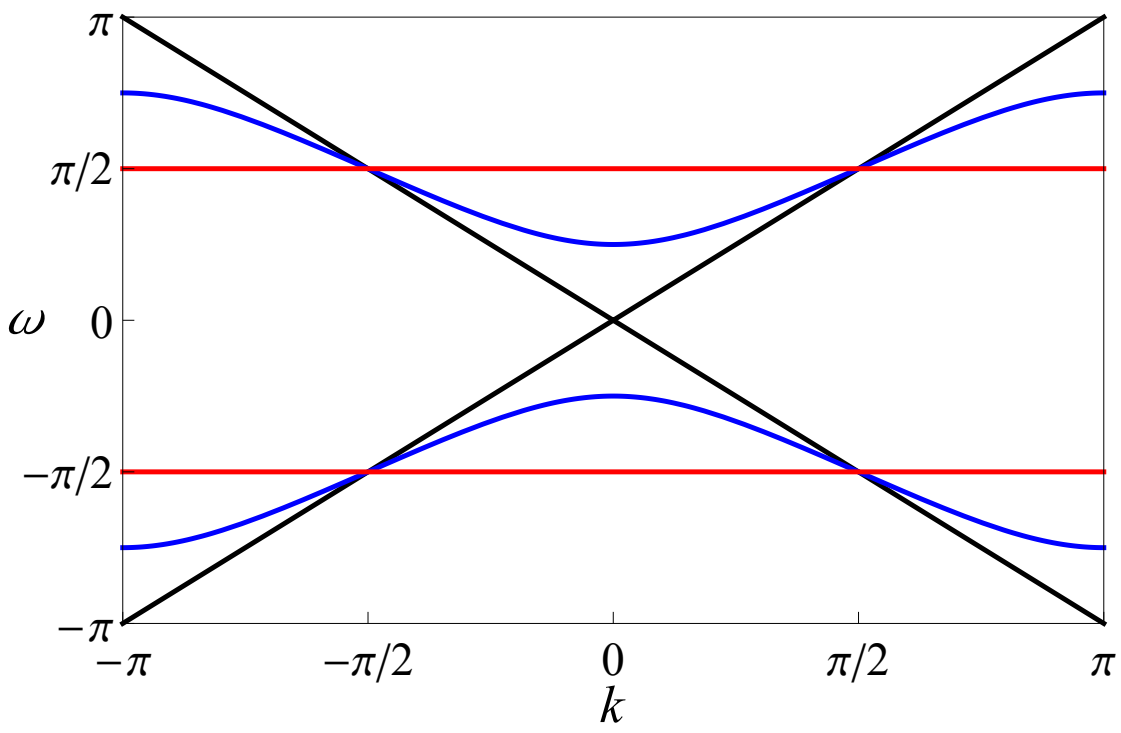

Figure 2.2: The dispersion relation $\omega(k)$ for different values of $\theta$: $\theta = 0$ (no gap, black solid line), $\pi/4$ (finite gap, blue line), $\pi/2$ (flat bands, red lines). Here, $\varphi = \varphi_1 = 0$.

#### 2.1.1.3 Symmetries of DTQW

For arbitrary unitary coin operators the quantum walk (2.3) possesses *bipartite* or *sublattice symmetry*, since even/odd sites are connected to odd/even sites only. The bipartite lattice symmetry implies that the spectrum is invariant under frequency shifts $\omega \to \omega + \pi$ with the following transformation rules for eigenvectors:

$$\{\omega \ , \ \hat{\psi}_n\} \to \{\omega + \pi \ , \ (-1)^n \hat{\psi}_n\}. \tag{2.9}$$

Note that any arbitrary spatial disorder in the coin operators is preserving the sublattice symmetry.

For site-independent angles $\varphi_n \equiv \varphi$ and $\varphi_{1,n} \equiv \varphi_1$ the quantum walk (2.3) possesses an additional *particle-hole symmetry*, which implies that the spectrum is invariant under frequency shifts $\omega \to -\omega + 2\varphi$ with the following transformation rules for the eigenvectors:

$$\{\omega \ , \ \psi_n^+ \ , \ \psi_n^-\} \to \{-\omega + 2\varphi \ , \ \psi_n^{+*} \exp\left(2i(\varphi_2 - \sum_{m=-\infty}^{n} \varphi_{1,m})\right) \ , \ \psi_n^{-*} \exp\left(-2i \sum_{m=-\infty}^{n-1} \varphi_{1,m})\right)\} \tag{2.10}$$

For more information about symmetries of DTQW, particularly in presence of noise, see [126].

### 2.1.2 DTQW as Floquet open system

DTQW evolve through a sequence of periodically applied coin and shift operators at discrete time steps thus making the dynamics periodic in time. This is an indicator that DTQW is a Floquet driven system. The full unitary evolution operator reads

$$\hat{\psi}(t+1) = \hat{V}\hat{\psi}(t) = \hat{T}\hat{S}\hat{\psi}(t), \tag{2.11}$$

which can be implemented as time-periodic Hamiltonian evolution over a unit (in appropriate units) time step

$$\hat{V} = e^{-i\int_0^1 H_{\text{eff}}(t)\mathrm{d}t}, \tag{2.12}$$

where we put $\hbar = 1$ and imply time-ordered exponential integration. There are multiple ways to introduce $H_{\text{eff}}$. An instructive and practical way is to implement the coin operator component through a sequence of Pauli matrix $\delta$-pulses, and the shift operator-associated Hamiltonian,

$$H_{\text{eff}}(t) = \begin{cases} aH_{c_i} & t_{i-1} \le t \le t_i : \quad i = 1,2,3,4 \\ H_T & t_4 < t \le 1 \end{cases}, \tag{2.13}$$

where $t_0 = 0$, $at_i = 1$ and $t_i \to 0$ for $i = 1,2,3,4$. Using decomposition of the coin matrix (2.1)

$$\begin{aligned}\hat{U} &= \begin{pmatrix} a & b \\ c & d \end{pmatrix} = e^{i\varphi}\begin{pmatrix} e^{i\varphi_1}\cos\theta & e^{i\varphi_2}\sin\theta \\ -e^{-i\varphi_2}\sin\theta & e^{-i\varphi_1}\cos\theta \end{pmatrix} \\ &= e^{i\varphi}\begin{pmatrix} e^{i\frac{(\varphi_1+\varphi_2)}{2}} & 0 \\ 0 & e^{-i\frac{(\varphi_1+\varphi_2)}{2}} \end{pmatrix}\begin{pmatrix} \cos\theta & \sin\theta \\ -\sin\theta & \cos\theta \end{pmatrix}\begin{pmatrix} e^{i\frac{(\varphi_1-\varphi_2)}{2}} & 0 \\ 0 & e^{-i\frac{(\varphi_1-\varphi_2)}{2}} \end{pmatrix} \\ &= e^{i\sigma^{(0)}\varphi}e^{i\sigma^{(3)}\frac{(\varphi_1+\varphi_2)}{2}}e^{i\sigma^{(2)}\theta}e^{i\sigma^{(3)}\frac{(\varphi_1-\varphi_2)}{2}},\end{aligned} \tag{2.14}$$

where $\sigma^{(\alpha)}$ are Pauli matrices with $\alpha = 0$ corresponding to the identity matrix. Thus we have

$$H_{c_1} = -\sigma^{(0)}\varphi,\ H_{c_2} = -\sigma^{(3)}\frac{(\varphi_1 + \varphi_2)}{2},$$
$$H_{c_3} = -\sigma^{(2)}\theta,\ H_{c_4} = -\sigma^{(3)}\frac{(\varphi_1 - \varphi_2)}{2}. \quad (2.15)$$

The Hamiltonian representation of $H_T$ is only local in $k$-space (quasi-momentum space). Using $\hat{\psi}_n = e^{-ikn}\hat{\psi}_k$, we get in $n$-space

$$\hat{T} = \begin{pmatrix} |n\rangle\langle n+1| & 0 \\ 0 & |n-1\rangle\langle n| \end{pmatrix} = \int \mathrm{d}k \begin{pmatrix} e^{-ik} & 0 \\ 0 & e^{ik} \end{pmatrix} |k\rangle\langle k|$$
$$= \int \mathrm{d}k e^{-i\sigma^{(3)}k} \otimes |k\rangle\langle k|. \quad (2.16)$$

Here the translation and mixture of orbital degrees of freedom encoded by $k$ are coupled to spin degrees of freedom, indicating spin-orbital coupling [108]. In $k$-space we simply get

$$\hat{T} = e^{i\sigma^{(3)}k}, \qquad H_T = \sigma^{(3)}k. \quad (2.17)$$

Note that this Hamiltonian corresponds to the zero-mass Dirac Hamiltonian [127].

The periodically driven nature of DTQW leads to many associated properties. For example, DTQW have a variety of Floquet topological phases [128, 107, 129]. For more information on DTQW as a Floquet system see [130].

### 2.1.3 Basic physical properties of DTQW

One of the key characteristics of DTQW compared to their classical counterpart, random walks, is the ballistic transport regime in absence of disorder. In Fig. 2.3 we see the spatial profile of the probability distribution of DTQW evolved from a symmetric state localized on a single site in the center of the system. The interference structure of the wave packet, as well as the ballistic transport, is intimately connected to the typical quantum superposition nature of single-particle quantum systems, also manifesting itself in DTQW [93].

Another key feature of quantum systems, entanglement, is also present in DTQW. By constructing initial conditions of different symmetry classes, one may construct both bosonic- and fermionic-like entangled DTQW systems [131]. Moreover, one may study entanglement even in terms of "single-particle" quantum walks by considering the Hilbert space as the product of the coin and the position spaces [132, 133]. In particular, it is shown that the coin-space entanglement of DTQW with $\theta = \pi/4, \varphi = \varphi_1 = \varphi_2 = 0$ (so-called Hadamard coin) with local initial conditions always lead to saturation of the entanglement entropy to the same value. Other many-body properties of DTQW include entanglement-controlled transport [134] and the fast growth of both classical and quantum correlations [135]. In two-body disordered interacting DTQW, emergence of two characteristic localization scales, enhancement of localization length due to interaction, and signs of many-body localization formation were reported [136].

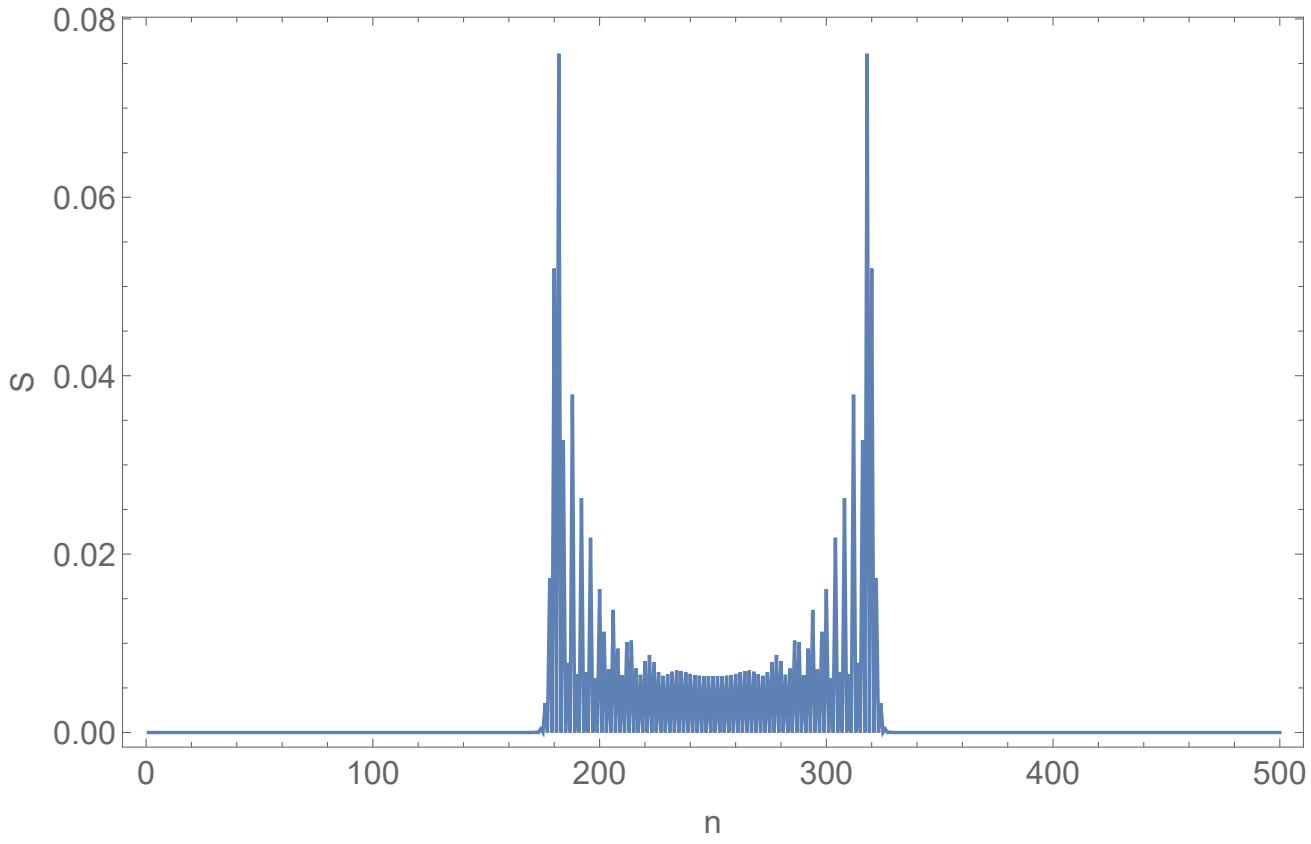

Figure 2.3: Profile of the norm $S_n = |\psi_{n,+}|^2 + |\psi_{n,-}|^2$ of the wave packet evolved for $t = 100$ starting from the initial conditions localized at the center of the system with the spinor $(1/\sqrt{2}, i/\sqrt{2})$. The coin paramters are $\theta = \pi/4$, $\varphi = \varphi_1 = \varphi_2 = 0$.

Other properties, including Anderson localization [137, 138] (section 2.2), nonlinear delocalization [139] (section 2.3), and support of moving and stationary breathers [140] (section 2.4) are described in the further sections.

### 2.1.4 Discrete time and DTQW as a computational toolbox

DTQW provide a unique framework for computationally demanding numerical studies. As we discussed in the previous, as well as the following sections, DTQW are physically representative of quantum systems and nonlinear maps. This comes with an additional advantage of being discrete both in space in time. Indeed, consider a single time step of DTQW evolution. The first part consists of applying the coin (2.1) to every spinor component. As those "coins" commute, numerically, this corresponds to the independent application of coins to certain parts of the wave function, which allows complete parallelization and vectorization. Moreover, the application of a 2x2 matrix is effectively a number of multiplications, additions, and local memory accesses, which are all cheap operations for modern computers. The second step, application of the shift operator (2.2), is even more basic and consists either of simple registry shifts or re-indexing of an array depending on the approach to the wave function storage.

Note, that these operations do not include any approximation, and thus are numerically exact, i.e. the only source of error is the finite algebra of the computers' float numbers. This especially plays a decisive role for long-time computations. In Hamiltonian systems, to maintain reasonable accuracy and conservation of the integrals of motion, one has to apply demanding numerical schemes with fine discretization in time (for example, see [141]). In DTQW, the only source of errors in any values, including integrals of motion, is the relative cut-off value of the computer algebra. For example, as we discuss in Sec. 2.3.4, we were able to conduct calculation up to evolution time $t = 2 \cdot 10^{12}$, while only accumulating an error of order $10^{-4}$, which roughly correspond to the evolution time multiplied by the computer's round-off error.

Straightforward parallelization, locality, and simple structure of the propagation algorithm allow implementing evolution using GPU resources allowing for quick sampling of multiple configurations of a particular problem. Additional optimization comes from noticing that the evolution of DTQW on the set of even sites is decoupled from that on the set of odd sites. Consequently, if the initial conditions are located exclusively on one of those (for example, completely lo-cal initial conditions that are often used), the cost of numerical propagation is halved.

Those benefits might first look to be somehow diminished for certain applications by the fact that the dimension of the Hilbert space is necessary doubled compared to one of a Hamiltonian system of the same length, but no additional local spinors degrees of freedom. But this additional overhead is completely canceled by the previously mentioned symmetry that leads to factorization of evolution on odd and even sites on double times (so-called stroboscopic sublattice factorization). This is even the case when we depart from the single particle case and start increasing the number of particles. This requires additional growth of the spinor space size [136]. In such case, the Hilbert space size growth as $2^m L^m$, where $m$ is the number of particles, and $L$ is the linear size of the system( compared to $L^m$ for Hamiltonian systems with trivial unit cells). But as the number of particles grows, the number of symmetry sectors also growth as $2^m$, leading to a collection of $2^m$ subspaces each of the same dimension $L^m$. Note that similarly to the Hamiltonian case, additional reduction of the many-body space size can be achieved by bosonic/fermionic symmetrization.

## 2.2 Anderson Localization in Discrete-Time Quantum Walks

In this section that follows [138] we study the impact of disorder in any of the angles on the DTQW dynamics. For the brief overview of Anderson localization, see Sec. 1.2.1. We find novel Anderson localized phases, and rigorously derive scaling relations for the weak and strong disorder regimes, and close to symmetry-related values in the spectrum.

The section is organized as follows: in subsection 2.2.1 we present the model, elaborate the quantum-mechanical dynamic equations. In subsection 2.2.2 we present numerical results on the localization length dependence on the model parameters. In subsection 2.2.3 we perform analytical derivations of the localization length in the limit of weak and strong disorder, and at symmetry points in the spectrum. We discuss the results in subsection 2.2.4.

### 2.2.1 Model

#### 2.2.1.1 Disorder and transfer matrix approach

We turn to the disordered case where any of the quantum coin angles $(\varphi, \varphi_1, \varphi_2)$ or $\theta$ are assumed to be uncorrelated random functions of the quantum particle position $n$. In this case, the *transfer matrix approach* is useful for both numerical and analytical approaches of computing the localization length. With Eq.(2.3) it follows

$$
\begin{aligned}
e^{-i\omega}\psi_{+,n} &= e^{i[\varphi_{1,(n-1)}+\varphi_{n-1}]}\cos\theta_{n-1}\psi_{+,(n-1)} \\
&+ e^{i[\varphi_{2,(n-1)}+\varphi_{n-1}]}\sin\theta_{n-1}\psi_{-,(n-1)}, \\
e^{-i\omega}\psi_{-,n} &= e^{-i[\varphi_{2,(n+1)}-\varphi_{n+1}]}\sin\theta_{n+1}\psi_{+,(n+1)} \\
&- e^{i[-\varphi_{1,(n+1)}+\varphi_{n+1}]}\cos\theta_{n+1}\psi_{-,(n+1)}.
\end{aligned}
\tag{2.18}
$$

The usual transfer matrix for a 1D lattice with two components per lattice site and nearest neighbor coupling is expected to have rank 4. However, the special structure of the shift operator (2.2) allows reducing the transfer matrix rank to

2. This can be observed with a redefinition of the two component field $\hat{\Psi}_n = (\psi_{+,(n)}; \psi_{-,(n-1)})$, which then leads to the transfer matrix equation

$$\hat{\Psi}_n = \hat{T}_{n-1} \hat{\Psi}_{n-1}, \tag{2.19}$$

where the transfer matrix $T$ has a following form:

$$\hat{T}_n = e^{i\varphi_{1,n}} \begin{pmatrix} e^{i\omega + i\varphi_n} \sec\theta_n & e^{i\varphi_{2,n}} \tan\theta_n \\ e^{-i\varphi_{2,n}} \tan\theta_n & e^{-i\omega - i\varphi_n} \sec\theta_n \end{pmatrix}. \tag{2.20}$$

It follows that disorder in the external synthetic flux $\varphi_1$ does not lead to the localization of the quantum particle wave function, since such a disorder is only modifying the phase difference on neighboring sites, while keeping the amplitude ratio unchanged. However, uncorrelated disorder in all other quantum coin angles $\theta_n$ (kinetic energy) $\varphi_n$ (potential energy) and $\varphi_{2,n}$ (internal synthetic flux) will lead to Anderson localization, as discussed in what follows. We will use probability distribution functions

$$\mathcal{P}(x) = 1/(2W) \ , \ x_0 - W \leq x \leq x_0 + W, \tag{2.21}$$

and $\mathcal{P}(x) = 0$ elsewhere, where $x$ stands for any relevant angle, and $x_0$ is the corresponding first moment (average). The disorder strength is chosen to be $0 \leq W \leq \pi$ as beyond this limits it does not correspond to uniform distribution.

### 2.2.2 Numerical computation of the localization length

In this section, we numerically compute the localization length $L_{loc}(\omega)$ using the canonical approach elaborated e.g. in Ref. [142]. We start with a nonzero $\hat{\Psi}_0$ and repeatedly apply randomly generated matrices (2.20) to this wave function according to (2.19). We use $|\hat{\Psi}_n| = \sqrt{|\Psi_{+,n}|^2 + |\Psi_{-,n}|^2}$ and compute the Lyapunov exponent at the $N$th iteration as

$$\lambda_N = \frac{1}{N} \sum_{n=1}^{N} \ln\left(\left|\hat{\Psi}_n\right|\right) \ . \tag{2.22}$$

The localization length is then obtained as[142]

$$L_{loc} = 1/\lambda_N. \tag{2.23}$$

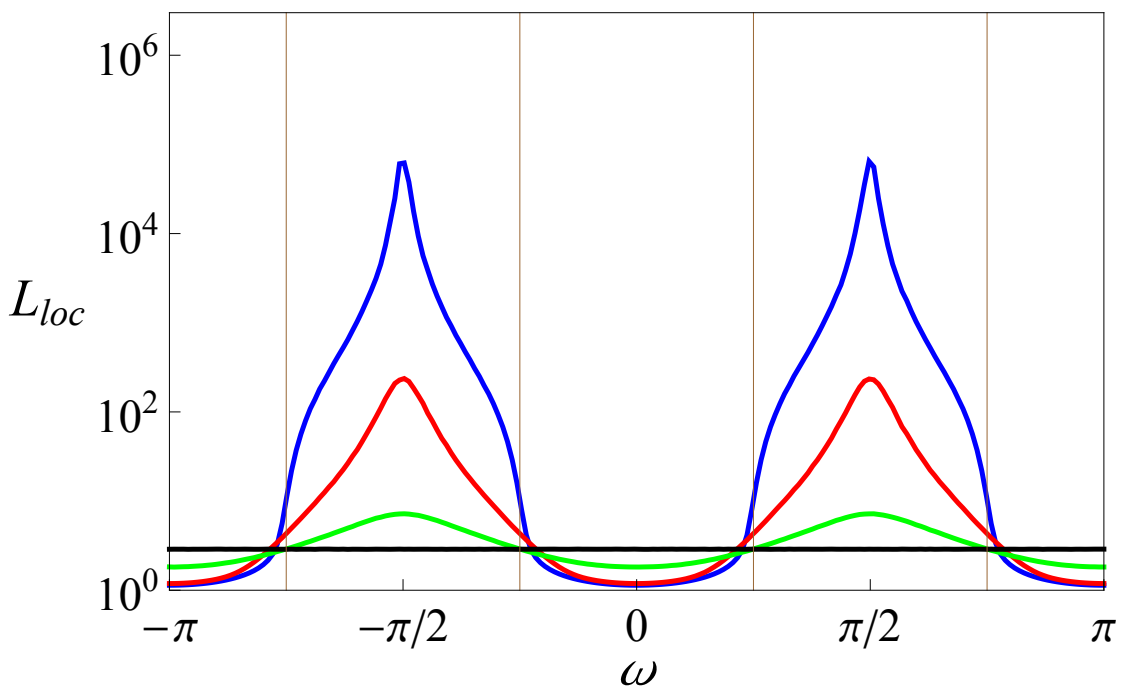

Figure 2.4: The dependence of the localization length on the frequency $\omega$ for disorder in potential field $\varphi$. From top to bottom at $\omega = \pi/2$: $W = \pi/20, \pi/5, \pi/2, \pi$ (blue, red, green). Here $\theta = \pi/4$. The brown vertical lines indicate the boundaries of the allowed bands (2.6).

In order to ensure convergence, we used $N = 10^6 - 10^9$ iterations. The validity of the approach was cross checked by direct diagonalization for large finite systems. We further note that for a disorder which is weak as compared to the gap of the band structure of the ordered case, the density of states deep in the gap will be strictly zero. Nevertheless, the transfer matrix approach will generate a certain (finite) localization length, which corresponds to some additional fictious defect state with a corresponding frequency.

#### 2.2.2.1 Disorder in $\varphi$

We remind that the angle $\varphi$ can be interpreted as a potential energy. The corresponding disorder is similar to *diagonal disorder* for tight binding Hamiltonians[142]. Without loss of generality we can take $x_0 = 0$ in (2.21) as non-zero $x_0$ corresponds to a global gauge. We observe that the localization length $L_{loc}$ is always finite for any finite but nonzero strength of the disorder $W$, see Fig. 2.4.

For weak disorder $W \ll \pi$ we find that $L_{loc}$ is large as the frequency $\omega$ is inside the allowed bands of the ordered case (2.6) and decreases rapidly as the frequency $\omega$ moves inside the gaps, with an anomalous enhancement of $L_{loc}$ at the band centers $\omega = \pm\pi/2$. As the strength of disorder increases the localization length variations diminish, and remarkably $L_{loc}$ becomes independent of $\omega$ for $W = \pi$. Variation of $\theta$ does not qualitatively changes the outcome. However, in the special case $\theta = \pm\pi/2$ the localization length vanishes $L_{loc} = 0$. Indeed, the eigenstates are then still compactly localized in full accordance with (2.8), while the eigenfrequencies simply become $\omega_n = \pm\pi/2 + \varphi_n$.

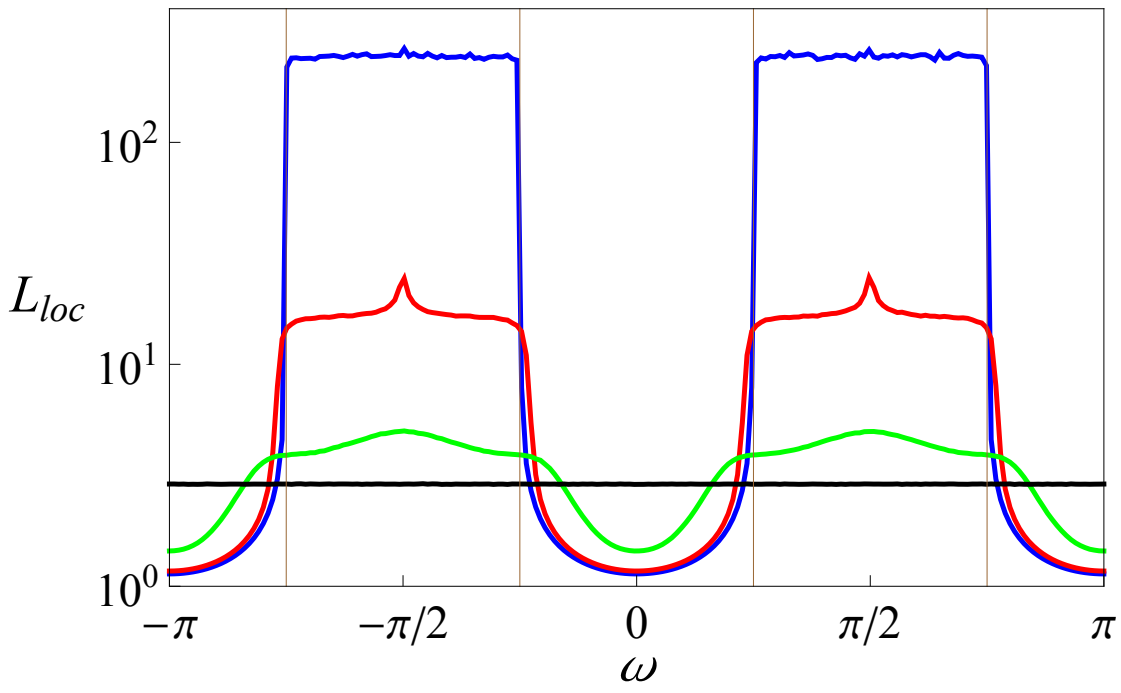

Figure 2.5: The dependence of the localization length on the frequency $\omega$ for disorder in $\varphi_2$. From top to bottom at $\omega = \pi/2$: $W = \pi/20, \pi/5, \pi/2, \pi$ (blue, reg, green). Here $\theta = \pi/4$. The brown vertical lines indicate the boundaries of the allowed bands (2.6).

#### 2.2.2.2 Disorder in $\varphi_2$

We remind that the angle $\varphi_2$ can be interpreted as an internal synthetic flux. Without loss of generality we can take $x_0 = 0$ in (2.21). We observe that the localization length $L_{loc}$ is always finite for any finite but nonzero strength of the disorder $W$, see Fig. 2.5. For weak disorder $W \ll \pi$ the localization length $L_{loc}$ is almost independent of $\omega$ inside the bands of the ordered case, with a small peak in the center of each band ($\omega \simeq \pi/2$). The localization length inside this peak can double tis value as compared to the plateau values outside the peak, see Fig. 2.6. However, according to our computations, the localization length stays finite at the peak center for finite disorder strength (inset Fig. 2.6). For strong disorder $W = \pi$ the localization length is frequency independent, see Fig. 2.5. Variation of $\theta$ does not qualitatively changes the outcome. However, in the special case $\theta = \pm\pi/2$ the localization length vanishes $L_{loc} = 0$. The eigenstates are then still compactly localized in full accord with (2.8), while the eigenfrequencies simply become $\omega_n = \pm\pi/2 + \varphi$.

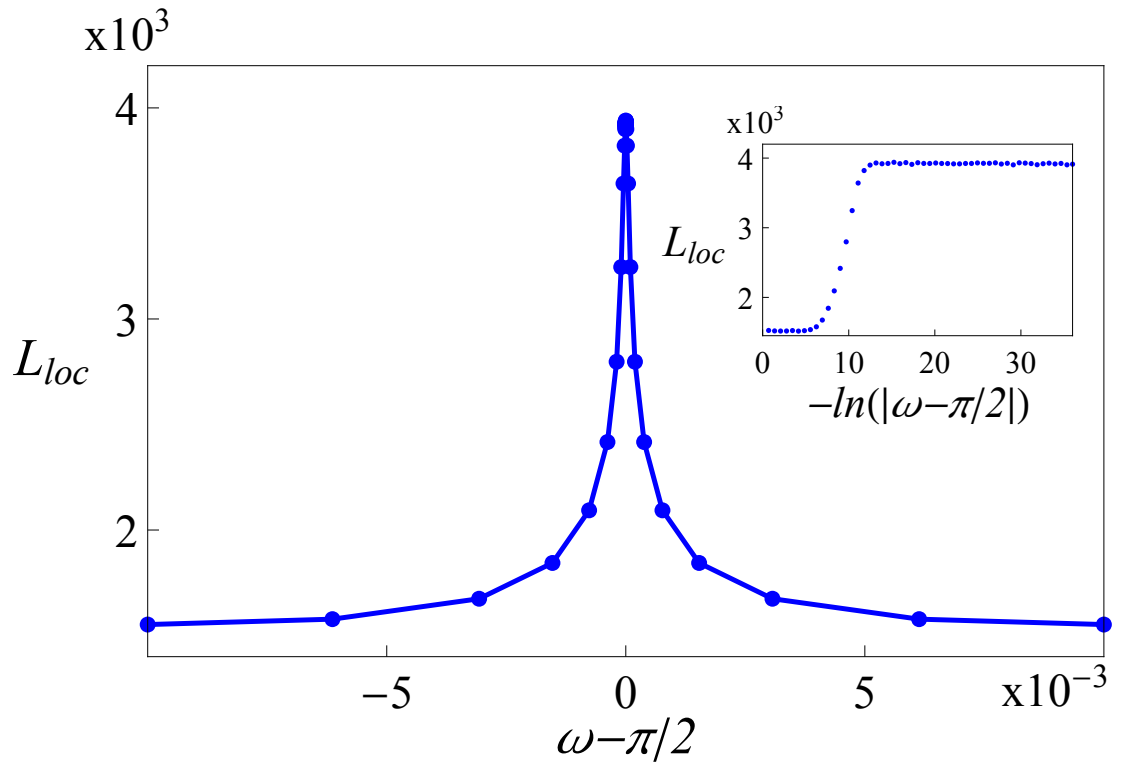

Figure 2.6: The localization length for weak disorder in $\varphi_2$ near the band center $\omega = \pi/2$ ($\theta = \pi/4$). Here $W = \pi/50$. Symbols – results of computations, lines are guiding the eye. Inset: same but resolving the frequency dependence of the localization length on a logarithmic scale close to the band center.

#### 2.2.2.3 Disorder in $\theta$

We remind that the angle $\theta$ can be interpreted as a kinetic energy of a quantum particle, which controls the band width. At variance to the previous cases, the localization length diverges logarithmically at the band centers $\omega = \pm\pi/2$ [109], and results, in general, depend on the average $x_0 = \theta_0$ in (2.21). While the divergence is barely seen in Fig. 2.7, it becomes evident in the inset in Fig. 2.8(a).

A further logarithmic divergence of the localization length is observed at $\omega = 0, \pi$ for the special case of zero average $\theta_0 = 0$, see Fig. 2.9.

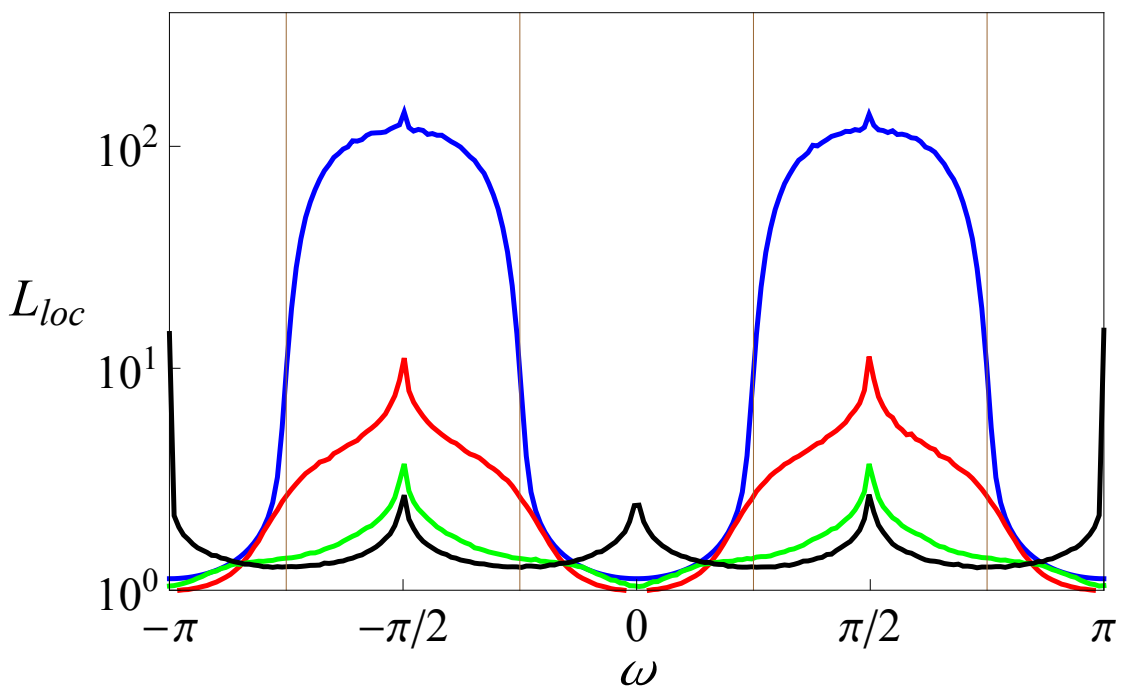

Figure 2.7: The dependence of the localization length on the frequency $\omega$ for disorder in $\theta$. From top to bottom at $\omega = \pi/2$: $W = \pi/20, \pi/5, \pi/2, \pi$ (blue, red, green). Here $\theta_0 = \pi/4$. The brown vertical lines indicate the boundaries of the allowed bands (2.6).

### 2.2.3 Analytical results on the localization length

In order to provide the analytical results for the localization length $L_{loc}$ for different types of disorder, we use methods of stochastic equations for the phase and amplitude of the wave function which has been previously used successfully for a 1D tight-binding model with diagonal and off-diagonal disorder [143, 144, 145]. These methods allow us to systematically study the dependence of the localization length on all relevant parameters.

#### 2.2.3.1 Weak disorder

For weak disorder $W \ll \pi$ we rewrite the transfer matrix in the following form: $\hat{T} = \hat{T}_0 + \hat{T}_d$, where the matrix $\hat{T}_0$ is the transfer matrix of the discrete time quantum walk in the absence of disorder. $\hat{T}_0$ contains the average angle $\theta_0$, and the average values of the angles $\varphi, \varphi_2$ can be zeroed without loss of generality. The matrix $\hat{T}_d$ randomly changes from site to site.

As a next step we choose a basis in which $\hat{T}_0$ is diagonal. The corresponding

unitary matrix

$$\hat{S} = \begin{pmatrix} \tan\theta_0 & i\xi \\ -i\xi & \tan\theta_0 \end{pmatrix}, \tag{2.24}$$

where $\xi = \sin\omega\sec\theta_0 - \sin k$, and $k$ is determined by dispersion relationship for a fixed value of $\omega$ in (2.6). In the new basis the transfer matrix $\hat{T}_0$ is written as

$$\hat{\tilde{T}}_0 = \hat{S}^{-1}\hat{T}_0\hat{S} = \begin{pmatrix} e^{ik} & 0 \\ 0 & e^{-ik} \end{pmatrix}. \tag{2.25}$$

Similarly the disorder-dependent part of the transfer matrix $T_d$ results in $\hat{\tilde{T}}_d = \hat{S}^{-1}\hat{T}_d\hat{S}$:

$$\hat{\tilde{T}}_d = \begin{pmatrix} \alpha_n & \beta_n \\ \beta_n^* & \alpha_n^* \end{pmatrix}, \tag{2.26}$$

where the parameters $\alpha_n$ and $\beta_n$ are functions of the random quantum coin angles and the average $\theta_0$.

We obtain a stochastic equation for the wave function $\tilde{\Psi}_{+,n}$:

$$\tilde{\Psi}_{+,n} = [e^{ik} + \alpha_n]\tilde{\Psi}_{+,n-1} + \beta_n\tilde{\Psi}^*_{+,n-1}\,. \tag{2.27}$$

Introducing the amplitude $r_n$ and phase $\chi_n$ of the wave function $\tilde{\Psi}_{+,n}$ as $\tilde{\Psi}_{+,n} = r_n e^{i\chi_n}$, we arrive at

$$\frac{r_n}{r_{n-1}} e^{i[\chi_n - \chi_{n-1} - k]} = 1 + e^{-ik}\alpha_n + e^{i[-2\chi_{n-1} - k]}\beta_n. \tag{2.28}$$

Thus, if the frequency $\omega$ is located inside of the frequency band gap (see Fig. 2.2) the corresponding wave vector $k$ takes an imaginary value, and, therefore, one can conclude that the localization length $L_{loc}$ is bounded from above by $1/|Im(k)|$.

If the frequency $\omega$ is located in the allowed frequency range of the ordered case, the wave vectors $k$ take real values. For weak disorder $W \ll \pi$ the values of $\alpha_n$ and $\beta_n$ are small and of the order of $W$. Then, it follows that $r_n$ and $(\chi_n - \chi_{n-1})$ vary weakly from site to site. Replacing the discrete site variable $n$ by a continuous variable $u$ and replacing differences by differentials, e.g. $r_{n-1} \to r(u) - dr/du$, we arrive at the following differential equations:

$$\begin{aligned} \frac{d(\ln r)}{du} &= Re[\alpha(u)]\cos(k) + Im[\alpha(u)]\sin(k) \\ &+ Re[\beta(u)]\cos(2\chi + k) + Im[\beta(u)]\sin(2\chi + k)\ , \\ \frac{d\chi}{du} &= k - Re[\alpha(u)]\sin(k) + Im[\alpha(u)]\cos(k) \\ &+ Im[\beta(u)]\cos(2\chi + k) - Re[\beta(u)]\sin(2\chi + k)\ . \end{aligned} \tag{2.29}$$

For uncorrelated disorder, we solve Eqs. (2.29) by using a standard perturbation analysis. In particular, we integrate the second equation in (2.29), insert the result into the first equation, expand up to second order terms in $\alpha(u)$ and $\beta(u)$ and discard fast oscillating terms. After a final averaging over disorder we obtain an exponential increase of the amplitude of wave function, $< \ln(r) >= u/L_{loc}$ with the localization length

$$L_{loc} = \frac{4}{< |\beta(u)|^2 >}. \tag{2.30}$$

Here $< |\beta(u)|^2 > \equiv N^{-1} \lim_{N\to\infty} \sum_{n=1}^{N} |\beta_n|^2$. Note that the perturbation analysis and Eq.(2.30) are not valid if the wave vector $k$ is close to the special points $k = 0, \pm\pi/2, \pm\pi$.

**Disorder in $\varphi$.** For disorder in $\varphi$, the random transfer matrix $\hat{T}_d$ takes a diagonal form:

$$\hat{T}_d = \sec\theta_0 \begin{pmatrix} e^{i\omega}[e^{i\varphi_n} - 1] & 0 \\ 0 & e^{-i\omega}[e^{-i\varphi_n} - 1] \end{pmatrix}. \tag{2.31}$$

Rotating this matrix to the new basis we obtain the parameter $\beta_n$ as

$$\beta_n = \frac{\tan\theta_0}{\sin k}[\cos k \sin\varphi_n - \sin\omega \sec\theta_0 (1 - \cos\varphi_n)]\ .$$

With Eq.(2.30) this leads to the final result,

$$L_{loc} = \frac{4\sin^2(k)\cot^2(\theta_0)}{(W^2/3)\cos^2 k + \sin^2\omega \sec^2\theta_0 (W^4/20)}. \tag{2.32}$$

We find that the localization length $L_{loc} \sim 1/W^2$. However, for $\omega = \pm\pi/2$ this scaling is replaced by $L_{loc} \sim 1/W^4$, which leads to a strong enhancement of the localization length. This is the explanation for the observed anomalous enhancement of the localization length in Fig. 2.4. In addition, the special gapless case $\theta_0 = 0, \pi$ yields complete delocalization $L_{loc} \to \infty$, as can be also easily observed from the original equation (2.18).

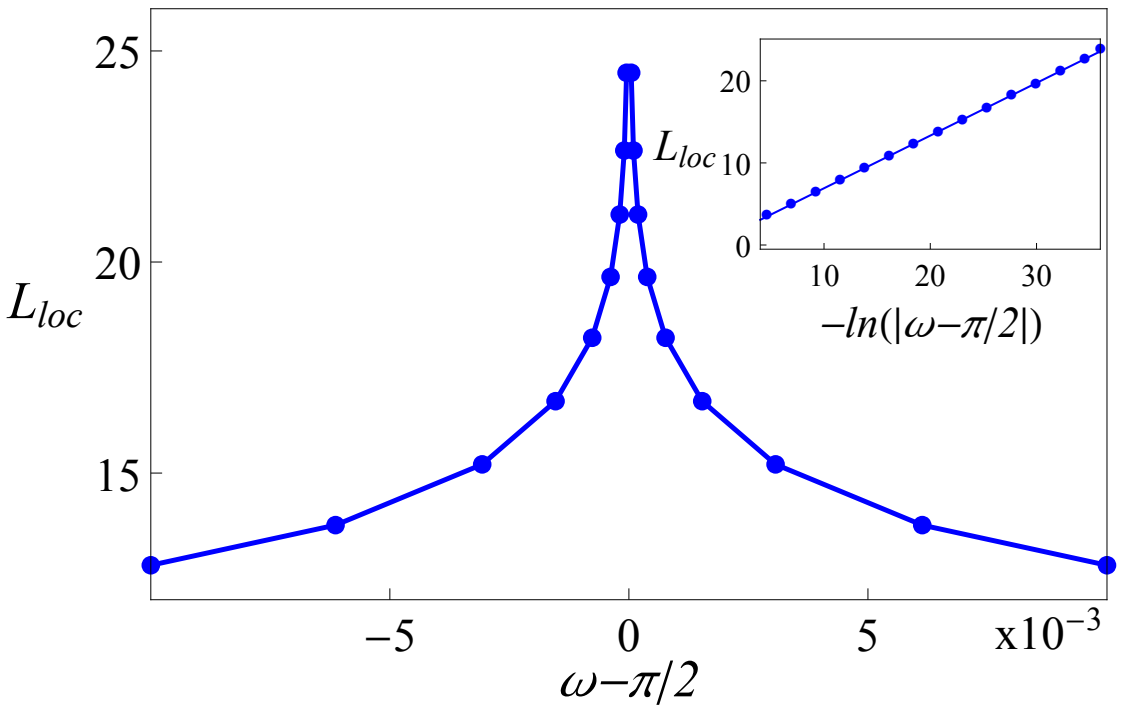

Figure 2.8: The numerically calculated dependence of the localization length $L_{loc}$ on the characteristic frequency $\omega$ near the band center $\omega = \pi/2$ for disorder in $\theta$. The average angle $\theta_0 = \pi/4$ and the strength of disorder $W = \pi/2$. Symbols – results of numerical computations, lines guide the eye. Inset: resolving the frequency dependence of the localization length on a logarithmic scale close to the band center. Symbols – results of numerical computations. The straight line is a linear fit of the data.

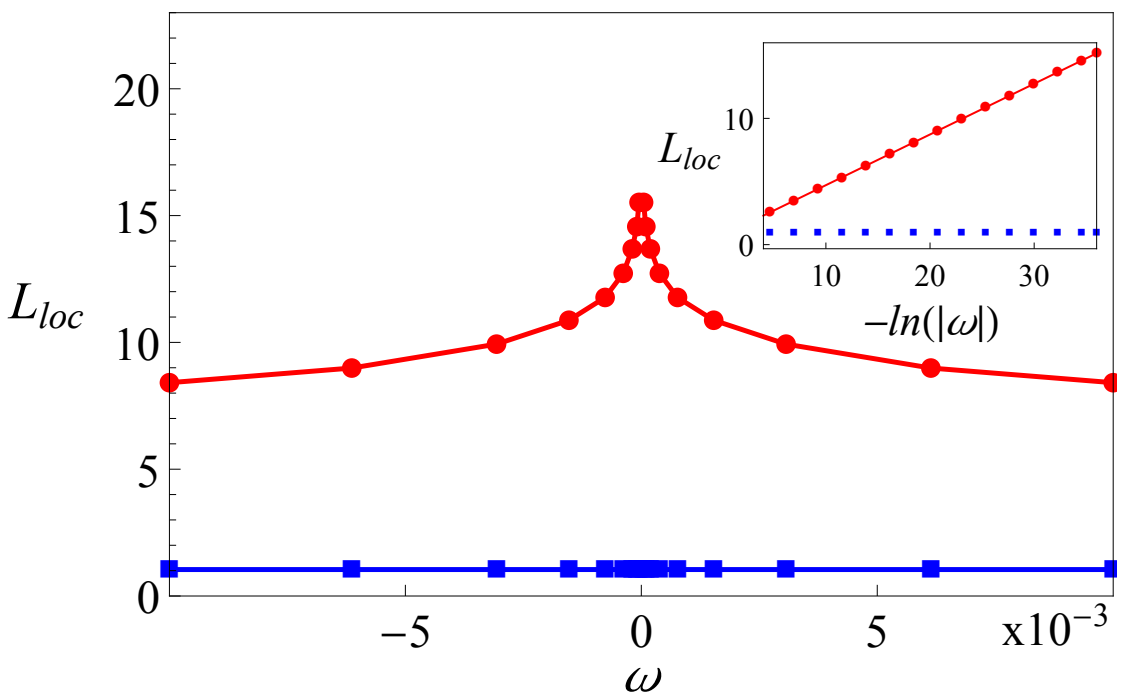

Figure 2.9: The numerically calculated dependence of the localization length on the frequency close to $\omega = 0$ for disorder in $\theta$. The strength of disorder $W = \pi/2$, and the average angles $\theta_0 = 0$ (red circles) and $\theta_0 = \pi/4$ (blue squares). Symbols – results of numerical computations, lines guide the eye. Inset: same but resolving the frequency dependence of the localization length on a logarithmic scale close to $\omega = 0$. The straight line is a linear fit of the data.

These features are in a good agreement with the numerical computations from the previous subsection (see Fig. 2.10). In particular, the analytical result (2.32) is in excellent agreement with the computed dependency of $L_{loc}(\omega = \pi/2)$ on the strength of disorder for different values of $\theta$, as shown in Fig. 2.10.

We proceed with estimating the localization length at the boundaries of the spectrum $\omega(k)$ of the ordered case, by choosing e.g. the limit $k \ll 1$ ($\omega \approx \theta_0$). Using (2.32) we obtain $L_{loc} = 12k^2/(\tan^2\theta_0 W^2)$. On the other hand, for $\omega$ values located inside of gap, $L_{loc} \simeq 1/|k|$. Both equations can be satisfied by the following scaling of the localization length at the boundaries of the spectrum:

$$L_{loc} = \eta \tan^{-2/3}(\theta_0) W^{-2/3} \,, \tag{2.33}$$

where $\eta$ is an unknown prefactor of order of one. In Fig. 2.11 we compare the numerically calculated dependence of $L_{loc}(\omega = \theta)$ on $W$ with the analytical prediction (2.33) for various values of $\theta_0$. We find excellent agreement with just one fitting parameter $\eta = 1.36$ for all cases.

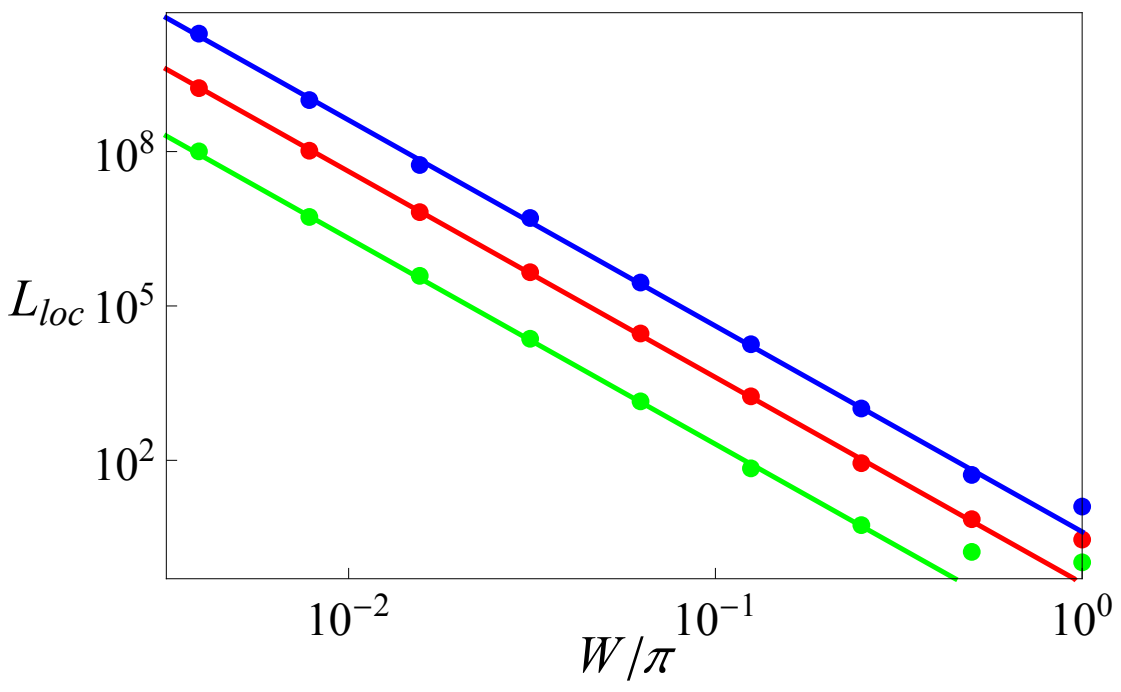

Figure 2.10: The localization length as a function of the disorder strength for disorder in $\varphi$ for $\omega = \pi/2$ (band center). Solid lines – the analytical result (2.32), symbols – numerical computations, Here $\theta_0 = \pi/8, \pi/4, 3\pi/8$ from top to bottom. The predicted scaling $L_{loc} \sim 1/W^4$ is observed.

**Disorder in $\varphi_2$.** For disorder in $\varphi_2$ the random transfer matrix $\hat{T}_d$ has only nonzero off-diagonal terms

$$\hat{T}_d = \tan\theta_0 \begin{pmatrix} 0 & e^{i\varphi_{2,n}} - 1 \\ e^{-i\varphi_{2,n}} - 1 & 0 \end{pmatrix} . \tag{2.34}$$

Rotating this matrix to the new basis we obtain the parameter $\beta_n$ as

$$\beta_n = -i\varphi_{2,n} \tan\theta_0 \ .$$

With Eq.(2.30) this leads to the final result

$$L_{loc} = \frac{12}{\tan^2\theta_0 W^2} \ . \tag{2.35}$$

The localization length is *independent* of $\omega$ for frequencies $\omega$ inside the bands, which explains the observed plateaus in Fig. 2.5. The predicted scaling with $W$ and $\theta_0$ is in excellent agreement with computational results as shown in Fig. 2.12, with numerical prefactor being $6$ instead of $12$.

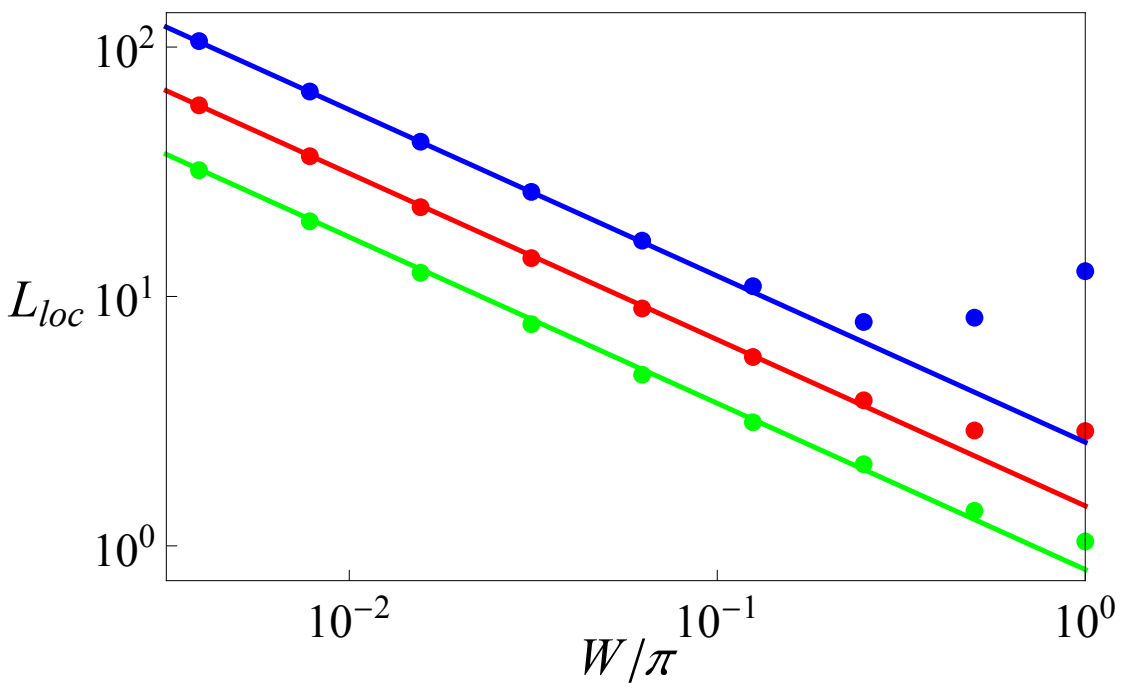

Figure 2.11: The localization length as a function of the disorder strength for disorder in $\varphi$ for $\omega = \theta_0$ (band edge). Solid lines – the analytical result (2.33) with $\eta = 1.36$, symbols – numerical computations. Here $\theta_0 = \pi/8, \pi/4, 3\pi/8$ (blue, red, green) from top to bottom. The predicted scaling $L_{loc} \sim 1/W^{2/3}$ is observed.

**Disorder in $\theta$.** For disorder in $\theta$ the random transfer matrix $\hat{T}_d$ takes the following form:

$$\hat{T}_d = \frac{\theta_n}{\cos^2\theta_n} \begin{pmatrix} e^{i\omega}\sin\theta_0 & 1 \\ 1 & e^{-i\omega}\sin\theta_0 \end{pmatrix}. \tag{2.36}$$

Rotating this matrix to the new basis we obtain the parameter $\beta_n$ as

$$\beta_n = \frac{\sin\omega}{\sin k\cos\theta_0}\theta_n \ .$$

With Eq.(2.30) this leads to

$$L_{loc} = \frac{12\sin^2 k\cos^2\theta_0}{W^2\sin^2\omega}. \tag{2.37}$$

We find that the localization length scales as $L_{loc} \sim 1/W^2$ similar to the previous cases. At the band edge $k \simeq 0$ ($\omega \simeq \pm\theta$) the localization length scales similar to the case of $\varphi$ disorder as $L_{loc} \simeq (\tan^2\theta_0 W^2)^{-1/3}$.

However, at the band center we observed a logarithmic divergence of the localization length from numerical computations, see Fig. 2.8. The divergence of the localization length at the precise band center was derived in Ref. [146]. This follows from the fact, that the parameter $\alpha_n$ in Eq. (2.26) strickly vanishes at the band center. Therefore $\alpha_n$ is a higher-order perturbation term and can be neglected close to the band center as well. Eq. (2.27) is then reduced to

$$\tilde{\Psi}_{+,n} = e^{ik}\tilde{\Psi}_{+,n-1} + \beta_n \tilde{\Psi}^*_{+,n-1}. \tag{2.38}$$

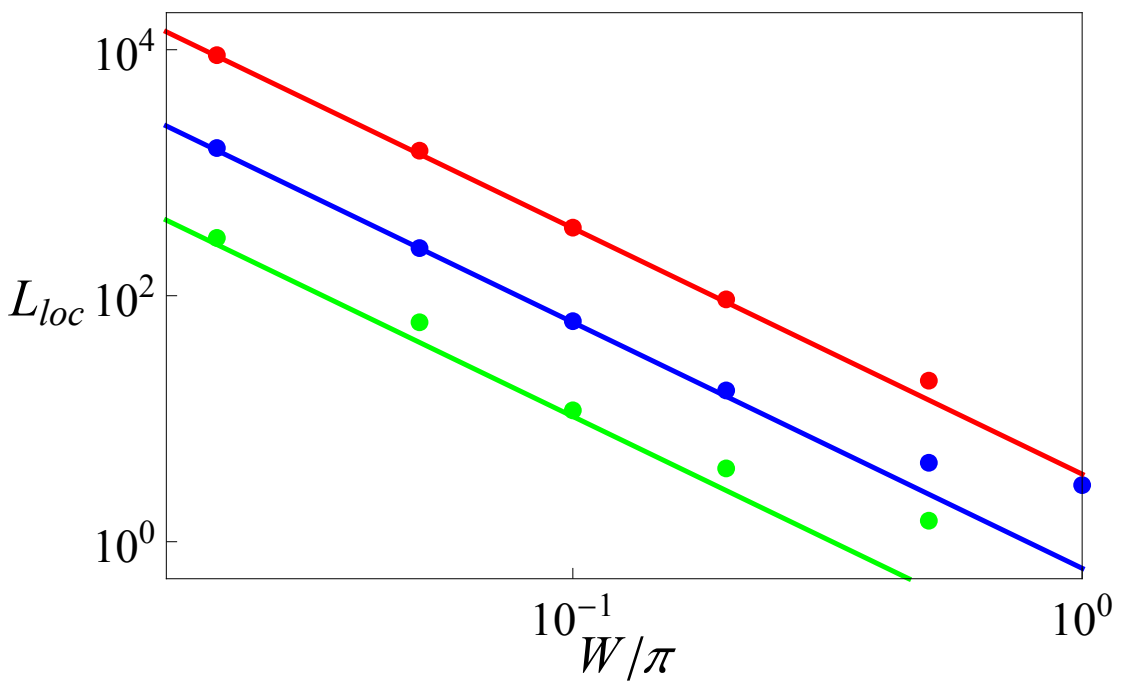

Figure 2.12: The mean localization length inside the band as a function of the disorder strength for disorder in $\varphi_2$. Solid lines – the analytical result (2.35) with an additional fitting parameter $\eta = 0.5$, symbols – numerical computations. Here $\theta_0 = \pi/8, \pi/4, 3\pi/8$ from top to bottom. The predicted scaling $L_{loc} \sim 1/W^2$ is nicely observed.

The corresponding differential equations (2.29) modify into

$$\begin{aligned} \frac{d(lnr)}{du} &= \beta(u)\sin[2\chi], \\ \frac{d\chi}{du} &= \pi/2 + \delta\omega - \beta(u)\cos[2\chi], \end{aligned} \tag{2.39}$$

where $\delta\omega$ is frequency detuning. Excluding $\beta(u)$ we find $< \ln r > = < \delta\omega \tan(2\chi) > u$, and, therefore, the corresponding localization length is $L_{loc} = [< \delta\omega \tan(2\chi) > ]^{-1}$. In order to compute the average, we introduce a new variable $z = 2\ln[\tan(\chi - \pi/4)]$ and rewrite the second equation in (2.39) as

$$\frac{dz}{du} = 4(\delta\omega)\cosh z + 4\beta(u), \tag{2.40}$$

with $L_{loc} = [< (\delta\omega)\sinh(z/2) >]^{-1}$. In order to find the average value of $z$ we transfer from the stochastic equation (2.40) to the corresponding Fokker-Planck equation for the probability $P(z)$, which satisfies

$$\frac{16W^2}{\cos^2\theta_0}\frac{d^2P(z)}{dz^2} - 4(\delta\omega)\frac{d}{dz}[\cosh(z/2)P(z)] = 0, \tag{2.41}$$

with the normalization condition $\int_0^{2\pi} d\chi P(z) = 1$. It follows that $L_{loc}{}^{-1} = \delta\omega\int dz \sinh(z/2)P(z)$. In the limit $\delta\omega \ll W^2$ we obtain a logarithmic enhancement of the localization length as

$$L_{loc} = A\ln|\frac{\delta\omega}{W^2}|\ ,\ A = \frac{12\cos^2\theta_0}{W^2}\ . \tag{2.42}$$

This dependence on $\theta_0$ and $W$ agrees perfectly with the numerical data in Fig. 2.13 (up to a numerical prefactor). Notice that the logarithmical divergence resembles a well-known Dyson-Wigner singularity reported previously in the electronic transport of 1D disordered tight-binding chain in the presence of off-diagonal disorder [143, 144, 145, 147].

#### 2.2.3.2 Strong disorder

We start with noting that the linear transfer matrix equations (2.19), which define a linear two-dimensional map, can be equivalently rewritten as a 1D map, which is, however, nonlinear. We introduce the variable

$$y_n = \frac{\hat{\Psi}_{+,n}}{\hat{\Psi}_{-,n}}, \tag{2.43}$$

in order to rewrite the two-component wave function $\hat{\Psi}_n$ as

$$\hat{\Psi}_n = A_n \begin{pmatrix} y_n \\ 1 \end{pmatrix}\ . \tag{2.44}$$

This definition allows one to reduce the 2D map defined by the transfer matrix (2.20) to a 1D map:

$$\begin{gathered} y_{n+1} = M(y_n) = e^{2i\varphi_{2,n}}\frac{c_n e^{i\lambda_n}y_n + 1}{y_n + c_n e^{-i\lambda_n}}, \\ c = 1/\sin(\theta_n),\ \lambda_n = \omega + \varphi_n - \varphi_{2,n}. \end{gathered} \tag{2.45}$$

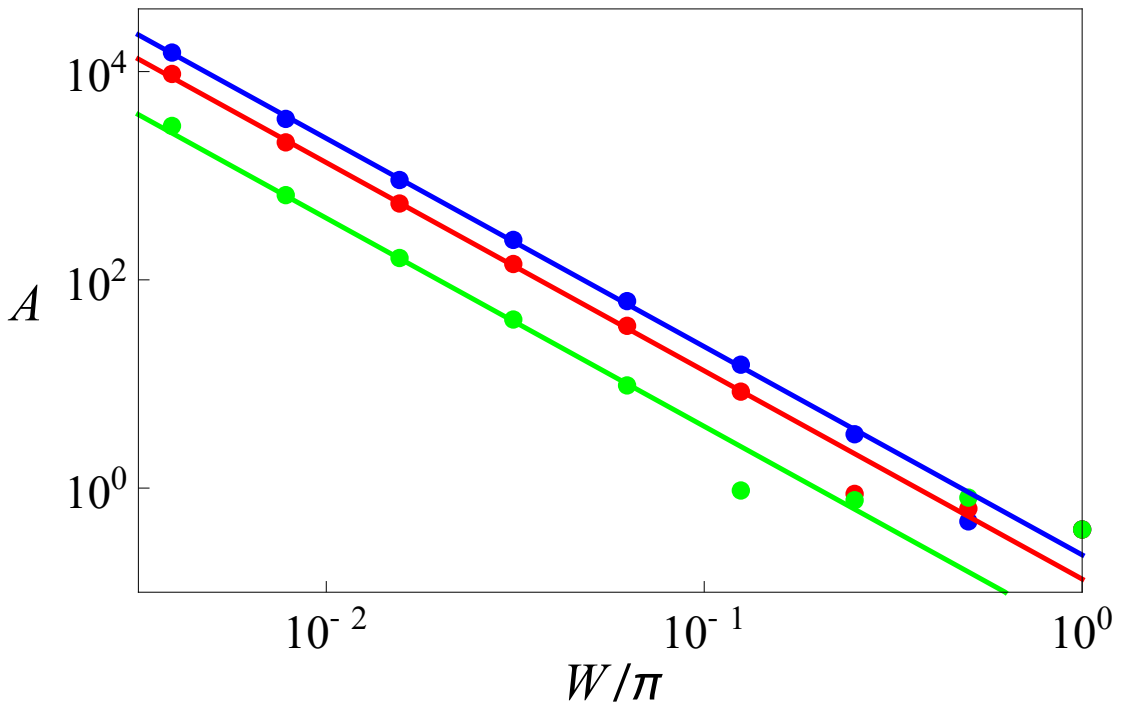

Figure 2.13: The numerically (symbols) and analytically (solid lines) calculated dependence of the coefficient $A = \eta\frac{12\cos^2\theta_0}{W^2}$ (see Eq. (2.42)) on the strength of disorder $W$, with the numerical fitting parameter $\eta = 0.22$. Here $\theta_0 = \pi/8, \pi/4, 3\pi/8$ from top to bottom. The predicted scaling $L_{loc} \sim 1/W^2$ is observed.

The complex variable $y_n$ takes random values, and is characterized by a stationary probability distribution $P(y)$ [148, 149]. Taking into account that the absolute value of the two-component wave function shows an exponential increase as $n$ goes to infinity, we obtain the localization length $L_{loc}$ as

$$1/L_{loc} = \lim_{N\to\infty}\frac{1}{N}\sum_{n=1}^{N}\ln\left(\left|\hat{\Psi}_{n+1}\right|/\left|\hat{\Psi}_n\right|\right). \tag{2.46}$$

By making use of (2.19) we obtain

$$1/L_{loc} = \left\langle \ln\left(\left|\hat{T}\hat{\Psi}\right|/\left|\hat{\Psi}\right|\right)\right\rangle$$
$$= \int \mathrm{d}\mu(\zeta)\int_0^\infty \mathrm{d}\rho\int_{-\pi}^{\pi}\mathrm{d}\phi\ P(\rho e^{i\phi})\cdot\ln\left(\left|\hat{T}(\zeta)\begin{pmatrix}\rho e^{i\phi}\\ 1\end{pmatrix}\right|/\sqrt{1+\rho^2}\right), \tag{2.47}$$

where $y = \rho e^{i\phi}$ and $\zeta$ is a random angle ($\varphi$, $\varphi_2$, or $\theta$), and $\mu(\zeta)$ is its measure.

The map (2.45) reduces the absolute value $\rho$ if $\rho > 1$ and increases it if $\rho < 1$. Thus $\rho = 1$ for the stationary distribution of $P(y)$. Then $P(y)$ has the following

form:

$$P(y) = p(\phi)\delta(\rho - 1). \tag{2.48}$$

The dynamics of the phase $\phi_n$ (which is defined $\mathrm{mod}\ 2\pi$) is determined by the following stochastic equation:

$$\begin{aligned} \phi_{n+1} = m(\phi_n) = 2\arg(\kappa(\lambda_n, \phi_n)) + 2\varphi_{2,n} - \phi_n \ , \\ \kappa(\lambda_n, \phi_n) = \sin(\theta_n) + e^{i(\lambda_n + \phi_n)} \ . \end{aligned} \tag{2.49}$$

This dynamic equation is reduced to an integral equation for the distribution $p(\phi)$:

$$p(\phi') = \int_{<\zeta>-W}^{<\zeta>+W} \frac{\mathrm{d}\zeta}{2W} \int_{-\pi}^{\pi} \mathrm{d}\phi p(\phi)\delta(\phi' - m(\phi)) \ . \tag{2.50}$$

**Disorder in $\varphi$ and $\varphi_2$.** We, first, consider disorder in $\varphi$. We present results for the case of the strongest disorder $W = \pi$. As $\omega$ only appears in combination $\omega + \varphi$, and integration in (2.50) is over the whole period in this case, one may disregard $\omega$ by shifting variables, thus making localization length independent of $\omega$. $\phi_2$ is a fixed constant, which allows to eliminate it in a similar way. This yields

$$p(\phi') = \frac{1}{2\pi} \int_{-\pi}^{\pi} \mathrm{d}\lambda \, p\left(2\arg(\kappa(\lambda, 0)) - \phi' \ \right). \tag{2.51}$$

This equation is satisfied by the uniform solution

$$p(\phi) = \frac{1}{2\pi}. \tag{2.52}$$

Substituting (2.48) and (2.52) with the transfer matrix (2.20) simplifies (2.47),

$$\frac{1}{L_{loc}} = \iint_{-\pi}^{\pi} \frac{\mathrm{d}\varphi \, \mathrm{d}\phi}{8\pi^2} \ln\left(\frac{1 + \sin^2\theta + 2\sin\theta\cos(\varphi + \phi)}{\cos^2\theta}\right). \tag{2.53}$$

Integrating separately over the logarithm of the numerator (which strictly vanishes) and the denominator, we finally arrive at

$$L_{loc} = -\frac{1}{\ln(|\cos(\theta)|)}. \tag{2.54}$$

Thus, the localization length is independent of $\omega$ and is determined only by the value of $\theta_0$. Exactly the same results will hold for the strongest disorder in $\varphi_2$ as well, and equation (2.54) again applies. In Fig. 2.14 we plot the analytical result (2.54) and compare to numerical computations using the transfer matrix approach, with excellent agreement.

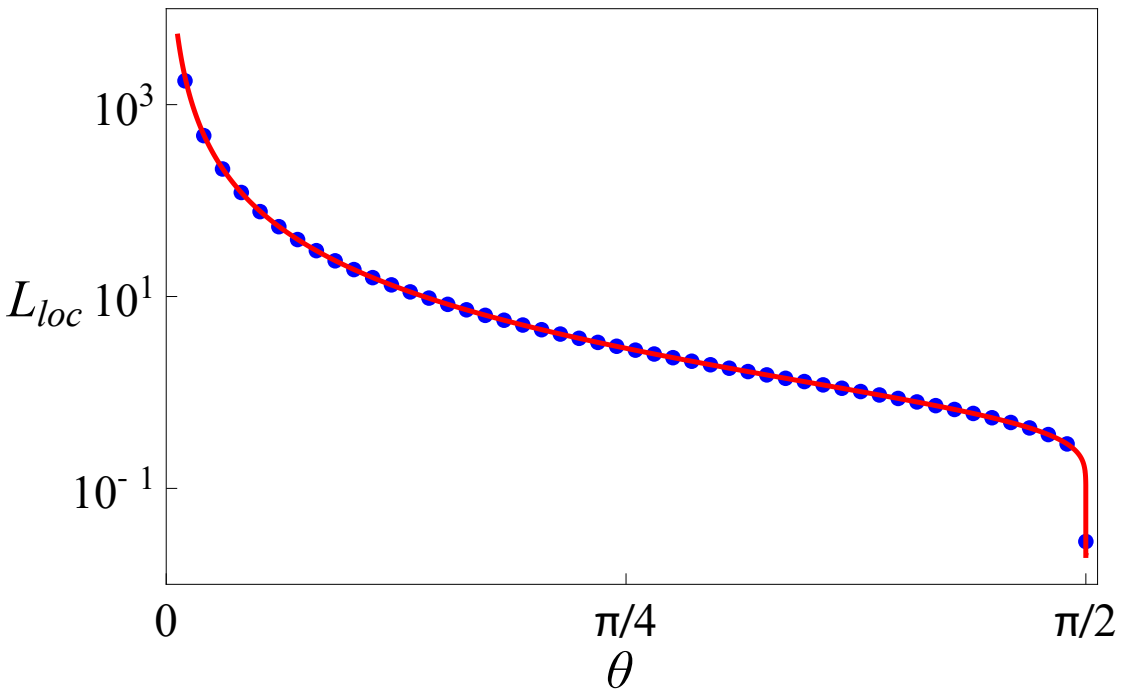

Figure 2.14: Localization length under maximal disorder in $\phi$. Red solid line corresponds to the analytical result (2.54), blue dots are numerical.

**Disorder in $\theta$.** In this subsection we analyze the singular behavior of $L_{loc}(\omega = 0)$ and $L_{loc}(\omega = \pm\pi/2)$ for disorder in $\theta$. We start with $\omega = 0$ (without loss of generality we choose $\varphi = \varphi_2 = 0$). Then, (2.49) reduces to a single equation

$$\kappa = \sin(\theta) + e^{i\phi} \; . \tag{2.55}$$

It follows that $\phi = 0$ is a fixed point of Eq. (2.49), and, therefore, $p(\phi) = \delta(\phi)$ solves Eq.(2.50). Substituting this into (2.47) we obtain

$$1/L_{loc} = \frac{1}{2W} \int_{\theta_0 - W}^{\theta_0 + W} \mathrm{d}\theta \ln |\cot(\pi/4 - \theta/2)|$$

to arrive at

$$\begin{aligned} 1/L_{loc} = \frac{1}{2W} \Big| & Cl_2\left(\pi/2 + \theta_0 - W\right) + Cl_2\left(\pi/2 - \theta_0 + W\right) \\ & - Cl_2\left(\pi/2 + \theta_0 + W\right) - Cl_2\left(\pi/2 - \theta_0 - W\right) \Big| \; , \end{aligned} \tag{2.56}$$

where $Cl_2(x)$ is the Clausen function of the 2nd order (see Ref.[150]). Thus, we find *delocalized states* in two particular cases: either for $\theta_0 = 0$ with arbitrary disorder strength, or for the case of the strongest disorder $W = \pi$ and any value of $\theta_0$, as shown in Fig. 2.15.

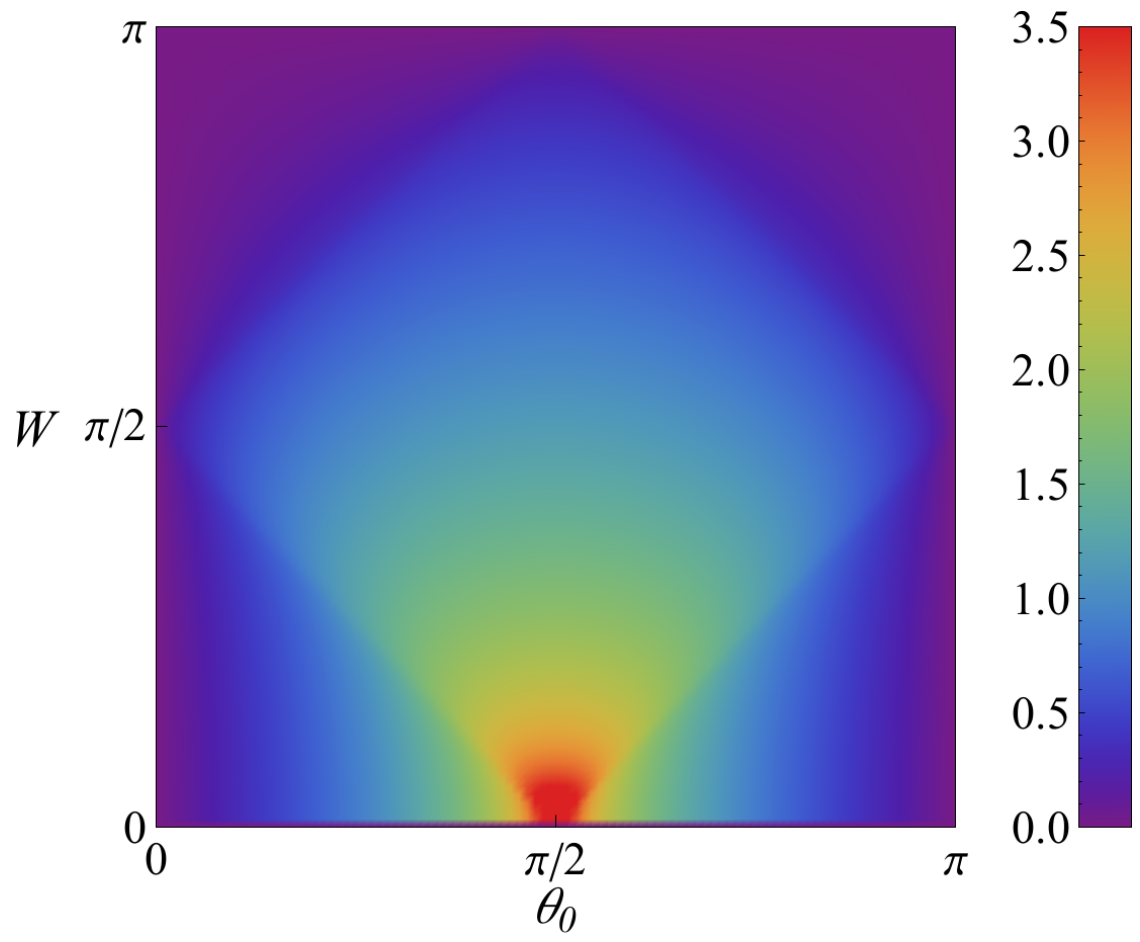

Figure 2.15: The inverse localization length $1/L_{loc}(\omega = 0)$ as a function of $\theta_0$ and $W$ according to Eq. (2.56). The inverse localization length strictly vanishes on the lines $\theta = 0, \pi$ and $W = \pi$.

Next, we consider the case $\omega = \pi/2$. Eq. (2.49) is reduced to

$$\kappa = \sin(\theta) + e^{i(\pi/2+\phi)} . \tag{2.57}$$

We find $m(\pi/2) = 3\pi/2$, $m(3\pi/2) = \pi/2$ as a period-two limit cycle solution of the map (2.49). Thus

$$p(\phi) = \frac{1}{2}\delta(\phi - \pi/2) + \frac{1}{2}\delta(\phi - 3\pi/2) . \tag{2.58}$$

Substitution of this into (2.47) yields $1/L_{loc} = 0$ for any set of parameters. We arrive at the conclusion that the localization length strictly diverges at $\omega = \pm\pi/2$, in agreement with our results for weak disorder Eq. (2.42).

### 2.2.4 Discussion

In conclusion, we have theoretically (numerically and analytically) analyzed the discrete-time quantum walk in the presence of spatial disorder. The dynamics of the system is determined by four angles of a quantum coin operator, i.e. $\varphi, \varphi_1, \varphi_2, \theta$ (2.1). In the absence of spatial disorder the dynamics of the quantum walk is characterized by the dispersion relation, i.e. the dependence of the characteristic frequency $\omega$ on the wave vector $k$ (2.6). The spectrum $\omega(k)$ contains two bands, and is tuned by varying the angle $\theta$. For $\theta = 0$ a gapless spectrum occurs, while for $\theta = \pm\pi/2$ the spectrum consists of two gapped flat bands. The equations, the spectrum, and the eigenvectors are invariant under two symmetry operations: bipartite and particle-hole symmetries.

Disorder in the external synthetic gauge field $\varphi_1$ does not impact the extended nature of the eigenstates, and does not destroy the mentioned above two symmetries. However, disorder in any of the remaining three angles $\theta, \varphi, \varphi_2$ enforces Anderson localization of the eigenstates. In particular, disorder in the kinetic energy angle $\theta$ leads to a logarithmic divergence of the localization length for particular values of the eigenfrequency $\omega$, while again keeping the bipartite and particle-hole symmetries untouched. Disorder in the onsite energy angle $\varphi$ and the internal synthetic flux angle $\varphi_2$ is destroying the particle-hole symmetry, and yields finite localization length for all allowed eigenfrequencies $\omega$. Remarkably we obtain that strongest disorder $W = \pi$ in $\varphi$ and $\varphi_2$ yields Anderson localized random eigenstates with a unique localization length, which depends only on $\theta$, but does not change for different eigenfrequencies $\omega$. This is possible because the space of eigenfequencies is compact and confined to the spectrum of a phase of a complex number residing on the unit circle.

We derive various scaling laws in the limit of weak and strong disorder, and obtain excellent agreement with numerical results using a transfer matrix approach. These results underline the richness of the considered system, which makes it not only attractive for application reasons, but also an ideal playground for various extensions including the impact of many body interactions, mean field nonlinearities, and flat-band physics, to name a few.

## 2.3 Wave Packet Spreading with Disordered Nonlinear Discrete-Time Quantum Walks

In this section, following [139], we study the evolution of DTQW with both disorder and mean-field-like nonlinearity incorporated in coin. To address the fundamental question of whether wave packet nonlinear spreading slows down or continues discussed in subsection 1.2.3 we use DTQW with a heavy emphasis on their numerical effectiveness (see subsection 2.1.4). Their highly efficient coding implementation is the key to address suitable hard computational problems with Hamiltonian dynamics and extending beyond Hamiltonian computational limits. We peek beyond previous horizons set by the CPU time limits for systems of coupled ordinary differential equations. We obtain results for unprecedented times up to $2 \times 10^{12}$ and thereby shift the previous Gross-Pitaevskii limits by four decades.

The section is organized as follows: in subsection 2.3.1 we explain the setup, in subsection 2.3.2 we give the main results. In subsection 2.3.3 we derive the equations of motion in normal modes bases, and thus show the connection and relevance of our results to previously studied systems. In subsection 2.3.4, we provide some details on the numerical approach. In the end, we discuss the results in subsection 2.3.5.

### 2.3.1 Setup

As the coin matrix, here we choose a particular realization of the general coin (2.1),

$$\hat{U}_n = \begin{pmatrix} \cos\theta & e^{i\varphi_n}\sin\theta \\ -e^{-i\varphi_n}\sin\theta & \cos\theta \end{pmatrix}, \tag{2.59}$$

with two angles $\theta$ (kinetic energy) and $\varphi_n$ (site-dependent internal synthetic flux). We consider a strongly disordered DTQW with random uncorrelated angles $\varphi_n$ being uniformly distributed over the entire existence domain $[-\pi, \pi)$. As we showed in Sec. 2.2.3.2, all eigenvectors are exponentially localized on the chain,

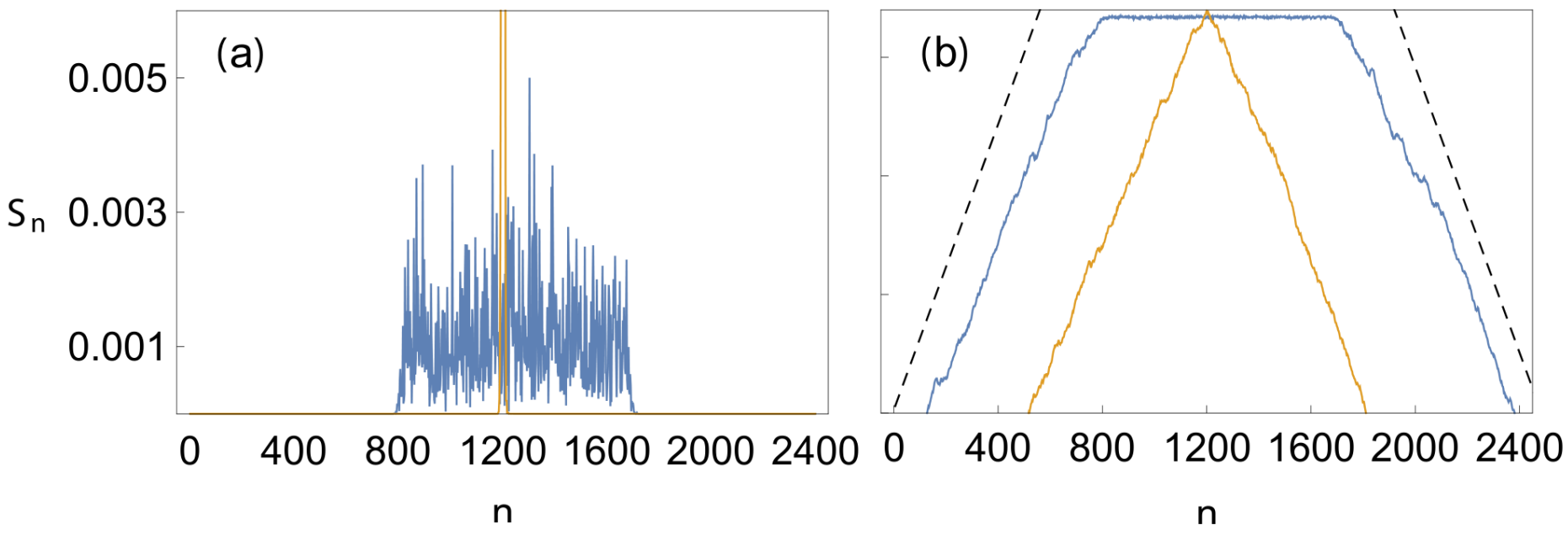

Figure 2.16: Wave packet density profiles $S_n$ for linear $g = 0$ (orange solid lines) and nonlinear $g = 3$ (blue solid lines) DTQWs at the time $t_f = 10^8$ (linear case) and $t_f = 2 \times 10^{12}$ (nonlinear case). (a) lin-lin plot. (b) log-normal plot. The black dashed lines indicate exponential decay with the corresponding localization length, and follow $\mathrm{e}^{-2|n-n_0|/L_{\mathrm{loc}}}$.

and localization length does not depend on eigenfrequency and reads (2.54)

$$L_{\mathrm{loc}} = -\frac{1}{\ln(|\cos(\theta)|)}. \tag{2.60}$$

Another remarkable feature is that for any value of the localization length – either small or large compared to the lattice spacing $\Delta n \equiv 1$ – the spectrum of the quasienergies of an infinite chain is densely filling the compact space of angles of complex numbers on a unit circle. Therefore, the density of states is constant, and gaps in the spectrum are absent.

Anderson localization is manifested through the halt of spreading of an evolving wave packet. In our direct numerical simulations, we choose $\theta = \pi/4$, which results in a localization length $L_{\mathrm{loc}} \approx 2.9$ and a typical localized Anderson eigenstate occupying about 10 lattice sites. We choose the initial state to be localized on $M$ sites:

$$\hat{\psi}_n(t=0) = \frac{1}{\sqrt{2M}}\{1, \mathrm{i}\}\ ,\ n = n_0, ..., (n_0 + M - 1)\ . \tag{2.61}$$

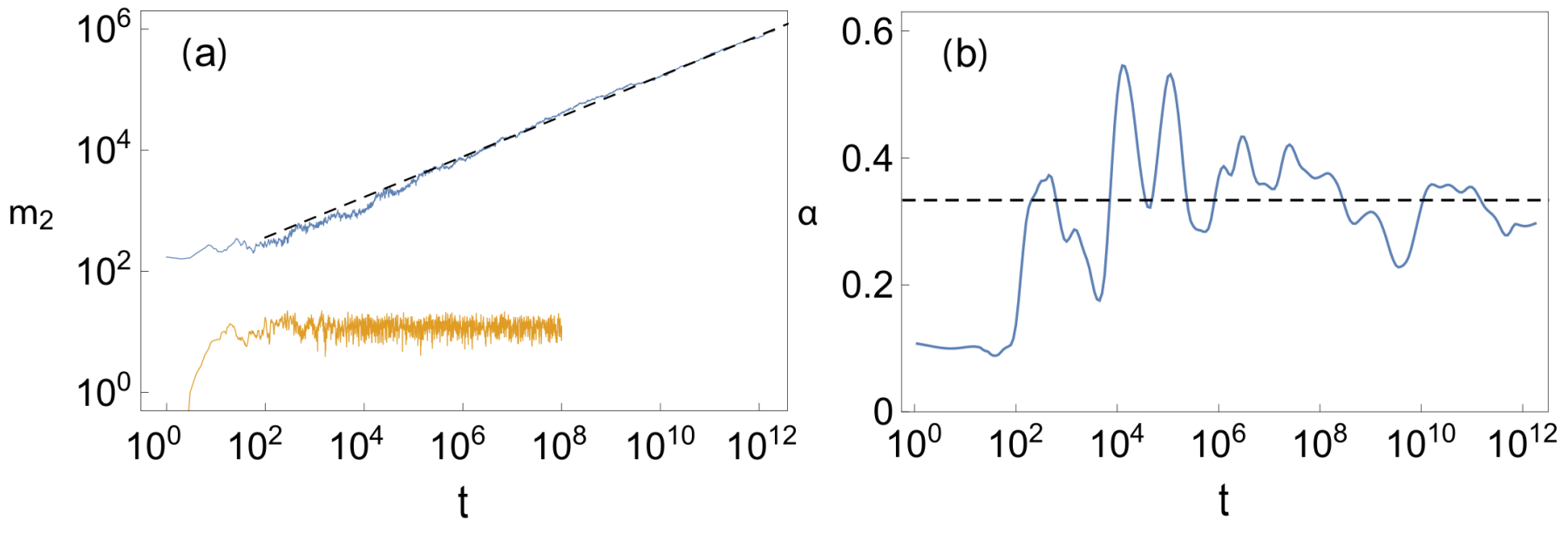

Figure 2.17: (a) $m_2(t)$ versus time in log-log scale for the linear $g = 0$ (orange solid lines) and nonlinear $g = 3$ (blue sold lines) case from Fig.2.16. (b) The derivative $\alpha(t)$ versus time for the data of the nonlienar run from (a) (blue line). Black dashed line corresponds to $\alpha = 1/3$.

### 2.3.2 Results

We evolve this state using Eq.(2.3) until $t = 10^8$ for a system of size $N = 2400$ and $M = 1$. The density distribution $S_n(t) = |\psi_n^-|^2 + |\psi_n^+|^2$ observed for such time is presented in Fig. 2.16(a) (orange solid lines). The distribution is clearly localized, with the width of a few localization lengths. The tails are exponentially decaying, with a slope which is well fitted using the localization length $L_{\text{loc}}$ in Fig. 2.16(b). To further quantify the halt of spreading, we compute the first moment $\bar{n}(t) = \sum_{n=1}^{N} nS_n(t)$, and then the central object of our studies – the second moment

$$m_2(t) = \sum_{n=1}^{N} (n - \bar{n}(t))^2 S_n(t) , \qquad (2.62)$$

The time dependence of $m_2$ is plotted in Fig.2.17(a). We observe a halt of the growth of $m_2(t)$ at $t \approx 10^2$, which together with the profile of the halted wave packet (see Fig. 2.16) is a clear demonstration of Anderson localization.

We now leave the grounds of linear DTQW and generalize the DTQW to a *nonlinear* unitary map by adding a density-dependent renormalization to the

angle

$$\varphi_n = \xi_n + gS_n \ , \tag{2.63}$$

where $g$ is the nonlinearity strength. We note that (2.63) conserves the total norm: $\sum_n \dot{S}_n = 0$.

Our main goal is to measure the details of subdiffusive wave packet spreading on large time scales with $S_n \ll 1$. Therefore, we use a low-density approximation for the quantum coin (2.59), which approximates the exponential factors of the coin keeping evolution unitarity:

$$e^{i\varphi_n} = e^{i\xi_n} \left( \sqrt{1 - g^2 S_n^2} + igS_n \right) . \tag{2.64}$$

The computational advantage of fast calculations of square roots as opposed to slow ones of trigonometric functions serves the purpose to further extend the simulation times. To guarantee unitarity of the evolution, we choose $M > 1$, such that $g^2 S_n^2 \ll 1$.

We evolve a wave packet with $g = 3$ and $M = 13$ and plot the density distribution at the final time $t_f = 2 \times 10^{12}$ in Fig.2.16 (blue solid line). This is a new record evolution time, beating old horizons by a factor of $10^4$. We observe a familiar structure of the wave packet: a homogeneous wide central part with clean remnants of Anderson localization in the tails (Fig. 2.16(b)). The width of the wave packet reaches about 900 sites and exceeds the localization length $L_{\mathrm{loc}}$ by a stunning factor of about 300. The time dependence of the second moment $m_2(t)$ is plotted in Fig.2.17(a). A clean and steady growth of the wave packet width is evident, and the linear fitting on the log-log scale (black line in Fig. 2.17(a)) indicates the universal $\alpha = 1/3$ value.

However, we note that a straightforward fitting with a single power law can yield misleading results, since it is not evident where the asymptotic regime (if any) will start. To study the asymptotic regime in detail we quantitatively assess it by applying standard methods of simulations and data analysis that we discuss at the end of this section in 2.3.4. We calculate the local derivatives on log-log scales to obtain a time-dependent exponent $\alpha(t) = \mathrm{d}\left[\ln(m_2)\right]/\mathrm{d}\left[\ln t\right]$. The resulting curve is plotted in Fig.2.17(b) and strongly fluctuates around the value 1/3.

In order to reduce the fluctuations' amplitudes, we evaluate $m_{2,n}(t)$ for $R = 108$ disorder realizations and compute the geometric average $\ln \overline{m}_2(t) = \sum_n \ln m_{2,n}(t)/R$. In Fig.2.18(a), the results are shown for various values of $g = 0.5, 1, 1.5, 2, 2.5$ up to the times $10^8$ (the corresponding values of $M$ are $5, 8, 8, 10, 10$). All curves approach the vicinity of $\alpha = 1/3$ with fluctuation amplitudes substantially reduced as compared to the single run in Fig.2.17(b). For $g = 0.5$ and $g = 2.5$ we extend the simulations up to the time $10^{10}$ in Fig.2.18(c) and observe a clear saturation of $\alpha(t)$ around 1/3 in Fig.2.18(d). In particular, the weakest nonlinearity value $g = 0.5$ is expected to show the earliest onset of asymptotic subdiffusion. Indeed, we find in this case $\alpha = 1/3 \pm 0.04$ starting with times $t \geq 10^7$.

For $g = 50$ we obtain an earlier onset of subdiffusion with $\alpha(t) > 1/3$, which then slowly converges to the asymptotic one with $\alpha = 1/3$ (Fig.2.18). We thus observe a crossover from weak to strong chaos similar to Ref. [151]. We have not observed any slowing down beyond the $\alpha = 1/3$ law in the whole range of newly accessible times.

### 2.3.3 Evolution of normal modes

It is instructive to rewrite the evolution equations in the basis of linear eigenmodes of the $g = 0$ case. We write the evolution equation using the full evolution operator $V$ (2.11),

$$|\hat{\psi}(t)\rangle = \hat{V}^t |\hat{\psi}(0)\rangle, \; t \in \mathbb{N},$$
$$\hat{V} = \hat{T} \otimes \sum_{n=1}^{N} \hat{S}_n. \tag{2.65}$$

In order to separate the nonlinear components of $\hat{V}$ in $|S_n| \to 0$, we consider a single coin operator on the site $n$. It can be factorized,

$$\hat{U}_n = \hat{Z}_n \hat{U}_n^{(0)} \hat{Z}_n^{-1},$$
$$\hat{Z}_n = \begin{pmatrix} \sqrt{1 - g^2 S_n^2} + igS_n & 0 \\ 0 & 1 \end{pmatrix}, \tag{2.66}$$

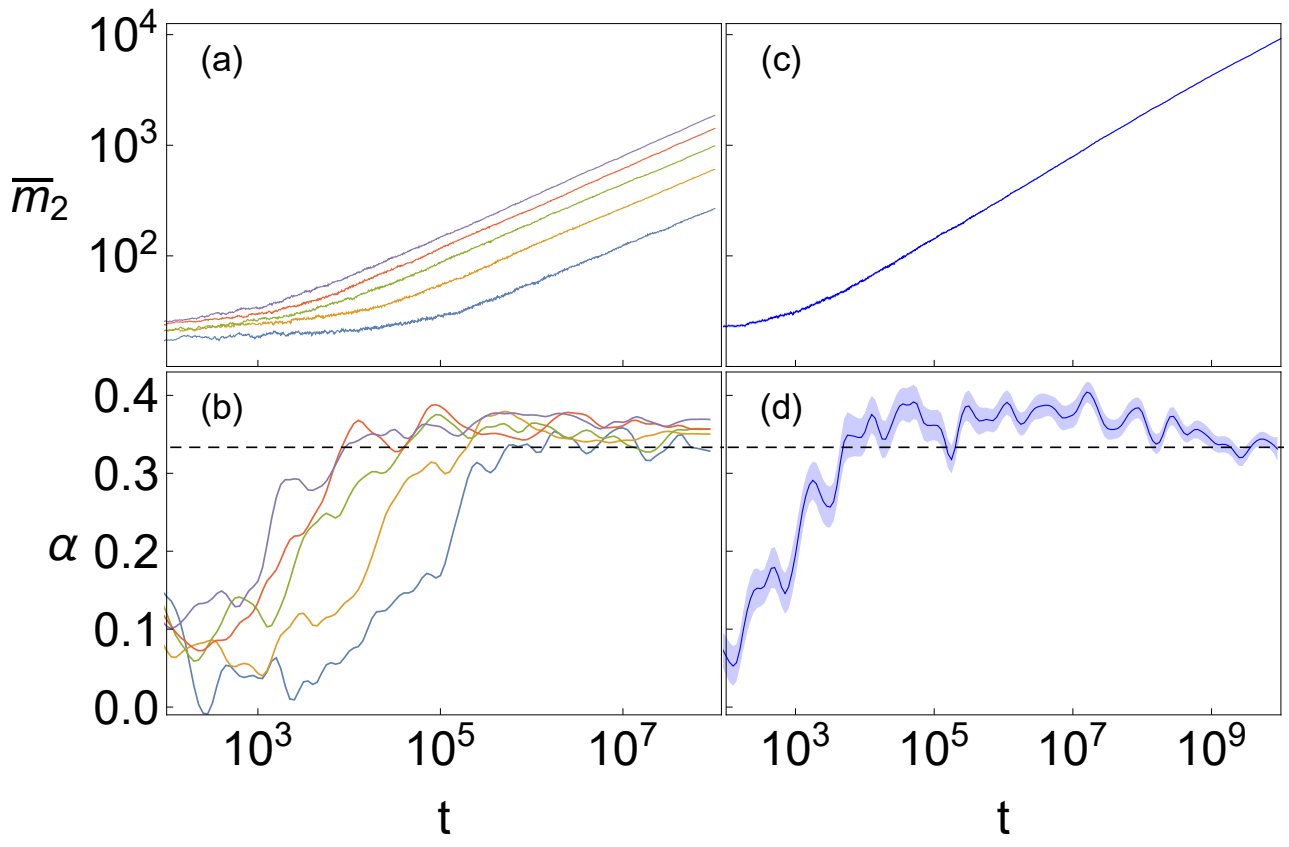

Figure 2.18: (a) The geometric average $\overline{m}_2$ versus time for 108 disorder realizations up to the time $t = 10^8$. The nonlinear parameter $g$ varies from bottom to top as $0.5$ (blue), $1$ (orange), $1.5$ (green), $2$ (red), $2.5$ (purple), $50$ (brown). (b) The derivative $\alpha(t)$ versus time for the data from (a) and same color codes. The horizontal dashed line corresponds to $\alpha = 1/3$. (c),(d) Same as in (a),(b) but for $g = 2.5$ (blue) and $g = 0.5$ (orange) and up to the time $t = 10^{10}$. Shaded areas indicate the statistical error.

where $\hat{U}^{(0)}$ is the local coin operator for zero nonlinearity. Evaluating this expression to separate the linear part and consecutive nonlinear terms with different nonlinear exponent yields

$$\begin{aligned} \hat{U}_n = \hat{U}_n^{(0)} + igS_n \sin\theta \begin{pmatrix} 0 & e^{i\xi_n} \\ e^{-i\xi_n} & 0 \end{pmatrix} \\ + \sin\theta \left(\sqrt{1 - g^2 S_n^2} - 1\right) \begin{pmatrix} 0 & e^{i\xi_n} \\ -e^{-i\xi_n} & 0 \end{pmatrix}. \end{aligned} \tag{2.67}$$

Using a Taylor expansion in the small nonlinearity limit, the evolution operator becomes a sum of a linear term, a second-order (in $\left|\hat{\psi}_n\right|$) term, and a sequence of $4k$-order terms $k \in \mathbb{N}$.

The full non-linear map reads

$$\hat{V} = \hat{V}^{(0)} + ig\sin\theta \sum_{k=1}^{N} S_k \hat{v}^{(k)} + \mathcal{O}\left(g^2 S_k^2\right),$$
$$\hat{v}^{(k)} = \hat{T} \otimes \left( \begin{pmatrix} 0 & e^{i\xi_n} \\ e^{-i\xi_n} & 0 \end{pmatrix} |n\rangle\langle n| \right). \tag{2.68}$$

We will now explicitly evaluate the first-order (in nonlinearity) non-negligible term only.

Let us consider the evolution in the linear-limit eigenmode basis $\hat{\psi}_n(t) = \sum_\alpha a_\alpha(t)\tilde{\psi}_{\alpha,n}$,

$$\hat{V}^{(0)}|\tilde{\psi}_{\alpha,k}\rangle = e^{i\omega_\alpha}|\tilde{\psi}_{\alpha,k}\rangle, \quad \alpha = 1, 2, \ldots 2N. \tag{2.69}$$

The evolution equations read

$$a_\alpha(t+1) = a_\alpha(t)e^{i\omega_\alpha}$$
$$+ig\sin\theta \sum_{\alpha_1,\alpha_2,\alpha_3}^{2N} I_{\alpha,\alpha_1,\alpha_2,\alpha_3} a_{\alpha_1}(t)^* a_{\alpha_2}(t) a_{\alpha_3}(t), \tag{2.70}$$

with the overlap integral

$$I_{\alpha,\alpha_1,\alpha_2,\alpha_3} = \sum_{k=1}^{N} \langle\tilde{\Psi}_{\alpha,k}|\hat{v}^{(k)}|\tilde{\Psi}_{\alpha_3,k}\rangle \left\langle \tilde{\Psi}_{\alpha_1,k}|\tilde{\Psi}_{\alpha_2,k} \right\rangle. \tag{2.71}$$

The structure of these equations is strikingly similar to the ones obtained from Hamiltonian dynamics [74, 73], including one of DNLS (1.24). In particular, we obtain cubic nonlinear terms on the rhs of the asymptotic expansion (2.70). Taking in account one-dimensionality of the system, the prediction of a subdiffusive exponent $\alpha = 1/3$ follows from Ref.[74, 73].

### 2.3.4 Numerical approach

Let us discuss the details of simulations and data analysis. We directly propagate the evolution equation (2.65). The initial conditions are uniformly spread over several neighboring sites to guarantee the positivity and unitarity of the coins (due to the square root term in coins).

The only source of the numerical error is the round-off errors of the finite dimensional computer algebra. We estimate the error by means of the total packet norm: it is equal to $1$ for initial conditions and is should to be constant due to unitary evolution. The relative value of the error never exceeded $10^{-4}$ in simulations.

In all the calculations we use $\theta = \pi/4$. The system size $N$ is between $2000$ and $2500$. The results which include ensemble averaging employ around $10^2$ realizations of the random field $\{\xi_n\}$. The total evolution times reach up to $2 \times 10^{12}$ time steps, which exceeds the maximum previously gained limits for such analysis to the best of our knowledge.

The averaged curve of the second moment $m_2(t)$ is smoothed with the locally weighted regression smoothing (LOESS) algorithm [152, 153]. The power-law exponent is then calculated as the two-point derivative of the smoothed data. To verify the smoothing procedure and exclude overfitting, we also performed fitting with Hodrick-Prescott filter [154], Gaussian convolution smoothing, and local 4th order polynomial fitting with an analytical derivative. We found all methods generating results within the statistical margin of error. The LOESS approach turned out to be more robust than other options. When averaging we estimate the error as the standard error of the mean.

To additionally speed up the simulations and reach larger time limits, we used GPU computing with CUDA language.

### 2.3.5 Discussion

DTQW represent very useful unitary map toolboxes which allow for extremely fast quantum evolution, in particular, due to covering finite times with one step (jump), and due to the fast(est) realization of a transfer/hooping/interaction on a lattice. We used a disordered version to obtain Anderson localization with the spectrum being dense and gapless and with compact support. The resulting localization length $L_{\text{loc}}$ does not depend on the eigenvalue of an eigenstate and can be smoothly changed in its whole range of existence using one of the control parameters of the DTQW. We then generalize the map to a nonlinear disordered

DTQW and study the destruction of Anderson localization. Wave packets spread subdiffusively with their second moment $m_2 \sim t^{\alpha}$ and the universal exponent $\alpha = 1/3$. The record time $t_f = 2 \times 10^{12}$ is reached, which exceeds old horizons by 3-4 orders of magnitude. The size of the wave packet reaches $\approx 300 L_{\text{loc}}$. The normalized strength (or better weakness) of the nonlinear terms in the DTQW reaches $0.01/\pi \approx 0.003$. No slowing down of the subdiffusive process was observed. Therefore chaotic dynamics appears to survive in the asymptotic limit of decreasing wave packet densities.

We also expect DTQW to be useful in the future also for exploring other hard computational tasks, e.g. subdiffusion in two-dimensional and even three-dimensional nonlinear disordered lattices, and many-body localization in interacting quantum settings.

## 2.4 Almost compact moving breathers with fine-tuned discrete-time quantum walks

In this section, we utilize a nonlinear generalization of DTQW to study numerically the dynamics of solitary type excitations in discrete lattices with great efficiency. We consider a fine-tuned case of DTQW, where dispersion relation has two flat bands. In this case, transport is inhibited and compact localized states can form (2.8). Nonlinearity leads to the appearance of stable moving and stationary nonlinear excitations which are superexponentially localized. The system further allows for fully compact bullet excitations moving with the maximum velocity, which may also form an intriguing interacting gas with unusual scattering properties.

By making use of direct numerical simulations and complementary generalized Newton schemes, we arrive at a plethora of nonlinear DTQW breathers. The breathers have superexponential tails, can be stationary, but also moving with a dense set of velocities. Analytical results fit well with numerical observations.

The section is organized as follows: in subsection 2.4.1 we briefly discuss discrete breathers, in subsection 2.4.2 we show the setup, in subsection 2.4.3 we study the resulting dynamics, and subsection 2.4.4 provides final discussion.

### 2.4.1 Discrete breathers

Discrete breathers (DBs) [155, 156] are generic time-periodic and spatially localized solutions of broad classes of nonlinear Hamiltonian network equations [157]. Discrete breathers have been observed and studied in a fascinating and broad setting of physical realizations which cover several decades of temporal and spatial scales (see [156] for a detailed review). If the band structure of the linear part of the Hamiltonian network is not degenerated, then DBs are localized in space either following an exponential decay (for analytical band structure functions) or an algebraic one (for non-analytical ones) [158]. If the band structure is degenerated and consists of flat band(s) only, DBs localize superexponentially fast [159]. For specific local symmetries of flat bands, DBs can even maintain the compactness of the corresponding compact localized eigenstates of the linear Hamiltonian network equations [160].

The generic appearance of DBs in lattice structures comes at a price. The absence of a continuous translational invariance (which is replaced at best by some discrete one) makes travel hard. Indeed, multiple attempts to obtain lossless traveling discrete breathers suffered from facing resonances between the velocity $v$ of a moving DB candidate, and phase velocities of small amplitude plane waves [156]. These – usually unavoidable – resonances produce nondecaying tails. Relief could be only obtained by choosing generalized discrete nonlinear Schrödinger equations. Their global phase (or simply gauge) symmetry allows finding stationary DBs supported by just one harmonics in time. Then, tailless moving DBs could be obtained for a discrete and non-empty set of velocity values $v$ [161, 162]. In contrast, systems with continuous translational invariance not only allow for Galilean or Lorentz boosting of solitary excitations if their stationary parents exist but also permit the occurrence of completely compact moving solitary excitations in nonlinear partial differential equations which have a missing linear dispersive part [163]. Interestingly, G. James reported recently on an attempt to find traveling DBs in a strongly nonlinear discrete nonlinear Schrödinger chain with a missing linear dispersive part [164]. However, even in this case nondecaying tails were observed.

### 2.4.2 Setup

For our purposes of DB study, we choose the simplest and generic version of the coin matrix (2.1) controlled by only one parameter $\theta$:

$$\hat{U}_n = \begin{pmatrix} \cos\theta_n & \sin\theta_n \\ -\sin\theta_n & \cos\theta_n \end{pmatrix}, \tag{2.72}$$

where the angles $\theta_n = \theta + \lambda S_n$ are nonlinear functions of $\hat{\psi}_n$ with the norm density per site $S_n = (|\psi_{+,n}|^2 + |\psi_{-,n}|^2)$. Note, that for finite systems with $N$ sites periodic boundary conditions are used. The evolution is unitary and preserves the total norm of the wave function $S = \sum_{n=1}^{N} S_n$. As discussed in subsection 2.1.4, the model enjoys stroboscopic sublattice factorization (SSF), i.e. the evolution on even and odd sites decouples locally if iterated over two time units. Indeed, it follows from Eqs. (2.3) that $\hat{\psi}_n(t+2) = F\left(\hat{\psi}_{n-2}(t), \hat{\psi}_n(t), \hat{\psi}_{n+2}(t)\right)$. For even $N$, it follows that the dynamics on even and odd sites decouples completely, similar to the two distinct sides of a Möbius band with no twists. Instead, for odd $N$ the even and odd site dynamics is globally coupled: any excitation on an even site will explore all even sites until it reaches the boundary where it turns odd and vice versa. This geometry is similar to the Möbius band with a nontrivial twist.

In the linear regime $\lambda = 0$, the coin operators turn identical and space-index independent. This leads to dispersion relation (2.6),

$$\cos(\omega) = \cos(\theta)\cos(k). \tag{2.73}$$

We note that for the particular choice $\theta = \pi/2$ the resulting band structure is composed of two flat bands $\omega = \pm\pi/2$. In this case, compact localized eigenstates exist (2.8), and the quantum walk dynamics is quenched resulting in the halt of any propagating wave. Adding back the nonlinear terms with $\lambda \neq 0$, any possible observed transport will be entirely due to the nonlinear terms which lead to an interaction between the compact localized states. For the remaining part of this section, we will use $\theta = \pi/2$ only.

### 2.4.3 Results

#### 2.4.3.1 Numerical observations

We study the evolution of a single-site initial condition $\hat{\psi}_n(0) = \frac{\delta_{n,n_0}}{\sqrt{2}}\{1,1\}$ with the wave function norm $S = 1$, launched at site $n_0$. We use a system with size $N = 2000$ and periodic boundary conditions, and evolve up to time $t = 10^4$. The resulting evolution of the norm density distribution is shown in Fig.2.19 for $\lambda = 0.1$. Part of the excitation remains in a relatively narrow core region, which spreads, albeit very slowly. Our main observation is that the core emits solitary type excitations at various times, which then continue to travel separately at various velocities. These objects are obvious candidates for traveling discrete breathers. In the following, we analyze them in detail.

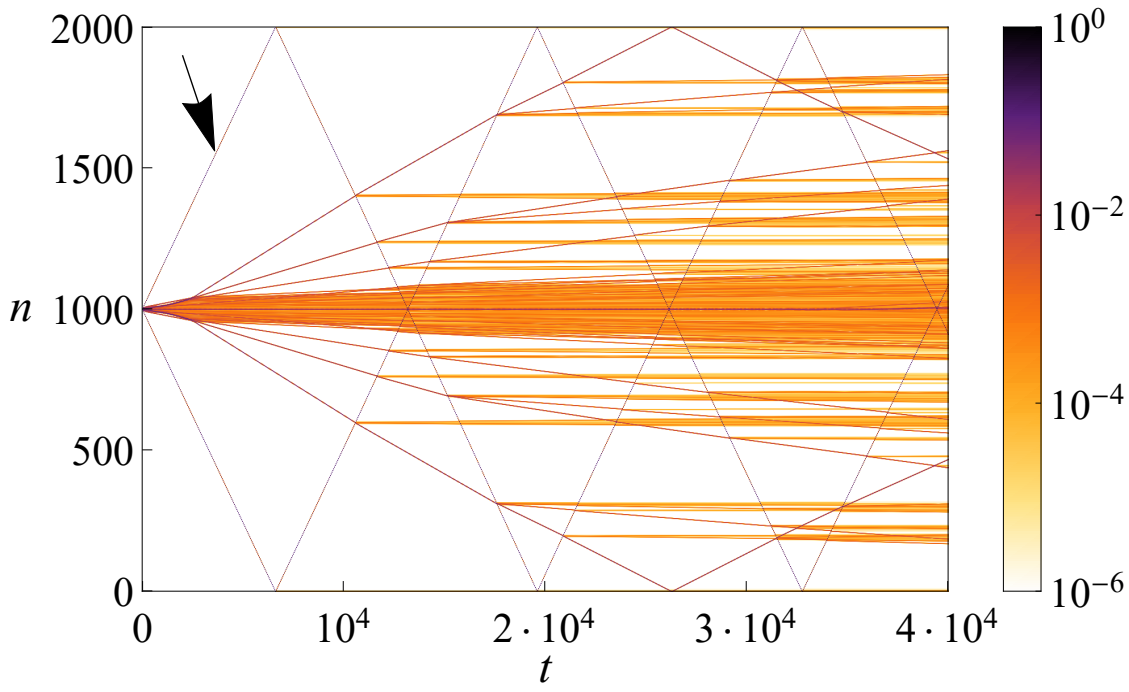

Figure 2.19: Typical temporal and spatial pattern of the nonlinear DTQW dynamics. Norm density $S_n$ is plotted as a function of time and coordinate with color coding on a logarithmic scale. $\lambda = 0.1$, with initial condition $\hat{\psi}_n(0) = \frac{\delta_{n,n_0}}{\sqrt{2}}\{1,1\}$ and $n_0 = 1000$. The arrow indicates the moving solitary excitation analyzed in Fig. 2.20.

#### 2.4.3.2 Cut and paste procedure

Since the moving solitary type excitations are well separated in space, we apply a 'Cut and Paste' procedure. We (i) evolve the system up to an appropriate cutting time $t_c$, (ii) identify the position $n_c$ of the core of a single isolated excitation, (iii) obtain the distance $l_c$ from the core, at which the wave function amplitudes decay to noise levels $\sim 10^{-3}$, and (iv) put the wave function amplitudes to zero for all sites $n < n_c - l_c$ and $n > n_c + l_c$. With a trivial re-shifting of temporal and spatial coordinates, we continue the evolution of the single moving object and its analysis. In particular, we repeat the 'Cut and Paste' procedure several times in order to allow the excitation to converge to a state which is almost radiationless, i.e. which is not leaving weakly excited sites behind. In order to get a real-valued object, we take an absolute value of each component of $\hat{\psi}$ and repeat the procedure.

An example is shown in Fig. 2.20. It corresponds to an originally found moving solitary object with approximate speed $v \approx 1/37$ indicated by the ar-row in Fig.2.19. The profile of the norm density distribution $S_n(t_c)$ is shown in Fig.2.20(a). It is almost compactly localized. To zoom into the tails, we plot $\log_{10} S_n(t_c)$ in the inset of Fig. 2.20(a) and observe tails which are decaying faster than exponential, i.e. superexponential (more details below). Due to the small velocity of that solitary excitation it is more instructive to observe the evolution by following the time dependence of the wave function amplitude at a given site. Using the SSF symmetry, we monitor the evolution of $\hat{\psi}_n$ for a few particular even sites at even times. The time-dependence of $|\psi_{\pm,n}|^2$ for the right-moving excitation from Fig.2.20(a) is shown in Fig. 2.20(b) for six consecutive even sites $n = 40, 42, 44, 46, 48, 50$ and for even times $t = 2t_e$. We observe equidistantly shifted curves for each of the wave function components, which indicate a motion at constant speed $v$, and a symmetry between the evolution of both wave function components:

$$\psi_{\pm,n+2}(t + \frac{2}{v}) = \psi_{\pm,n}(t) \ , \ \psi_{+,n}(t + \frac{1}{v}) = -\psi_{-,n}(t) \ . \tag{2.74}$$

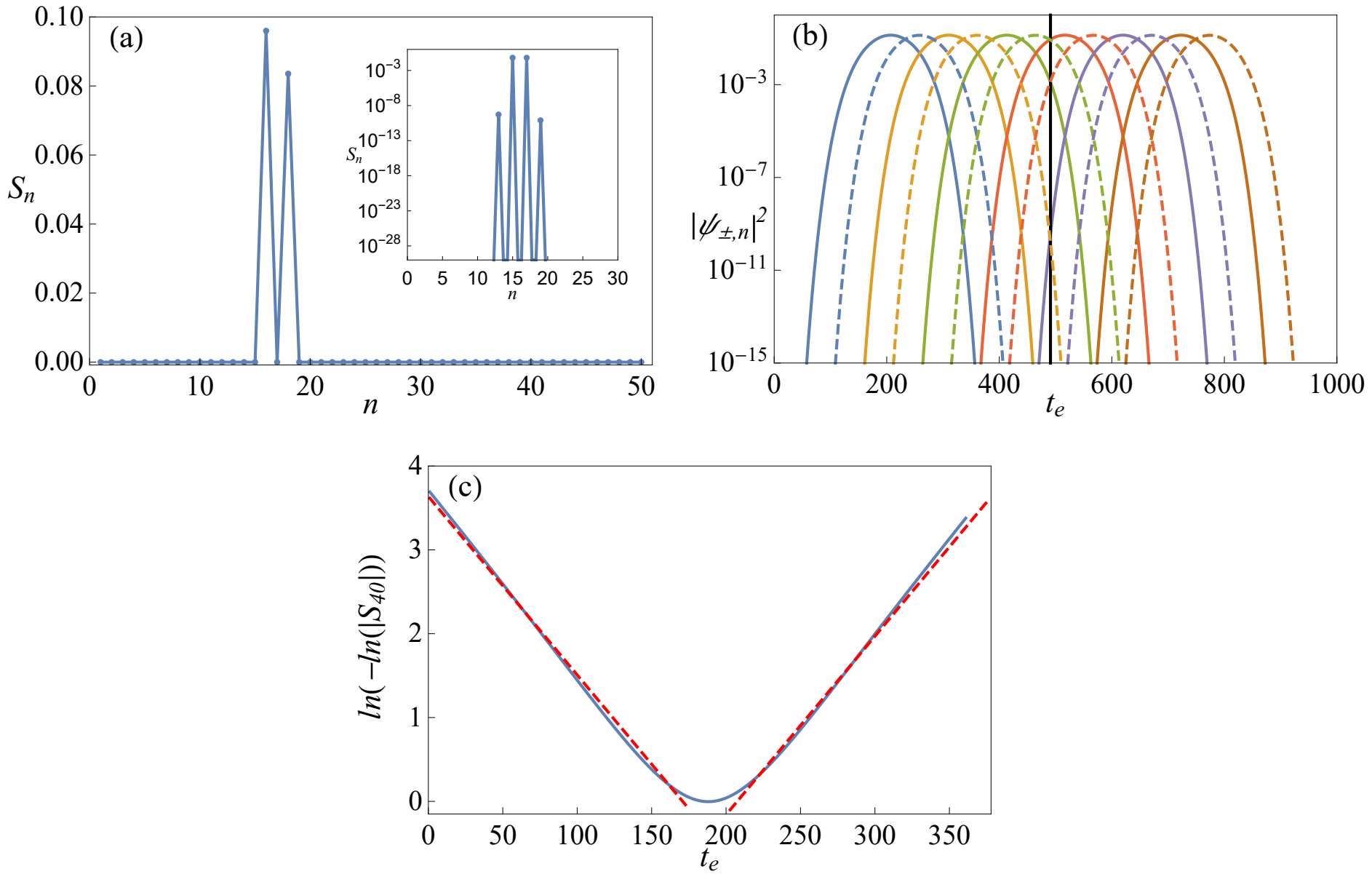

Figure 2.20: Analysis of a moving solitary excitation with total norm $S = 0.175$ and $\lambda = 0.1$. (a) Snapshot of the norm density distribution $S_n(t)$ versus $n$. (b) The time evolution of $|\psi_{\pm,n}|^2$ for $n = 40, 42, 44, 46, 48, 50$ (from left to right). Solid lines – $|\psi_{+,n}|^2$, dashed lines – $|\psi_{-,n}|^2$. The vertical black line guides the eye for the observation of a symmetry between both (see text for details). (c) $S_{40}$ versus even time $t_e$ on double logarithmic scale (blue solid line). Red dashed lines show the predicted superexponential decay (see Eqs. (2.82) and (2.83)).

Together with faster than exponential tail decay, it follows that the rear tail profile of that right moving solitary excitation satisfies the inequalities,

$$|\psi_{+,n}(t)| \ll |\psi_{-,n}(t)| \ll |\psi_{+,n+2}(t)| \ . \tag{2.75}$$

Note, that it is straightforward to generalize these inequalities to the front tail of a right-moving solitary excitation, and to left moving excitations as well. We will use these conditions below to obtain analytical results. Finally, we plot the time dependence of the norm $S_{40}(t_e)$ in Fig.2.20(c) on a double logarithmic scale. A superexponential decay of both the front and the rear tails is clearly observed.

#### 2.4.3.3 Tail analysis

In order to describe the tails of a moving solitary excitation, let us expand Eq.(2.3) to the first order in $\lambda S_n \ll 1$:

$$\psi_{\pm,n}(t+1) = -\lambda S_{n\mp1}(t)\psi_{\pm,n\mp1}(t) \pm \psi_{\mp,n\mp1}(t) \ . \tag{2.76}$$

Reducing the analysis to even times and sites (SSF symmetry) and accounting for the inequalities (2.75) in the rear tail, we arrive at

$$\begin{aligned} \psi_{+,2n}(2t) + \psi_{+,2n}(2t-2) &= -\lambda G\left[\psi_{-,2n}(2t-2)\right], \\ \psi_{-,2n}(2t) + \psi_{-,2n}(2t-2) &= \lambda G\left[\psi_{+,2n+2}(2t-2)\right], \end{aligned} \tag{2.77}$$

where $G[\psi] = |\psi|^2\psi$. To solve the above equations we use the traveling wave ansatz,

$$\psi_{+,2n}(2t) = (-1)^{t+n} g_r(2vt - 2n),$$
$$\psi_{-,2n}(2t) = (-1)^{t+n+1} g_r(2vt - 2n - m), \tag{2.78}$$

with a yet to be determined argument shift $m$. The real-valued rear tail function $g_r(x)$ satisfies the difference equations

$$g_r(y) - g_r(y - 2v) = -\lambda g_r^3(y - m),$$
$$g_r(y - m) - g_r(y - 2v - m) = -\lambda g_r^3(y - 2), \tag{2.79}$$

with $y = 2vt - 2n$. Since both equations have to deliver the same solution, we conclude that $m = 1$, confirming the validity of Eq. (2.74).

If the velocity of the moving solitary excitation with core position $y = n_c$ is small, i.e. $v \ll 1$, we can replace the difference equations (2.79) by a nonlinear differential equation with a discrete delay to describe the rear tail dynamics:

$$2vg_r'(y) = -\lambda g_r^3(y - 1) \; , \; y \ll n_c \; . \tag{2.80}$$

A straightforward generalization to the front tail dynamics yields

$$2vg_f'(y) = \lambda g_f^3(y + 1) \; , \; y \gg n_c \; . \tag{2.81}$$

Both rear and front tail differential equations (2.80) and (2.81) yield the superexponential decay solution,

$$g(y) = A \exp\left[-\alpha e^{|y|/\xi} + \beta|y|\right] \; , \; |y - n_c| \gg 1 \; . \tag{2.82}$$

Indeed, substituting (2.82) into (2.80),(2.81) we obtain the superexponential decay length $\xi$:

$$\xi = 1/\ln 3 \; , \; \beta = 1/(2\xi) \; , \; 2v\alpha/\xi = \lambda e^{-3\beta} A^2 \; . \tag{2.83}$$

The superexponential decay is plotted in Fig.2.20(c) and agrees well with the numerically observed front and rear tails.

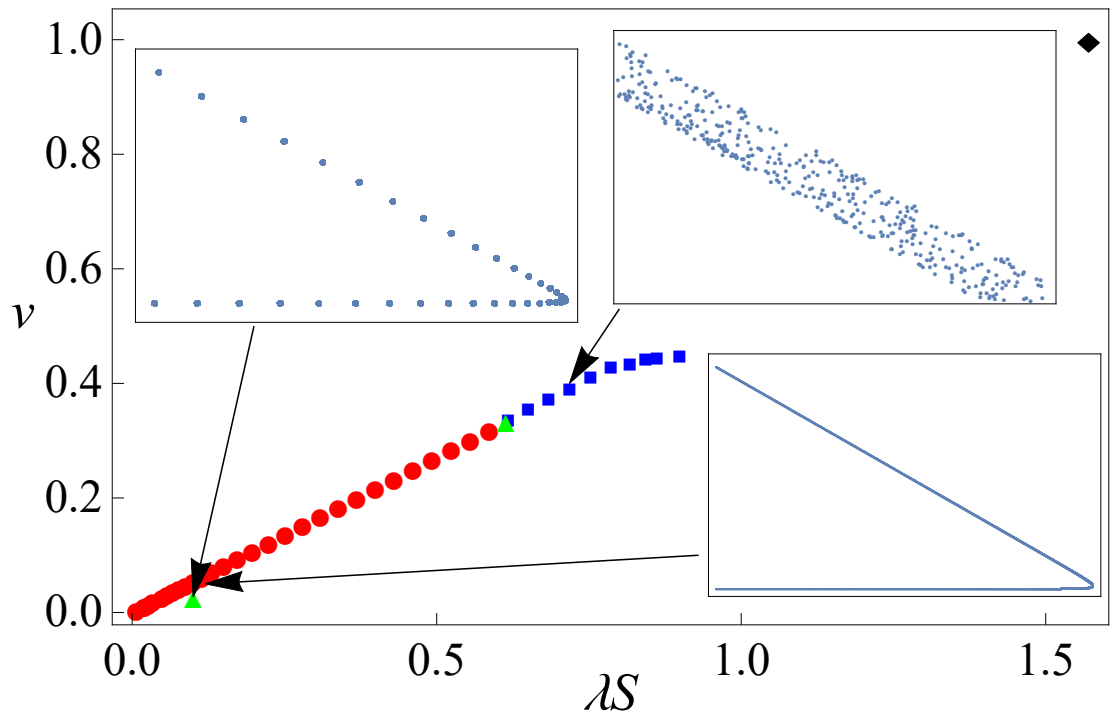

Figure 2.21: The dependence of the discrete breather velocity $v$ on the parameter $\lambda S$. Green triangles: Newton-generated periodic solutions; red circles: quasi-periodic solutions; blue squares – transient chaotic solutions. Black diamond – analytical bullet solution. Insets show the corresponding generalized Poincare section plots.

Both tail solutions have to be glued together in the core of the moving excitation, where the above analysis does not hold. Therefore, in general, $A$ could be different in the front and the tail for asymmetric profiles. However the numerical results in Fig.2.20(b,c) clearly indicate that the profiles are symmetric. Together with the reasonable assumption that the square amplitude in the tails is proportional to the total norm $S$ of the whole moving solitary excitation $|A|^2 \propto S$, we arrive at the scaling relation

$$v \propto \lambda S \ . \tag{2.84}$$

To test this relation, we perform a single "cut and paste' procedure to a number of moving solitary excitations as shown e.g. in Fig.2.19, measure their norm and velocity, and plot the result in Fig.2.21. We observe good agreement with (2.84) for $\lambda S \leq 0.7$ and corresponding velocities $v \leq 0.4$.

#### 2.4.3.4 Moving discrete breathers

For the range of parameters $\lambda S < 0.9$ we were able to obtain a number of moving discrete breathers applying once the procedure of "cut and paste". We characterize the internal dynamics of these objects by computing a generalized Poincare section in a co-moving frame: at each time $t$ we obtain the position $m$ of the largest value of $S_n(t)$ and plot $S_{m+2}$ versus $S_m$. The results are shown in Fig.2.22. We find three different types of discrete breathers: periodic, quasiperiodic, and chaotic.

Periodic moving discrete breathers are characterized by a rational value of their velocity, which leads to a finite number of points on the Poincare section as shown in Fig.2.22(a), where $v = 1/37$, $\lambda = 0.1$ and $S = 1$. The profile of the moving discrete breather is fully restored after 37 iterations and one additional shift along the lattice (up to a global phase).

Quasiperiodic moving discrete breathers have an irrational velocity with a Poincare section which forms a dense 1D line segment for an infinite number of iterations. An example is shown in Fig.2.22(b) for $v \approx 0.041$, $\lambda = 0.5$ and $S = 0.15$.

Chaotic moving discrete breathers generate a Poincare section which corresponds to a stripe segment with finite width and additional fine (potentially fractal) structure inside that, unlike the previous, case doesn't follow a regular path. An example is shown in Fig.2.22(c) for $v \approx 0.39$, $\lambda = 1.7$ and $S = 0.42$.

Chaotic moving discrete breathers cannot be cast into the form of a traveling wave solution (2.78), and, therefore, cease to be exact solutions. Instead, these objects are slowly losing the norm by radiating plane waves in their wake. These leftovers do not propagate further due to the flat band structure of the small amplitude equations. Thus, the chaotic-moving breathers are slowing down their speed $v$. The rate of that process is probably related to the thickness of the above stripe segments in the Poincare sections. Still, that rate can be very small, such that we observe traveling chaotic breathers over several thousands of lattice sites without a notable change of their speed.

Quasiperiodic moving discrete breathers, instead, can very well correspond to exact traveling wave solutions. However, it may well be that these objects are chaotic breathers with very narrow and thus undetected finite width of the stripe segments. It may also well be that these objects are, in fact, exact traveling waves but with rational values of their velocity $v$, which leads to a period, which is of the order of the simulation time. We are not aware of proper computational means to tell these different scenarios apart.

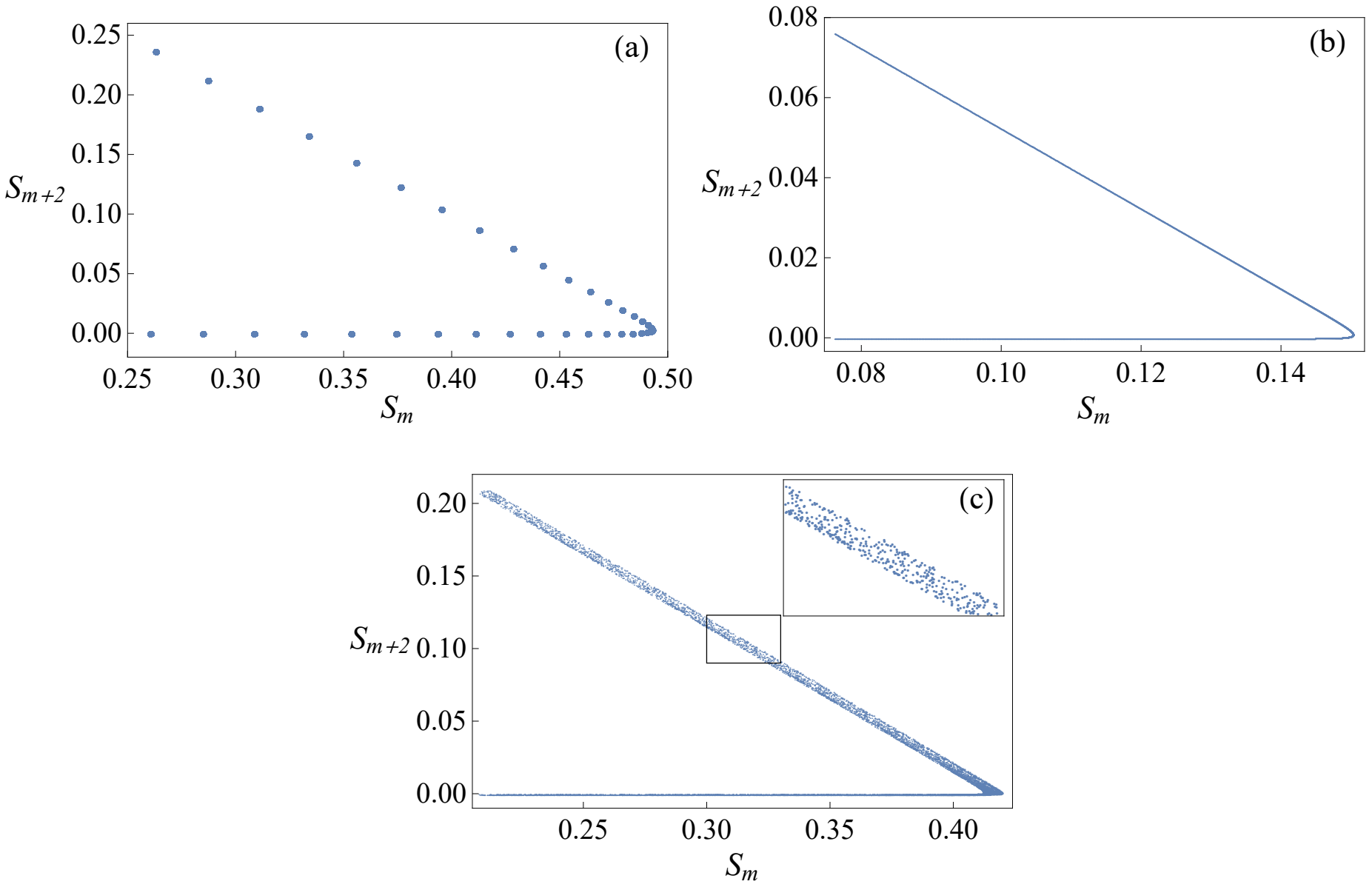

Figure 2.22: The Poincare sectioning of moving discrete breathers (see text for details). (a) Periodic discrete breather with rational velocity $v = 1/37$, $\lambda = 0.1$ and $S = 1$. (b) Quasiperiodic discrete breather with velocity $v \approx 0.041$, $\lambda = 0.5$ and $S = 0.15$. (c) Chaotic discrete breather with $v \approx 0.39$, $\lambda = 1.7$ and $S = 0.42$.

Periodic moving discrete breathers can be obtained with very high numerical precision using a generalized Newton scheme. For that we rewrite Eq.(2.3) in a compact form as a unitary map of the entire field $\hat{\psi} \equiv \left\{\hat{\psi}_n\right\}$. The action of one iteration in (2.3) is just a nonlinear unitary map $\hat{\psi}(t+1) = \mathcal{U}\hat{\psi}(t)$. A translation shift along the lattice by one lattice site will be denoted by $\mathcal{T}\left\{\hat{\psi}_n\right\} =$ $\left\{\hat{\psi}_{n+1}\right\}$. Then, a periodic moving breather is encoded by two integers $p, q$ and its frequency $\Omega$, such that the breather field satisfies

$$\left(\mathcal{U}^q - e^{i\Omega q}\mathcal{T}^p\right)\hat{\psi} = 0 \ , \ v = \frac{p}{q} \ . \tag{2.85}$$

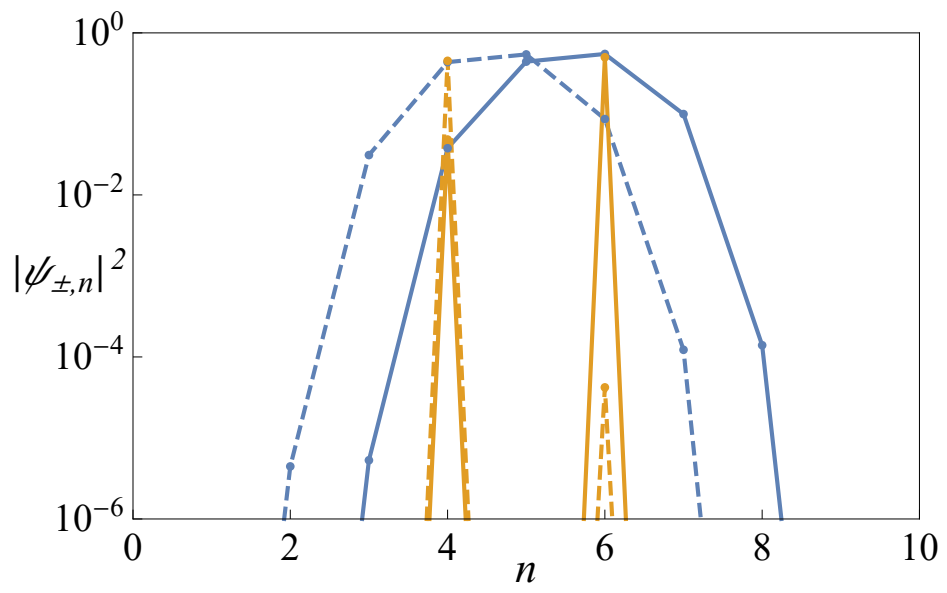

Figure 2.23: Snapshots of spatial distributions of $|\psi_\pm|^2$ for two periodic moving discrete breathers. Solid lines – $|\psi_{+n}|^2$, dashed lines – $|\psi_{-n}|^2$. These periodic solitary excitations correspond to values of $p = 1, q = 37$ (blue lines) and $p = 1, q = 3$ (orange lines).

Solutions of (2.85) can be obtained using standard Newton schemes which search for zeros of vector functions [165]. We found two such solutions. The first one was described above, and has $v = 1/37$, $q = 37, p = 1$ and $\Omega = \pi/37$. The second one has velocity $v = 1/3$, $q = 3, p = 1$ and $\Omega = 2\pi/3$. Both solutions are shown with green circles in Fig.2.21. The profiles are shown in Fig.2.23.

#### 2.4.3.5 Bullets

For the particular value of the nonlinearity $\lambda S = \pm\pi/2$ the equations (2.3) admit exact moving and compact excitations, which we coin *bullets*. These bullets have a nonzero field amplitude on only one lattice site on only one of the two wave function components. Its velocity $v = \pm 1$ (see Fig.2.21). A right-moving bullet is given by

$$\hat{\psi}_n(t) = (-1)^{st}\sqrt{S}\left\{e^{i\phi_+}, 0\right\}\delta_{n,n_0+t} \ , \tag{2.86}$$

with $s = \pm 1$ being the sign of $\lambda$. A left-moving bullet, respectively, is given by

$$\hat{\psi}_n(t) = (-1)^{st}\sqrt{S}\,\{0, \mathrm{e}^{i\phi_-}\}\,\delta_{n,n_0-t}\ . \tag{2.87}$$

Thus, the dynamics of a single bullet is characterized by two conserved quantities – the total norm $S = \pi/(2|\lambda|)$ and the phase $\phi_\pm$. Interestingly, the dynamics of *a gas* of left- and right-moving bullets shows the conservation of the numbers of left and right movers. Two bullets approaching each other will experience no interaction if their relative distance is odd, but will undergo an act of elastic scattering if that distance is even. After such an elastic reflection, a left mover is turned into a right mover (carrying its phase $\phi$ with it) and vice versa. The evolution of such a gas of left and right moving bullets is shown in Fig.2.24(a,b). The dynamics of the gas will lead to instabilities in the presence of a small noisy background. This background is generated during the computation due to roundoff errors. It is observed in Fig.2.24(a) at times $t \approx 500$. The time dependence of the smallest background norm, $S_{min}$, during that evolution is shown in Fig.2.24(c). The background intensity is growing exponentially fast up to times $t \approx 1400$. Yet, for larger times the system still shows up with an interacting gas of moving solitary excitations with various velocities, see Fig.2.24(a).

#### 2.4.3.6 Stationary discrete breathers

In addition to the above surfeit of moving discrete breathers, we also report on the existence of stationary discrete breathers with zero velocity. These objects are zeros of the nonlinear map

$$\left(\mathcal{U}^q - \mathrm{e}^{i\Omega q}\mathcal{I}\right)\hat{\psi} = 0\ , \tag{2.88}$$

where $\mathcal{I}$ is the unity operator. Below, we will consider the case $q = 1$ only. Solutions can be again searched for by using a generalized Newton scheme. The spatial profile of one of these solutions is shown in Fig.2.25(a). A double logarithmic plot of the profile in Fig.2.25(b) reveals its superexponential tails. Interestingly, the deviation of the stationary breather frequency $\Omega$ from $\pi/2$ yields a linear dependence on the control parameter $\lambda S$ (see Fig.2.25(c)).

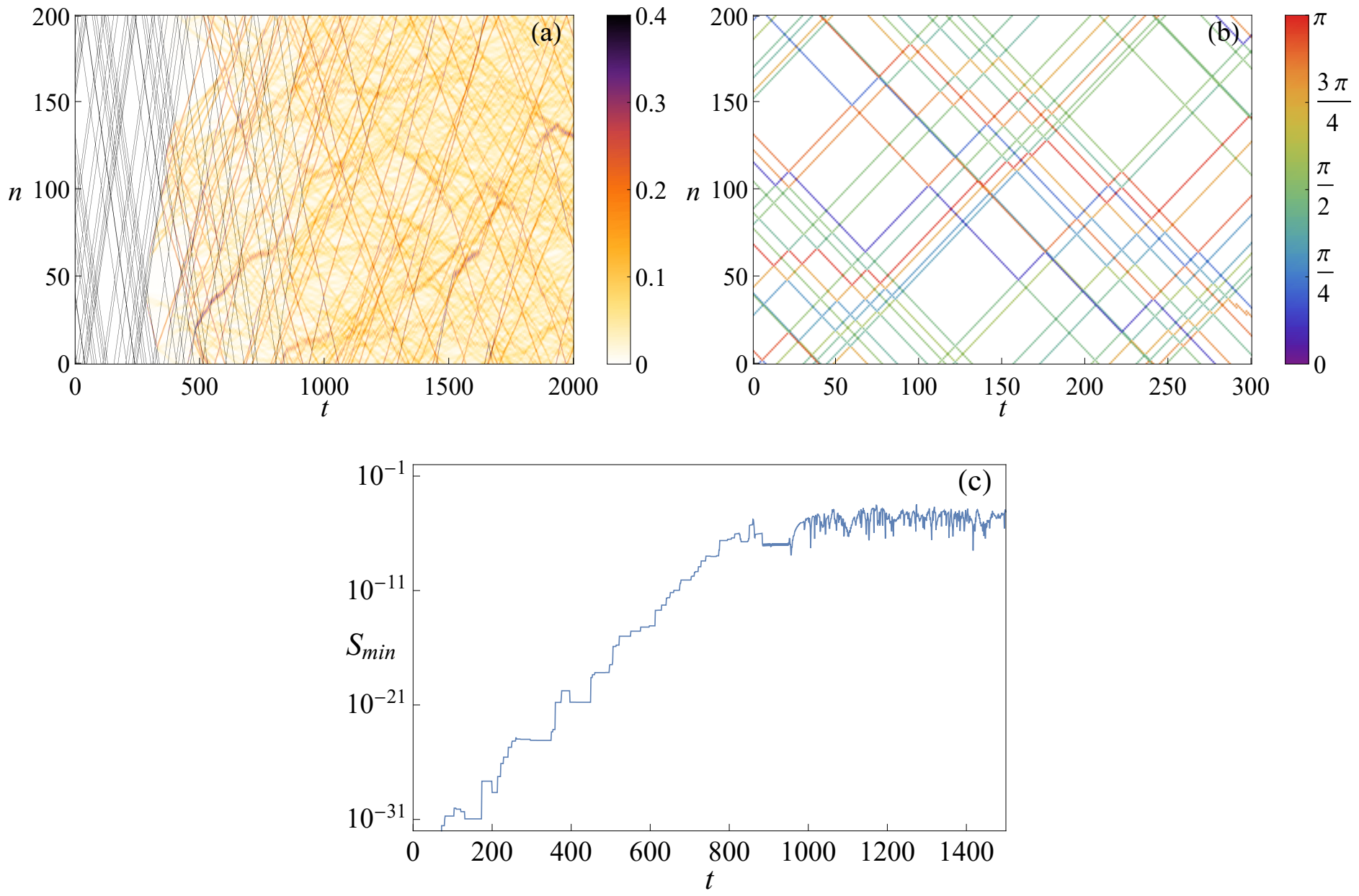

Figure 2.24: Evolution of a bullet gas with randomly chosen initial coordinates and phases. (a) Spatial and temporal dependence of the norm density, $S_n(t)$; (b) Bullet phases (modulo of $2\pi$) , $\phi_\pm$, as a function of space and time; (c) Time evolution of the smallest $S_{min}$.

To obtain an analytical solution for the tails of a stationary breather, we use the observation from Fig.2.25(a) that one of the two components is dominating in the tail, e.g. $|\psi_{+,n}| \gg |\psi_{-,n}|$ for the right tail of the breather. Note that the final result will be invariant on the choice of the tail. With the ansatz

$$\psi_{+,n}(t) = \exp[i\Omega t + \phi_+]g(n),$$

(2.89)

$$\psi_{-,n}(t) = \exp[i\Omega t + \phi_-]g(n + n_0),$$

and Eq. (2.76) we obtain $n_0 = 1$ and

$$2g(n)\cos\Omega = -\lambda g^3(n-1). \quad (2.90)$$

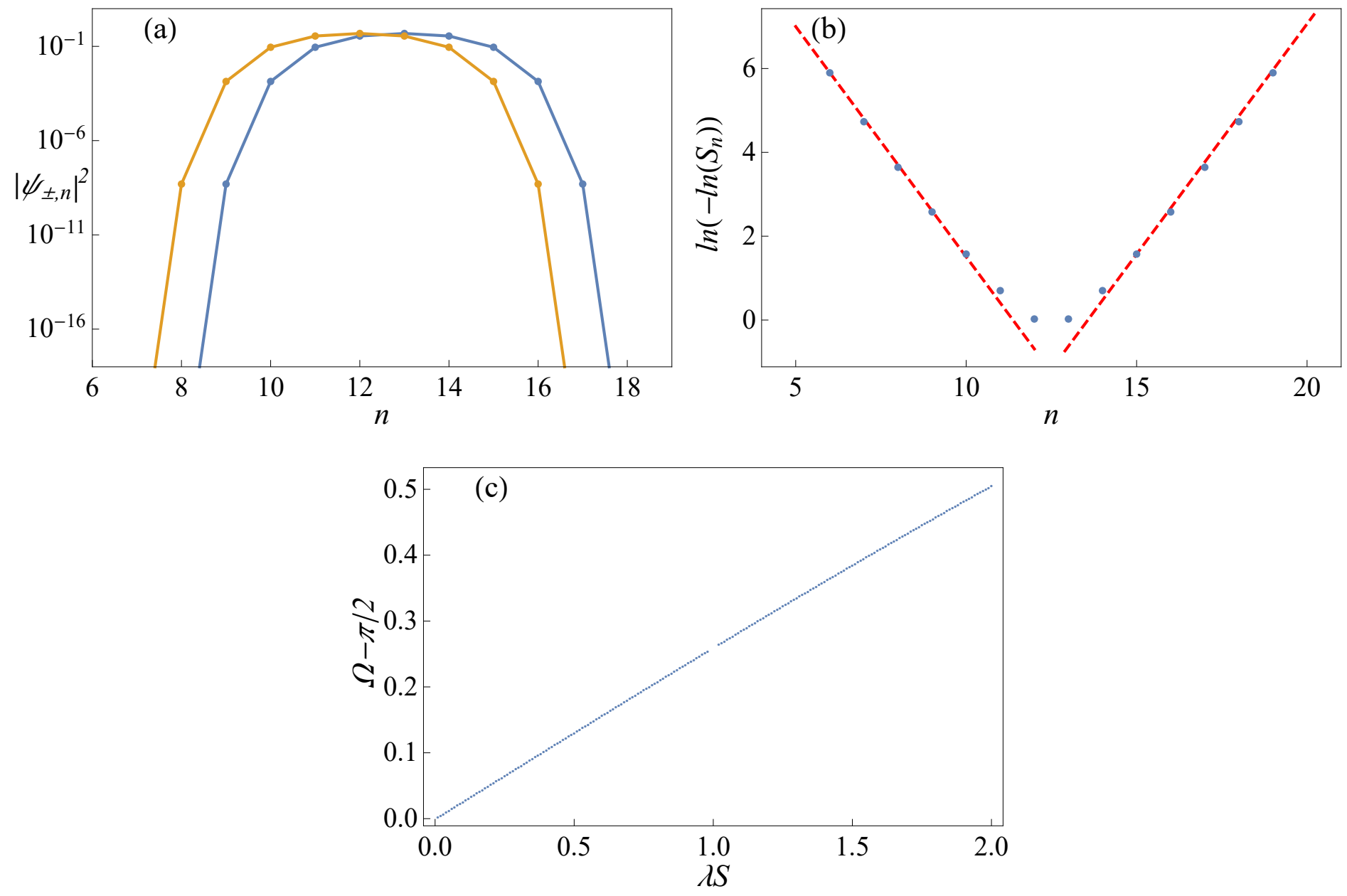

Figure 2.25: (a) A snapshot of a single frequency stationary breather solution. Orange line – $|\psi_{+n}|^2$, blue line – $|\psi_{-n}|^2$; (b) Spatial dependence of $S_n$ showing superexponential decay in its tails (the red dashed line corresponds to the solution of (2.90); (c) Scaling of the breather frequency with the parameter $\lambda S$.

The solution of the above nonlinear difference equation yields the superexponential decay $g(n) = A\exp\left[-\alpha e^{|n|/\xi}\right], \quad |n| \gg 1$ with $\xi = 1/\ln 3$. The frequency of the breather is determined by both the nonlinearity and its norm: $(\Omega - \pi/2) \propto \lambda S$, and this scaling is in a good accord with numerical analysis, see the Fig. 2.25(c).

### 2.4.4 Discussion

Discrete-time quantum walks are unitary maps defined on the Hilbert space of coupled two-level systems, which turn into a very efficient unitary toolbox for addressing a variety of problems which lack easy solutions in Hamiltonian settings. In particular, we studied the dynamics of excitations in a nonlinear discrete-time quantum walk, whose fine-tuned linear counterpart has a flat band structure. The linear counterpart is, therefore, lacking transport, with exact solutions being compactly localized. A solitary entity of the nonlinear walk moving at velocity $v$ is then shown to not suffer from resonances with small amplitude plane waves with identical phase velocity. That solitary excitation is localized stronger than exponential, due to the absence of a linear dispersion. We found a set of stationary and moving breathers with almost compact superexponential spatial tails. At the limit of the largest velocity $v = 1$ the moving breather turns into a completely compact bullet. Remarkably, these bullets can form an interacting gas with the bullet phases participating in the scattering process. It remains an interesting question as to whether such highly localized – and even compact – moving nonlinear solitary excitations may be used in applications involving discrete-time quantum walks.

# Chapter 3

# Non-Equilibrium States in Lindblad Systems

## 3.1 Introduction

A quantum system introduced to an environment generally ceases to be describable by a wave function. As usually we do not have a complete description of the environment, the state of the system becomes a mix of pure states, called density matrix

$$\varrho = \sum_i p_i |\Psi_i\rangle\langle\Psi_i|. \tag{3.1}$$

The evolution of the density matrix in the limit of zero coupling to the environment is described by the von Neuman equation,

$$\dot{\varrho}(t) = -i\,[H, \varrho(t)]\,, \tag{3.2}$$

where $H$ is the system's Hamiltonian. This description is no longer sufficient when a system is coupled to an external environment (bath). A class of generalizations of the von Neuman equation (3.2) that addresses dissipation is called quantum master equations (QME). This term originates from classical master equations, i.e. equations governing the evolution of probabilistic (often Markovian) systems. The classical part of QME is restricted to diagonal dynamics of the density matrix. Coupling between the diagonal and the off-diagonal parts through QME is what constitutes quantum coherence [166].

Among QME, the Gorini–Kossakowski–Sudarshan–Lindblad master equation [167, 168], or shortly, the Lindblad master equation, has a special place. Firstly, it is the only Markovian QME that follows certain restrictions that a priori constitute exact physical correctness – trace preservation and complete positivity (see the next subsection for details). Secondly, the Lindblad equation, along with the Redfield equation [169], is the typical outcome of system-bath weak-coupling approximation [170]. Thirdly, the Lindblad equation has a relatively simple structure that allows for easy implementation and computation [171]. The Lindblad equation reads

$$\dot{\varrho}(t) = \mathcal{L}\varrho(t) = \mathcal{L}_H\varrho(t) + \mathcal{L}_{\mathrm{dis}}\varrho(t) = -i\,[H, \varrho(t)] + \sum_{s=1}^{M} \gamma_s \left[ A_s \varrho(t) A_s^\dagger - \frac{1}{2}\{A_s^\dagger A_s, \varrho(t)\} \right]. \tag{3.3}$$

Here, $\mathcal{L}$ is the Lindbladian (super-)operator that consists of the Hamiltonian, $\mathcal{L}_H$, and the dissipative, $\mathcal{L}_{\mathrm{dis}}$, parts. The sum in the second line is over $M$ dissipative channels, where $M$ is limited by $N^2-1$, where $N$ is the dimension of $\varrho$. Operators $A_s$ are called dissipative (or jump) operators and they can be arbitrary matrices depending on the particular bath choice. They are accompanied by non-negative weights $\gamma_s$.

The role of dissipative operators $A_s$ is illustrative within the quantum trajectories (or quantum jumps) method. This is an approach to solve the Lindblad equation using its stochastic unraveling into a number of non-deterministic "trajectories" in terms of effective wave functions [172]. Within quantum jumps, each trajectory is continuously propagated with an effective Hamiltonian between random "jumps", when dissipative operators act on the wave functions (similarly to measurements) abruptly changing their states. The complete evolution in terms of the density operator $\varrho$ is recovered by a Monte-Carlo sampling over multiple quantum trajectories. Each quantum trajectory can be physically interpreted as the evolution of a system conditioned by specific realizations of measurements that the bath performs over the system [173]. Thus one may look at dissipative operators $A_s$ as at actions/measurements that environment imposes on the system.

### 3.1.1 Ways of derivation of the Lindblad equation

There exist two mainstream ways to derive the Lindblad equation that also correspond to two main roles of the Lindblad equation in dissipative quantum physics. Here we sketch out the core ideas, for more details see, for example, Ref. [14].

The first is microscopic, or also called heuristic method starts with the Hamiltonian

$$H = H_{\mathrm{S}} + H_{\mathrm{E}} + H_{\mathrm{I}}, \tag{3.4}$$

where $H_{\mathrm{S}}$, $H_{\mathrm{E}}$, and $H_{\mathrm{I}}$ and Hamiltonians of the system, environment, and interaction correspondingly. Then, a sequence of approximations is applied

- Born approximation utilizes an assumption that coupling between the system and the environment is weak, and thus, correlations between the system and the environment develop slowly compared to the typical time scales of the system. This allows to use $\varrho \approx \varrho_{\mathrm{S}} \otimes \varrho_{\mathrm{E}}$, where $\varrho_{\mathrm{S}}$ and $\varrho_{\mathrm{E}}$ are partial density matrices $\varrho_{\mathrm{E}} = \mathrm{tr}_{\mathrm{S}}\, \varrho$, $\varrho_{\mathrm{S}} = \mathrm{tr}_{\mathrm{E}}\, \varrho$. This is the first step which allows to decouple the system's evolution from the environment degrees of freedom.

- The Markov approximation implies that excitations transferred from the system to the bath decay quickly (have auto-correlation times much smaller than in-system times). This leads to the assumption $\dot{\varrho}(t) = \mathcal{F}\left(\varrho(t)\right)$, where $\mathcal{F}$ is a not yet specified functional. Using only Born and Markov approximations one arrives at the Redfield equation

$$\dot{\varrho}_{\mathrm{S}}(t) = -\int_0^\infty \mathrm{d}s\, \mathrm{tr}_E \left[H_I(t), \left[H_I(s), \varrho_{\mathrm{S}}(t) \otimes \varrho_E\right]\right], \tag{3.5}$$

  written in the interaction picture here. Importantly, while being in some sense more general, unlike the Lindblad equation, the Redfield equation does not preserve positivity of the density matrix, which may lead to unphysical results [174, 175].

- Rotating wave (secular) approximation asserts that if the typical system times $\sim 1/|E_i - E_k|$, $i \neq k$ of the system are large compared to the relaxation time of the open system they can be averaged out.

These approximations combined produce the Lindblad equation (3.3). Note that all these approximations imply weak coupling, while the rotating wave approximation also requires the environment to be "large" (to have a much denser energy spectrum).

The Lindblad equation is also the most general Markovian trace-preserving completely-positive master equation. In other words, it is the generator of any element of the set of quantum transformations forming the Markovian quantum dynamical semi-group [166]. Consider the set of transformations $\varrho \to C_t \varrho$ defined for positive times $t$ such that for any element of Hilbert-Schmidt space (also called von Neumann algebra) $\varrho$

1. Zero-time propagation corresponds to the identity operator,

   $\mathcal{C}_0 \varrho = \varrho$,

2. Evolution forward in time is always Markovian,

   $\mathcal{C}_{t+s} \varrho = \mathcal{C}_t \mathcal{C}_s \varrho$ , for any positive $t$ and $s$.

3. Transformation $\mathcal{C}_t$ is continuous in time $t$

4. Trace of the density matrix is preserved,

   $\mathrm{tr}(\mathcal{C}_t \varrho) = \mathrm{tr}(\varrho)$

5. Transformation is completely positive. Positivity means that positive operators (which density matrices always are) are transformed into positive operators, i.e. $\mathcal{C}_t \varrho \geq 0$ if $\varrho \geq 0$. Complete positivity is a stronger condition, requiring that if a system is considered together with an arbitrary uncoupled system (ancilla) in the maximally entangled state, transformation is still positive $\mathcal{C}_t\left(\mathbb{1}\otimes\varrho\right)\geq 0$ if $\varrho\geq 0$ for any $t$ and arbitrary-dimensional $\mathbb{1}$ (though it is sufficient to show this for $\mathbb{1}$ of the same dimension as $\varrho$ [176]).

Note that the maximally entangled state is chosen as it contains full information of the system.

Giving the set of these conditions, often shortly called Markovian TPCP (trace preservation and complete positivity), one can prove that the only possible family of generators of this dynamics is the Lindblad equation (3.3).

## 3.2 Technical Aspects of Propagation Large Lindbladian Systems

Computational studies of many-body quantum systems are limited by the so-called curse of dimensionality: the total length $L$ of description (number of parameters required to specify a state) of an isolated quantum system consisting of $N$ components (spins, qubits, ions, etc.), each one with $d$ degrees of freedom, scales as $L(N) \sim d^N$. To specify an *arbitrary* state of a system of $50$ qubits one needs $2^{50} \approx 10^{15}$ complex-valued parameters. In the case of open quantum systems, the complexity squares: to describe a density operator one needs $L(N) \sim d^{2N}$ real-valued parameters.

This is a known problem in modern data science – manipulations (or even simply storing) of data tensors becomes impossible when the data are stored in high-dimensional spaces. The attempts to break the curse led to the development of a variety of low-rank tensor approximation algorithms [177]. These algorithms are used now in signal processing, computer vision, data mining, and neuroscience [178]. The most robust algorithms are based on Singular Value Decomposition (SVD). One particularly efficient algorithm for multilinear algebra manipulations is the so-called Tensor-Train (TT) decomposition [179]. In physical literature, it is commonly referred to as Matrix Product State (MPS) (or Matrix Product Operator (MPO)) representation [180]. While these two names are used simultaneously (though in different fields of research), the underlying mathematical structure is the same [181]. The MPS/MPO/TT approach allows to reduce the description of many-body states to a linear scaling $L(N) \sim N$[179].

The MPS/MPO representation allows for effective propagation of quantum many-body systems in time by using the so-called time-evolving block decimation (TEBD) scheme [182, 183]. In short, this is a procedure to reduce the description of a state, obtained after every propagation step, to a given fixed length $L_{\mathrm{cut}}$. The accuracy of the propagation procedure is controllable through $L_{\mathrm{cut}}$. Some many-body systems "behave" well during the TEBD propagation and so the amount of the neglected information is tolerable (we are not going to discuss physical properties underlying such a "good behavior" here, for details see extensive literature on the subject, e.g., Ref. [180]). Important is that the MPO/TT-TEBD scheme can be used to propagate open systems [183] and thus, get in touch with the corresponding asymptotic states [184, 16]. It is crucial, therefore, to estimate the computational resources needed for the realization of this program.

In this section, following [185], we discuss technical aspects of the TEBD propagation scheme and benchmark our results on known physical models. In subsection 3.2.1 we discuss the algorithm, in subsection 3.2.2 we describe its massively parallel effective implementation, in subsection 3.2.3 we provide benchmarking results, and we conclude in subsection 3.2.4.

### 3.2.1 The TEBD algorithm

#### 3.2.1.1 Tensor-train decomposition and time-evolving block decimation propagation

We start with the TT representation of a $N$-dimensional complex-valued tensor $B^{i_1,i_2\dots i_N}$ with $i_k = 1, 2, \dots M$ [179],

$$B\left[i_1, i_2 \dots i_N\right] = \sum_{\alpha_1 \dots \alpha_{N-1}} \Gamma^{[1]i_1}_{\alpha_0,\alpha_1} \lambda^{[1]}_{\alpha_1} \Gamma^{[2]i_2}_{\alpha_1,\alpha_2} \lambda^{[2]}_{\alpha_2} \dots \Gamma^{[N-2]i_{N-1}}_{\alpha_{N-2},\alpha_{N-1}} \lambda^{[N-1]}_{\alpha_{N-1}} \Gamma^{[N]i_N}_{\alpha_{N-1},\alpha_N}. \tag{3.6}$$

One may interpret this structure as a "train" of $\Gamma$'s that encode local structure in each dimension, and $\lambda$'s that quantify correlations between them. Each $\Gamma^{[k]}$ is an array of $M$ matrices $r_{k-1} \times r_k$ with restrictions $r_j \leq M \max\left(r_{j-1}, r_{j+1}\right)$ with boundary conditions $r_0 = r_N = 1$. Thus, the dimensions of the matrices are $1 \times M,\ M \times M^2,\ M^2 \times M^3 \dots M^2 \times M,\ M \times 1$. This corresponds to the full representation with $M^N$ complex parameters. The construction of the TT representation involves SVD operations (their number is proportional to the number of the system components). When SVD is performed, one can keep only certain singular values based on the approximation criterion. One possibility is to discard all values smaller than a fixed number. An alternative approach is to introduce a so-called **bond dimension** $R$, a cut-off value such that on each bound

**Algorithm 1** : TEBD method implementation

**upload**: system & method parameters (N, $T_j[i_j, i'_j]$, $T_{j,j+1}[i_j, i'_j, i_{j+1}, i'_{j+1}]$, $dt$, $T_{\max}$, $R$), initial state ($\Gamma^{[j]i_j}_{\alpha_{j-1}\alpha_j}$, $\lambda^{[j]}_{\alpha_j}$)
**for** $t = 0$ **to** $T_{\max}$ **do**
    propagate all $\Gamma^{[j]i_j}_{\alpha_{j-1}\alpha_j}$ on $[t; t + dt/2]$
    propagate $\Gamma^{[j]i_j}_{\alpha_{j-1}\alpha_j}$, $\lambda^{[j]}_{\alpha_j}$, $\Gamma^{[j+1]i_{j+1}}_{\alpha_j\alpha_{j+1}}$ with odd $j$ on $[t; t + dt/2]$ ▷ 4-6 do with hard cutoff of the local bond dimension
    propagate $\Gamma^{[j]i_j}_{\alpha_{j-1}\alpha_j}$, $\lambda^{[j]}_{\alpha_j}$, $\Gamma^{[j+1]i_{j+1}}_{\alpha_j\alpha_{j+1}}$ with even $j$ on $[t; t + dt]$
    propagate $\Gamma^{[j]i_j}_{\alpha_{j-1}\alpha_j}$, $\lambda^{[j]}_{\alpha_j}$, $\Gamma^{[j+1]i_{j+1}}_{\alpha_j\alpha_{j+1}}$ with odd $j$ on $[t + dt/2; t + dt]$
    propagate all $\Gamma^{[j]i_j}_{\alpha_{j-1}\alpha_j}$ on $[t + dt/2; t + dt]$
**end for**
**save** results
**release** memory

$i$ only singular values $\lambda^{[i]}_j$, $j \leq R$, are kept and the rest are truncated. We use the latter option. Each such local approximation on the set of singular values $\{\lambda^{[i]}\}$ introduces a truncation error, $E_i(R) = \sum_{j>R}\left(\lambda^{[i]}_j\right)^2$. For more details, see [179] and [180].

The TT representation provides a basis for an approximate tensor propagation algorithms. Here we use the TEBD scheme. Consider a tensor flow governed by an evolution generator consisting only of operations acting on one or two adjacent dimensions,

$$\begin{aligned}\frac{d}{dt}B[i_1 \ldots i_N] &= \sum_j \sum_{i_j} T^{[1]j}_{i_j} B[i_1 \ldots i_j \ldots i_N] \\ &+ \sum_{j_1, j_2} \sum_{i_{j_1}, i_{j_2}} T^{[2]j}_{i_{j_1}, i_{j_2}} B[i_1 \ldots i_{j_1} i_{j_2} \ldots i_N],\end{aligned} \tag{3.7}$$

where operations $T^{[1]}$ and $T^{[2]}$ act only on the $i$-th component and a pair of components, respectively. We use a standard time discretization technique to iteratively integrate this equation (starting from some initial tensor). In terms of operators, the solution reads

$$B(t + dt) = L(dt)B(t) = \exp\left[\left(\sum_j \hat{T}^{[1]j} + \hat{T}^{[2]j}\right) dt\right] B(t). \tag{3.8}$$

As $T$ operators generally do not commute, we have to approximate the matrix exponents. To minimize the error, it is convinient to separate the operators into groups as large as possible, such that all the operators belonging to one group commute with each other. All 1D operators commute by default, and 2D acting on odd/even pairs commute within their oddity groups. We use modified second order Suzuki-Trotter decomposition [180]. As all operators are commuting by construction, corresponding computation can be parallelized. Each two-index operator may include a cut-off if after the reorthogonalization, the number of singular value exceeds bound dimension $R$. Corresponding accumulated truncation error $E(t, R)$ is, then, calculated as a sum of local errors over all the operation during evolution up to time $t$. Computations are dominated by SVD, so resulting complexity is $\mathcal{O}(NR^3)$, where $R$ is the bond dimension. With $\mathcal{O}(N)$ cores available, it becomes $\mathcal{O}(R^3)$ and thus the computational task is perfectly scalable.

### 3.2.2 Implementation of the TEBD algorithm

The method described in subsection 3.2.1 is implemented as shown in Algorithm 3.2.1.1 using the C++ programming language. We found, that the matrix operations (mainly SVD) are the most time-consuming. In this regard, we employ the Armadillo software library integrated with highly optimized mathematical routines from the Intel Math Kernel Library to improve performance. Finally, Armadillo/MKL routines take about 50-80% of computation time during the propagation step, depending on the current system state.

The algorithm assumes performing a set of integration operations for individual components of the system at every time step. These operations are not independent but can be ordered according to their order for the organization of parallel computations. In particular, all one- and two-particle interactions can be performed completely in parallel.

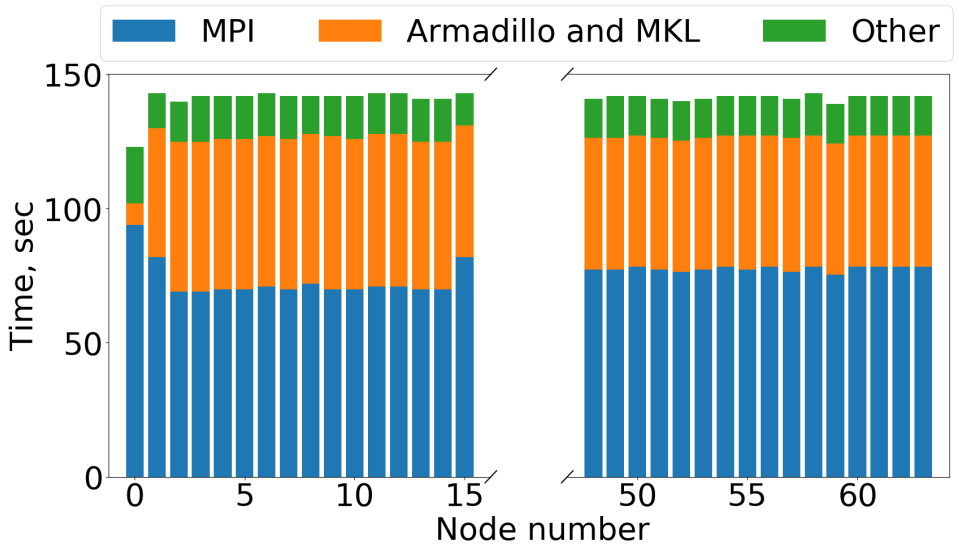

Figure 3.1: Distribution of computational and communication functions run time. 64 MPI-processes were executed on four nodes of the cluster.

The cluster parallelization is done by using the MPI technology. We apply the classic master-worker scheme for parallelization of the algorithm. For that, the single managing MPI-process (master) forms separate tasks for one– and two–particle interactions, monitors their dependencies from each other and readiness, distributes tasks to all other processes (workers) and accumulates the results.

All computational experiments have been done on a cluster with a $2 \times 8$-core Intel Xeon CPU E5-2660, 2.20GHz, 64 GB RAM, Infiniband QDR interconnect. The code was compiled with the Intel C++ Compiler, Intel Math Kernel Library and Intel MPI from the Intel Parallel Studio XE suite of development tools and the Armadillo library.

### 3.2.3 Benchmarking

As test-beds we used two models of open spin chains with next-neigbor couplings. We use a model from Ref. [184]. Let us consider a system with Hamiltonian

$$H = \sum_{i}^{N} \sigma_i^{(1)} \sigma_{i+1}^{(1)} + \sigma_i^{(2)} \sigma_{i+1}^{(2)} + \Delta\, \sigma_i^{(3)} \sigma_{i+1}^{(3)} + h_i \sigma_i^{(3)}, \tag{3.9}$$

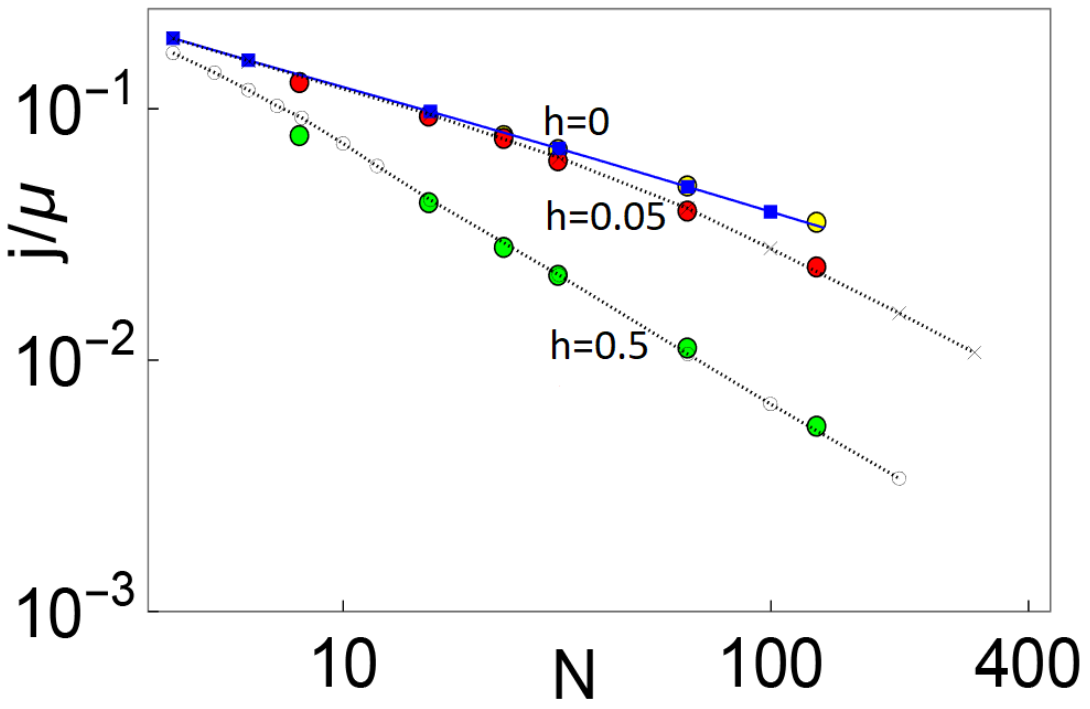

Figure 3.2: Scaling of the spin current $j$ through a disordered spin chain with $N$ spins for different values of disorder strength $h$. The results (big colored circles) are plotted on top of the results reported in Ref. [184]. The maximal size of the model system used in our simulations is $N = 128$. For every set of parameters, we performed averaging over $20$ disorder realizations. The propagation time step $dt = 0.1$ and bond dimension $R = 50$.

where $h_i$ are uniform independent values from $[-h, h]$. Open boundary conditions are implied. To describe dissipation we use four jump operators

$$\begin{aligned} A_1 &= \sqrt{1+\mu}\sigma_1^{(+)},\ A_2 = \sqrt{1-\mu}\sigma_1^{(-)}, \\ A_3 &= \sqrt{1-\mu}\sigma_N^{(+)},\ A_4 = \sqrt{1+\mu}\sigma_N^{(-)}. \end{aligned} \tag{3.10}$$

Parallel code was run on four computational nodes of the cluster (1 MPI-process per CPU core, 64 MPI-processes overall). Total computation time was $143$ s per realization. The resulted diagram for the distribution of computational and communication functions run time is presented in Fig. 3.1. It is shown that the calculations are fairly well balanced, which is an undoubted advantage of the parallelization scheme. However, MPI communications take a significant part of the computation time, while further increasing the number of cluster nodes used will not significantly speed up the calculations, which is a limitation of the scheme. Computational efficiency (ratio of computation time to total execution time) was $47\%$.

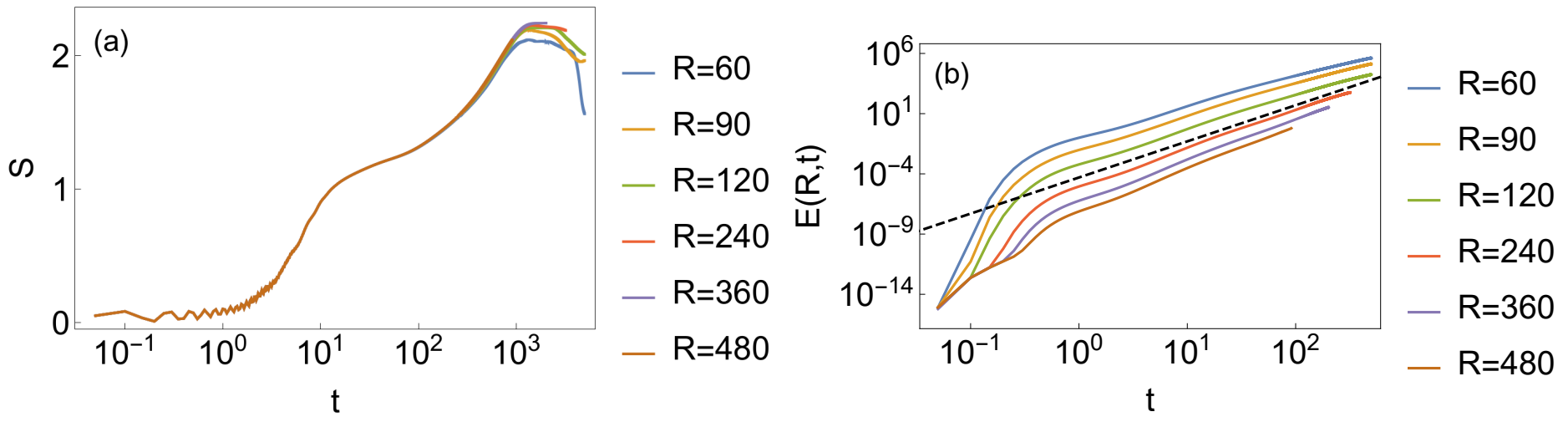

Figure 3.3: (a) Evolution of the operator entanglement entropy $S$ for a single disorder realization for the model from Sec. 3.3, for different values of bond dimension $R$. The propagation time step $dt = 0.1$ and the system size is $N = 128$. Note that for $R = 480$ we did not reach the asymptotic 'plateau' because it was not possible to numerically propagate system further (we hit the two-week limit). (b) Increase of the accumulated truncation error in time.

We find that it is possible to reproduce - with high accuracy – the results reported in Ref. [184] by using bond-dimension $R = 50$. On Fig. 3.2 we present a comparison of the results of the sampling we perform with our code (big circles; yellow, red and green) with the results by Žnidaric and his co-authors [184]. We use propagation step $dt = 0.1$ (which is sufficiently small to make the related error non-significant) and propagate every system up to $t = 10^4$, irrespectively of its size. For every value of $N$ and disorder strength $h$, we additionally performed averaging for 20 disorder realizations. Each realization took from 2 minutes to 2 hours depending on the size of the system and involved up to two nodes (for large system sizes, $N > 64$).

For the model from Sec. 3.3 we study the dynamics of the operator entanglement entropy $S = -\sum_i \lambda_i \log \lambda_i$ (for a fixed disorder realization) for different bond dimensions as shown on Fig. 3.3a. We found that $R_c = 360$ constitutes a threshold value after which the asymptotic entropy does not change upon further increase of the bond dimension. The calculation time for this value of bond dimension was four weeks of continuous propagation on four computing nodes. We also analyze the evolution of accumulated error $E(R, t)$ in this case. The operator entropy, which signals the arrival to the asymptotic state, is not accompanied by the saturation of the error. The latter continues to grow in a power-law manner, see the dashed line in Fig. 3.3b. This means that MPO states – even with $R = 480$ – are still different from the genuine asymptotic state of the model (which is the zero-value eigenelement of the corresponding Lindbladian).

### 3.2.4 Discussion

We presented a parallel implementation of the MPO-TEBD algorithm to propagate many-body open quantum systems. Parallelization is performed using the MPI technology and employs the master-worker scheme for computational tasks distribution. High-performance implementations of linear algebra from the Intel MKL were used to better utilize the computational resources of modern hardware.

The performance tests on the a cluster demonstrated that $64$ MPI processes running on four computational nodes is the optimal configuration for the model systems with $N = 128$ spins. As a next step, we plan to explore the possibility of further improvements of the parallelization by reducing the communications and increasing the efficiency of using computational resources. After that, we hope to reach the limit $N \simeq 400$ with the test-bed models.

## 3.3 Many-Body Localization in Open Quantum Systems

One of important questions (especially in the context of recent experiments [7, 64, 63]) concerns the impact of the interaction with an environment on the states of MBL systems on large time scales. This question has been addressed recently in a series of papers [186, 187, 188], where the action of the environment was modeled with a set of local dephasing operators in the framework of Lindblad master equation. The ultimate fate of the systems is plain: dephasing, whatever small, grinds *any* system – with or without many-body interactions, governed by an MBL or ergodic Hamiltonian – into an infinite temperature state [14]. On the way to the asymptotic state, however, systems with MBL and non-MBL Hamiltonians behave notably differently, e.g., they exhibit slow stretched exponential (MBL) vs exponential (non-MBL) relaxation of observables, see Refs [186, 187, 188]. This difference was detected in recent experiments [189]. However, the asymptotic state cannot be changed by modifying the system Hamiltonian since the identity (which is the density operator corresponding to the infinite temperature state) commutes with any Hamiltonian.

In this section, following [16], we show that, by introducing a special (though physically relevant) type of dissipation into a system already subjected to dephasing, we can drive the system into a new asymptotic state which bears detectable signatures of localization (or of its absence, depending on the system Hamiltonian). These signatures can be revealed by analyzing the population imbalance [186, 187, 188] (a quantity measured in experiments [7, 64, 189]), the operator space entanglement entropy [190, 188], and the mean spectrum gap ratio [60] of the steady state density operator. We perform it in two steps. First, we demonstrate that the proposed dissipation, when acting alone, can sculpt an asymptotic state with localization features. Next, we show that these features are robust to the action of dephasing. In subsection 3.3.1 we introduce the model, in subsection 3.3.2 we discuss the results. We clarify the numerical results that are not covered in 3.2 in subsection 3.3.2, and provide the discussion in 3.3.4.

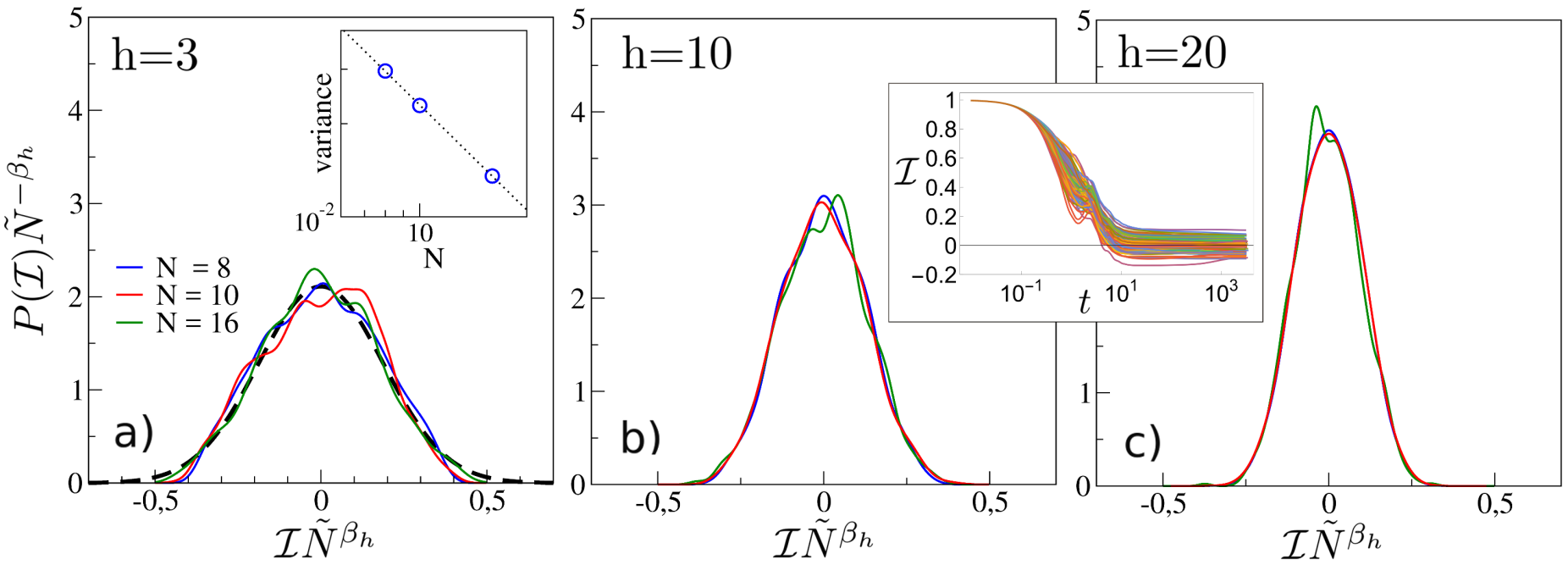

Figure 3.4: Probability density function $P(\mathcal{I})$ of the steady state imbalance $\mathcal{I}$ for different disorder strengths and system sizes. Dashed thick line on panel (a) is the distribution sampled with a constrained random partition for $N = 16$ (see text). Distributions are scaled with $\tilde{N}^{\beta_h}$, where $\tilde{N} = N/8$ and exponent $\beta_h$ has values $0.55$ for $h = 0.3$ (a) and $0.8$ for $h = 10, 20$ (b,c). Insets: (a) scaling of the distribution variance with $N$ for $h = 3$ (dashed line is the power-law $N^{2\beta_h}$) and (b-c) the time evolution of the imbalance for $10^2$ disorder realizations, $h = 20$ and $N = 32$, obtained with the TEBD propagation (not used for the histograms) [183]. The parameters are $\gamma_p = 0.1$, $U = J = 1$. Number $K$ of realizations is $10^4$ ($N = 8$) and $4 \cdot 10^3$ ($N = 10, 16$).

### 3.3.1 Model

We study a conventional interacting spinless fermion MBL model with Hamiltonain 1.16

$$H = -J \sum_{l=1}^{N} \left( c_l^\dagger c_{l+1} + c_{l+1}^\dagger c_l \right) + U \sum_{l=1}^{N} n_l n_{l+1} + \sum_{l=1}^{N} h_l n_l, \qquad (3.11)$$

The dissipation is captured with a master equation 3.3. We are interested in the asymptotic (steady state) density operator $\varrho_\infty = \lim_{t\to\infty} \varrho(t)$.

For local Hermitian dissipative operators (dissipators) $A_s^\dagger = A_s$, we have $\varrho_\infty = \mathbb{1}/L$ (if the asymptotic state is unique), where $\mathbb{1}$ is the identity operator in the half-filling subspace and $L = N!/[(N/2)!]^2$ is the number of accessible states. This is the case of local dephasing [1], $A_l = c_l^\dagger c_l$, $l = 1, ..., N$, considered in Refs. [186, 187, 188]. On the other hand, formally one could construct a non-Hermitian local operator $A^i$ such that $A^i|\phi_i\rangle = 0$, where $|\phi_i\rangle$ is the $i$-th eigenstate of the Hamiltonian $H$. Then the asymptotic state is $\varrho_\infty = |\phi_i\rangle\langle\phi_i|$ [191]. However, such dissipators are highly non-local and too disorder-specific to be practically relevant. Also note that all other single-site operators (except for the identity, which does not influence the Hamiltonian dynamics) do not confine the evolution to a sector with fixed number of particles. In Ref. [187] dissipation in the form of a single-particle loss operator, $A_l = c_l$, was considered. Evidently, the steady state in this case is the vacuum $|0^N\rangle$.

We choose dissipative operators which act on a pair of neighboring sites [191],

$$A_l = (c_l^\dagger + c_{l+1}^\dagger)(c_l - c_{l+1}), \quad \forall \gamma_l = \gamma_p. \tag{3.12}$$

This non-Hermitian operator tries to synchronize the dynamics on two sites by constantly recycling anti-symmetric out-of-phase mode into the symmetric in-phase one. A possible experimental realization of this dissipation is an array of superconducting microwave resonators [15].

As we want to consider the situation when both types of dissipation are affecting the system dynamics, the dissipative part of the generator in the Lindblad equation includes $N$ dephasing operators $c_l^\dagger c_l$, $l = 1, ..., N$, and $N - 1$ pairwise dissipators, acting with rates $\gamma_d$ and $\gamma_p$, respectively. The total number of dissipative channels is therefore $M = 2N - 1$.

We start the analysis from the limit $\gamma_d = 0$ and $\gamma_p = 0.1$. By using the Jordan-Wigner transformation, the system, Eqs. (3.11) , (3.12), can be mapped onto a model of $N$ spins confined to the manifold $S^z = \sum_{l=1}^N s_l^z = 0$ [192]. This relation allows us to implement the TEBD scheme generalized to matrix product operators [183] and propagate the model system with $N > 10$ to its steady state. See subsection 3.3.3 for further details. As the initial state we use $\varrho(0) = |\psi_0\rangle\langle\psi_0|$, $|\psi_0\rangle = |1010...10\rangle$. For small systems, $N \leq 10$, we find asymptotic density operator as a kernel of the Lindblad generator, $\mathcal{L}\varrho_\infty = 0$.

[1]Note that the use of the term *dephasing* differs in atomic physics and quantum optics, where it only affects non-diagonal components of the density matrix.

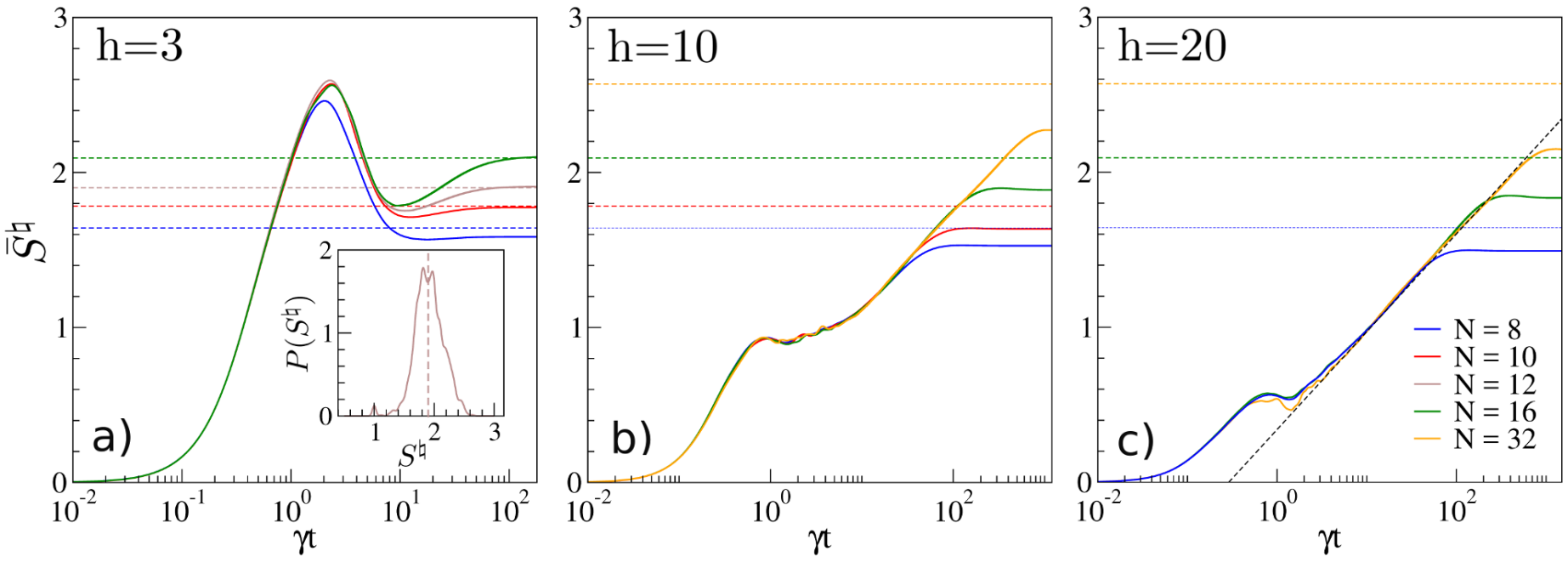

Figure 3.5: Averaged operator-space entanglement entropy $\bar{S}^{\natural}(t)$ of the density operator $\varrho(t)$ as function of time. Dashed lines are the values of the entropy for the infinite-temperature state. The dotted line on panel (c) is $\frac{1}{5}\log_2(t) + \text{const}$. Inset: The probability density function of the entropy of individual disorder realizations, for $h = 3$ and $N = 12$. The initial condition is $\varrho(0) = |\psi_0\rangle\langle\psi_0|$, $|\psi_0\rangle = |1010...10\rangle$. Other parameters are as in Fig. 3.4.

## 3.3.2 Results

### 3.3.2.1 Core indicators

*Imbalance*. The imbalance 1.21 is defined as

$$\mathcal{I}(t) = \frac{N_o(t) - N_e(t)}{N/2}, \tag{3.13}$$

where $N_o$ ($N_e$) is the number of fermions on odd (even) sites. This characteristics was measured in the recent experiments [7, 64, 189].

In the case of dephasing-driven dynamics, the asymptotic imbalance $\mathcal{I} = \lim_{t\to\infty}\mathcal{I}(t)$ is uniformly zero. When the dissipation is non-Hermitian, the steady state is disorder specific and asymptotic imbalance is a real-valued random variable. We sample its probability density function (PDF) $P(\mathcal{I})$ for different systems sizes and disorder strengths, see Fig. 3.4. The imbalance can be considered as a sum of $N/2$ random variables, $\xi_l = n_{2l-1} - n_{2l}$, $l = 1, ..., N/2$, where $n_s$ is the occupation of the $s$-the site. Since these variables are correlated, their sums are not subjected to the Central Limit Theorem (CLT) [193]. We check a scaling hypothesis $N^{-\beta_h}P(N^{\beta_h}\mathcal{I}[N])$, with the exponent $\beta_h$ being a function of disorder ($\beta_h = 1/2$ will correspond to the CLT case). Exponent values can be estimated by calculating variance of the PDF's for different $N$ and then fitting the obtained dependence with the power-law $N^{2\beta_h}$, see the inset in Fig. 3.4(a). We find $\beta_h \backsimeq 0.55$ for the ergodic regime and $\beta_h \approx 0.8$ for $h = 10, 20$. To get some insight, we consider a particular realization $\{n_1, n_2, ..., n_N\}$ as a result of uniform sampling from a set of $N$ independent and identically distributed random variables constrained by preservation of the total sum, $\sum_{s=1}^{N} n_s = N/2$, and the condition $\forall n_s < 1$ (no more than one particle per site). See subsection 3.3.3 for further details. The result of such sampling for $N = 16$ is in a good agreement with the imbalance PDF obtained for the ergodic regime, see dashed line on Fig. 3.4(a).

*Operator-space entanglement entropy (OSEE)*. This quantity was introduced by Prosen and Pižorn [194] as an operator generalization of the spatial entanglement entropy (defined for pure states). To calculate the OSEE, one should split the chain into two (equal in our case) parts and make the Schmidt decomposition of the density operator, $\varrho = \sum_k \sqrt{\mu_k} C_k \otimes D_k$, where the operators $C_k$ $(D_k)$ act non-trivially on the left (right) half only and form a complete Hilbert-Schmidt basis in the corresponding subspace. The normalized coefficients $\bar{\mu}_k$ define the entropy value $S^\natural = -\sum_k \bar{\mu}_k \log_2 \bar{\mu}_k$. For a pure state, $S^\natural$ is twice the standard entanglement entropy [195].

The OSEE is a practically relevant characteristics: low entropy of a state implies low complexity of the matrix product operator representation of the corresponding density operator [195]. For example, the entropy is lower for the infinite temperature state than for a pure state of high entanglement, which is directly opposite in the case of the von Neumann entropy. Therefore, we consider the operator entropy a better choice [188] to characterize $\varrho_\infty$ than the von Neumann entropy [186].

We find that, in the ergodic phase, the averaged (over the disorder) OSEE $\bar{S}^\natural(t)$ saturates to $S^\natural(\mathbb{1})$, see Fig. 3.5(a). This implies an *effective* thermalization of the system: At variance to the case of local dephasing [188], the individual realization entropy values are not all identical to $S^\natural(\mathbb{1})$ but are the distributed around it, see the inset in Fig. 3.5(a) The initial short-time evolution of the entropy follows the Hamiltonian path; it is a linear growth, which in the absence of the dissipation would saturate to the Page value [196], $S^\natural_{\mathrm{Page}} \backsimeq N-1$. After the time interval $t \gtrsim \gamma^{-1}$ the contribution of the dissipative part of the generator $\mathcal{L}$ becomes tangible and eventually brings the entropy down to an asymptotic value $S^\natural(\varrho_\infty) \ll S^\natural_{\mathrm{Page}}$.

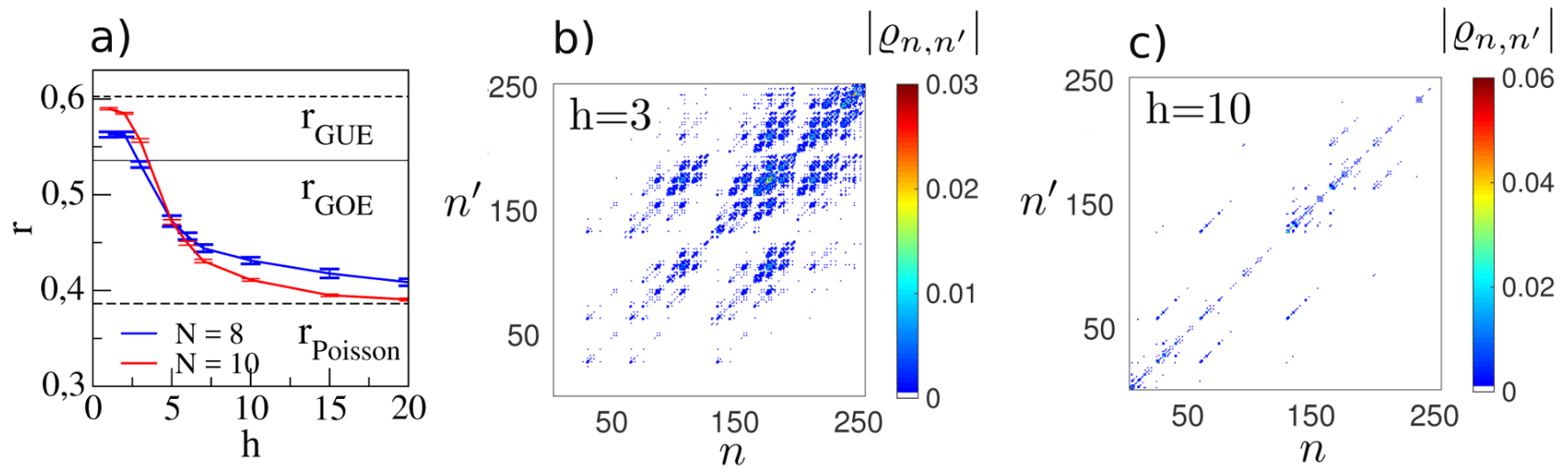

Figure 3.6: (a) Averaged ratio of consecutive level spacing $r$ of $\varrho_\infty$ as a function of disorder strength $h$. The ratio is sampled for chains with $N = 8$ and 10 sites and averaged (for every value of $h$) over $10^2$ disorder realizations. The error bars show the variance of the ratio. (b-c) Absolute values of the elements of the steady state density matrix for a single disorder realization and two different values of $h$. The matrices are expressed in the Fock basis (for the half-filling sector) which is sorted in the lexicographical order. Only elements with absolute value larger than $10^{-5}$ are shown. Other parameters are the same as in Fig. 3.4.

In the MBL phase, the averaged OSEE saturates to values below $S^\natural(\mathbb{1})$, see Figs. 3.5(b-c). This can be explained by generalizing the argument used in Ref. [50] for the Hamiltonian MBL systems. While in the ergodic phase all – even distant – sites (spins) are ’tied’ by the conservation of the total particle number (total spin), in the MBL phase the correlations are short-ranged and restricted by the localization length. Therefore, the entanglement has to be short-ranged in the MBL phase. It is noteworthy that, similar to the entanglement entropy in the Hamiltonian case [55, 56, 57, 58], a relaxation of the OSSE to its asymptotic value is marked by a logarithmic growth, $S^{\natural}(t) \propto \log(t)$ – a feature found before with local dephasing [188].

*Ratio of consecutive level spacing for the steady state density operator.* According to the quantum chaos theory [197], Poisson and Wigner-Dyson distributions of the spacing $\delta_j = E_{j+1} - E_j$, where $\{E_j\}$ are eigenvalues of Hamiltonian sorted in ascending order, correspond to regular (integrable) and chaotic (non-integrable) quantum systems. Similarly, we can expect Poisson and Wigner-Dyson distributions for MBL and ergodic many-body Hamiltonians respectively [60, 61]. However, these indicators assume a uniform level density which is rarely the case with physical Hamiltonians. To circumvent this problem, Oganesyan and Huse considered the distribution of the ratios $r_j = \min[\lambda_j, \lambda_j^{-1}]$, $\lambda_j = \delta_j/\delta_{j-1}$, which do not depend on the local density of states [60]. It follows that spectral averages of $r$ yield $r_{\text{Poisson}} \backsimeq 0.386$ for Poisson random variables, $r_{\text{GOE}} \backsimeq 0.536$ for Gaussian orthogonal (GOE), and $r_{\text{GUE}} \backsimeq 0.603$ for Gaussian unitary (GUE) ensembles [198].

In another context, Prosen and Žnidaric proposed to quantify the non-equilibrium steady state density operators in terms of their level spacing distributions [199]. They found that the transition from integrability to non-integrability (where integrability of density operators, as was defined in Ref. [199] as "the existence of an algebraic procedure for their construction in finitely many steps", e.g., by using the matrix product state ansatz [200]) corresponds to the Poisson-to-GUE transition. We follow this idea and find that in the ergodic phase the spectrum of the steady state density operator yields $r$ values close to $r_{\text{GUE}}$, while in the limit of strong localization it approaches $r_{\text{Poisson}}$, see Fig. 3.6(a). This correspondence improves with increasing $N$. The structure of the density matrices $\varrho_\infty$ is notably different in the ergodic and strong localization regimes, see Figs. 3.6(b-c): While in the ergodic phase matrices exhibit a well-developed off-diagonal structure and thus a relatively high purity and interference pattern, in the deep MBL regime they have near the diagonal structure, with a few "hot spots" (a similar structure was found before with a dissipative single-particle model [201]).

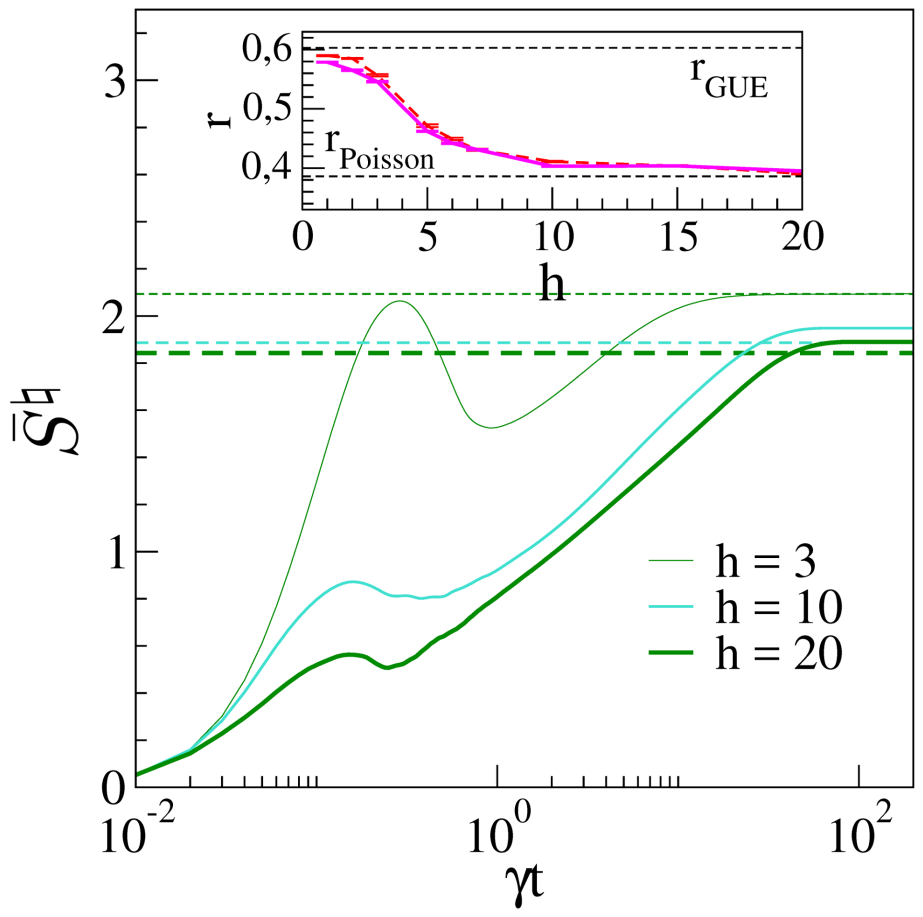

Figure 3.7: Time evolution of the averaged operator-space entanglement entropy $\bar{S}^{\natural}(t)$ when both types of dissipation are acting simultaneously. Dashed lines correspond to asymptotic values obtained with the pairwise dissipation, see Fig 3.6. The system size is $N = 16$. The inset shows averaged ratio of consecutive level spacing $r$ of $\varrho_\infty$ ($N = 10$) as a function of disorder strength $h$ for $\gamma_{\rm d} = 0$ (thick solid line) and $\gamma_{\rm d} = 0.1$ (thick dashed line). Other parameters are the same as in Fig. 3.4.

#### 3.3.2.2 Combined action of the pairwise dissipation and local dephasing

Now we study the interplay between the dephasing and the pairwise dissipation. In this case, the steady state density operator is the solution of the operator equation $\mathcal{L}_{\mathcal{H}}\varrho_\infty + \mathcal{L}_{\rm d}\varrho_\infty + \mathcal{L}_{\rm p}\varrho_\infty = 0$, where two last super-operators are parametrized with rates $\gamma_{\rm d}$ and $\gamma_{\rm p}$.

We argue that a steady state obtained with the pairwise dissipation is stable with respect to the action of dephasing. Namely, the quantifiers of the steady state density operator $\varrho_\infty$ change continuously with the increase of $\gamma_{\rm d}$. This conjecture is based on the notion of stability introduced for many-body dissipative systems with no faster than linear (in time) growth of the support of initially localized operators [202, 203]. This seems to be the case for $h = 10, 20$ as it has been detected with the OSEE, see Figs. 3.6(b-c). To validate the conjecture, we calculate the OSEE and the mean gap ratio $r$ for $\gamma_\mathrm{d} = \gamma_\mathrm{p} = 0.1$, Fig. 3.7. While the asymptotic operator entropy changes slightly in the localization phase (and remains constant in the ergodic phase), changes of the mean gap ratio values are smaller than the sampling errors (see the inset in Fig. 3.7). Therefore, by introducing the pairwise dissipation into a system already subjected to decoherence, we can restore localization features and distinguish between ergodic and MBL Hamiltonians.

### 3.3.3 Numerical methods

To calculate characteristics of the model systems of a size $N \leq 10$, we numerically solve the operator equation $\mathcal{L}\varrho_\infty = 0$. We, first, expand the density operator, model Hamiltonian $H$, and jump operators $A_s$, over the basis of the $SU(N)$ group generators, thus transforming the original Lindblad equation into a system of $L^2 - 1$, $L = \binom{N}{N/2}$, non-homogeneous real-valued linear equations. We next solve the obtained system of equations by using the standard routines developed for large and sparse systems of linear equations (LAPACK). After finding the solution vector, we sum up the obtained $L^2 - 1$ real coefficients, by multiplying them with the matrices of the $SU(N)$ generators, and obtain the steady state density operator in the matrix form. Having this matrix, we can calculate all needed quantifiers and expectation values of relevant observables. For model systems with $N > 10$, we implement the MPO presentation of the density operator and use the TEBD propagation algorithm discussed before in section 3.2. For $N = 10$ and $10^2$ individual disorder realizations, we compare results of the numerically exact spectral propagation (by diagonalizing generator $\mathcal{L}$ and implementing its dual basis and eigenvalues to calculate $\varrho(t)$ at any instant of time) and of the TEBD propagation; the relative error for the imbalance and entropy did not exceed $10^{-4}$ for all three values of $h$ used in the paper.

Finally, to sample the PDF of the imbalance, $P(\mathcal{I})$, where $\mathcal{I} = \sum_{l=1}^{N/2}[n_{2l-1} - n_{2l}]$, by using a constrained sampling, we need first to sample a set of $N$ random identically distributed variables, $\{n_1, n_2, ..., n_N\}$, which fulfills two conditions,

(i) $\sum_{s=1}^{N} n_s = N/2$ and (ii) $\forall n_s < 1$. First condition means that, by dividing all variables with $N/2$, we have to perform a sampling from unit simplex. For that we implement 'Kraemer algorithm'. Condition (ii) is fulfilled by using a simple rejection strategy, i.e., if a particular realization $\{n_1, n_2, ..., n_N\}$ does not fulfill the condition, it is droped and a new one is sampled.

### 3.3.4 Discussion

We proposed three quantitative identifiers of MBL in open systems. The imbalance statistics is accessible in experiments [7, 64, 189] but requires studying systems of different sizes, which limits applicability to systematic studies of classes of systems. The operator-space entanglement entropy indicates differences between phases both in the asymptotic limit and during the relaxation towards it, but is hard to experimentally measure. Nevertheless, it builds a strong physical connection to inherent dephasing processes. The level spacing of the asymptotic density operator $\varrho_\infty$ bridges MBL and quantum chaos theory [197, 199], but remains a purely theoretical tool.

There are three factors [and, correspondingly, three terms in the generator $\mathcal{L}$ of the master equation, Eq. (3.3)] contributing to the formation of the steady state density operator. The pairwise dissipation tries to built classical and quantum correlations between distant sites and reasonably week local dephasing is not able to wash them out. At the same time, the MBL mechanisms, induced by the Hamiltonian, try to restrict the correlations to the localization length. As a result of the balance between these three factors, an asymptotic state with localization "footprints" appears.

Future studies could consider the incorporation of the disorder into local rates $\gamma_p(s)$ and, ultimately, a creation of MBL states by dissipative means solely. Disordered pairwise dissipation acquires a relevance in the context of recent experiments with dissipatively coupled exciton-polariton condensate arrays [204].

## 3.4 Quasi-stationary states of open quantum systems

"*We are dealing with a stochastic process with an absorbing state (i.e., a state which cannot be left once the system got into it) so that the absorption happens with probability one. What is the distribution in the limit $t \to \infty$ provided that the absorption did not happen until time $t$?*". This excerpt from a paper by Yaglom [205] delivers a whole idea of quasi-stationary states (QS) laid out by Darroch and Seneta twenty-five years later [206]. In the Markov chain framework, the QS are related to the probability distributions over the set of states from which all the absorbing [207] ones are excluded. Some quasi-stationary distributions could be very sustainable, so that the time needed for their eventual 'evaporation' into the absorbing states is much longer than all the relevant observation time scales [208]. Such states can be nominated for metastable state [209]. They form a basis for many features of out-of-equilibrium dynamics including localization and thermalization [210].

In this section, following [211], we generalize the concept of QS to open quantum systems. We consider Markovian generator $\mathcal{L}$ of an open quantum evolution in the Lindblad form. In subsection 3.4.1 we derive generalized master equation and derive a projected master equation. In subsection 3.4.2, we propose an algorithm to sample quasi-stationary density operators, based on the idea of stochastic unraveling for master equations, and use two different models to illustrate our findings. In subsection 3.4.3, we apply the developed methodology to a testbed system consisting of a quantum generalization of a meta-population model often used as an example of a system with QS. We discuss the results in subsection 3.4.4.

### 3.4.1 Projected master equation

Consider a general Lindblad equation (3.3) defined on an $N$ by $N$ dimensional Hilbert-Schmidt space. Under general assumptions, there is a unique stationary state $\rho_{\text{st}}$, such that $\mathcal{L}\rho_{\text{st}} = 0$, which is an asymptotic state of evolution from any initial condition. In order to study quasi-stationary states and relaxation towards them, here we derive a generalized master equation that governs the evolution of the subsystem that excludes the stationary state. To exclude non-trivial entanglement with the stationary state, we will only consider pure stationary states. This will allow us to get a closed-form master equation.

Let us denote the stationary state $\rho_{\text{st}} = |0\rangle$. We define projection on the stationary state and its orthogonal complement,

$$P_0 = |0\rangle\langle 0|, \qquad P_\perp = \mathbb{1} - P_0. \tag{3.14}$$

The assumption of a unique pure stationary state entails [212]: (a) the stationary state is a dark state of each dissipative operator $A_i$, $\forall i : \; A_i P_0 = 0$; (b) $|0\rangle$ is an eigenstate of the $H$, and, consequently, the Hamiltonian does not couple the stationary state with the orthogonal complement space, $P_\perp H P_0 = 0$.

Using the properties (a),(b) and projecting the Lindblad (3.3) on the orthogonal complement subspace, one car derive the generalized quasi-stationary master equation for projected density matrix $\rho_\perp = P_\perp \rho P_\perp$,

$$\dot{\rho}_\perp = \mathcal{L}_\perp[\rho_\perp] - \frac{1}{2}\left\{\sum_i A_{d,i}^\dagger A_{d,i}, \rho_\perp\right\}. \tag{3.15}$$

It consists of two parts, the first one is standard Lindblad evolution in the orthogonal subspace,

$$\mathcal{L}_\perp[a] = -i[H_\perp, a] + \sum_i A_{\perp,i} a A_{\perp,i}^\dagger - \frac{1}{2}\left\{A_{\perp,i}^\dagger A_{\perp,i}, \rho_\perp\right\}, \tag{3.16}$$

where $H_\perp = P_\perp H P_\perp$ and $A_{\perp,i} = P_\perp A_i P_\perp$ are projected Hamiltonian and dissipative operators. The second part is a *decay* term with projected dissipative operators $A_{d,i} = P_0 A_i P_\perp$. It does not preserve the trace of the density matrix and describes the decay of the probability of the state to not yet be absorbed into the stationary state. While the master equation (3.15) is thus not anymore trace-preserving, it can be transformed to the form

$$\dot{\rho}_\perp = \tilde{\mathcal{L}}\rho_\perp = \phi(\rho_\perp) - \kappa\rho_\perp - \rho_\perp\kappa^\dagger, \tag{3.17}$$

where $\phi$ is completely positive (as standard Lindbladian generator) and $\kappa$ is any operator. This proves [176] that the master equation (3.15) generates a completely positive quantum map, thus $\rho_\perp$ always corresponds to a physical state. Such maps are known in the literature as non-conservative dynamical semigroups [166].

We define QS as an eigenstate of (3.15) that corresponds to the eigenvalue with the largest real part. The real part of that eigenvalue itself is a spectral gap, and thus, it defines the lifetime of the QS. At any moment of time, the probability of absorption to the stationary state (which is the same as the stationary state density matrix component) can be calculated as $\rho_{00}(t) = P_0\rho(t)P_0 = 1 - \mathrm{Tr}\{\rho_\perp(t)\}$. To complete the description of the full density matrix through this approach, we derive an equation governing stationary-orthogonal coherence evolution,

$$\begin{aligned}\dot{\rho}_{\perp 0} = &\left[-i(H_\perp + H_{00}\mathbb{1}) - \frac{1}{2}\sum_i \left(A^\dagger_{\perp,i}A_{\perp,i} + A^\dagger_{d,i}A_{d,i}\right)\right]\rho_{\perp 0}\\ &+ \sum_i A_{\perp,i}\rho_\perp A^\dagger_{d,i},\end{aligned} \tag{3.18}$$

where $\rho_{\perp 0} = P_\perp \rho P_0$. This vector equation is not closed and contains $\rho_\perp$ as non-uniformity. The corresponding equation for $\rho_{0\perp}$ is simply a Hermitian conjugation of (3.18). Equations (3.15) and (3.18) contain all the spectral properties of the initial Lindblad equation (3.3). All the eigenvalues of (3.15), combined with the eigenvalues of the uniform part of (3.18) (plus complex conjugated), and the zero eigenvalue corresponding to the stationary state, form the full spectrum of the Lindblad equation (3.3), see Fig.3.8.

### 3.4.2 Generalized quantum trajectories

Studying quasi-stationary dynamics may be numerically complicated if it is also meta-stable. Direct application of conventional methods to (3.3) does not separate the stationary state from the rest of the space. This is especially challenging if the system size is too large for the problem to be addressed through direct diagonalization. A common numerical approach in such cases is quantum trajectories method [172]. It is based on the stochastic unraveling of the Lindblad master equation for density matrices into a statistical average over multiple "trajectories", non-normalized wave function governed by effective non-Hermitian evolution with randomized quantum jumps due to dissipative channels. In this section, we derive a generalization of the QT method for the non-trace preserving master equations.

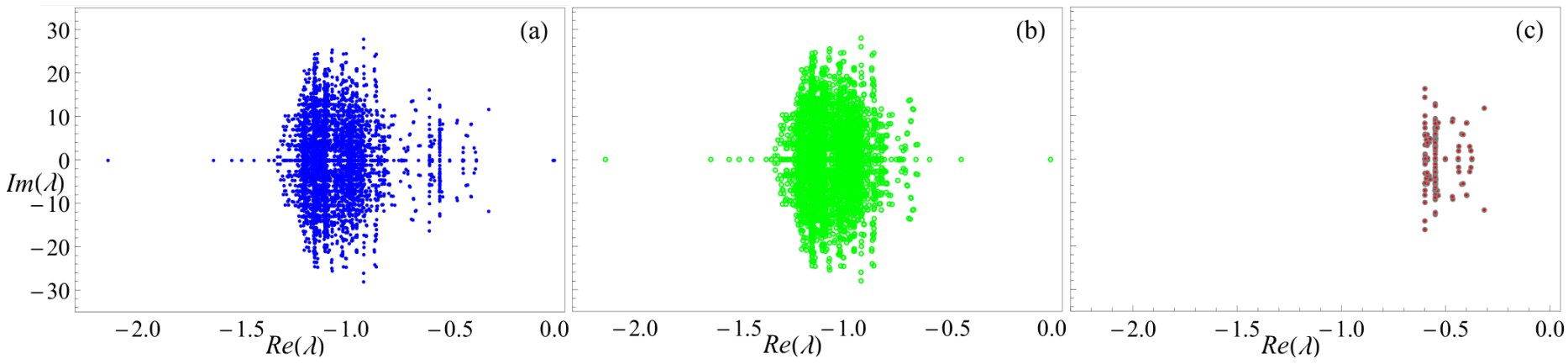

Figure 3.8: The real and imaginary parts of the Lindbladian eigenvalues $\{\lambda_i\}$ of the metapopulation model (3.25) with $c = 0.1, e = 0.1$ and XXZ Hamiltonian (3.4.3) with $a = 1.5, b = 2.3$, $n = 6$. (a) full Lindbladian (3.3); (b) orthogonal complement Lindbladian (3.15); (c) coherence (coupling the orthogonal and stationary) part of the Lindbladian.

The key element of QT is the decrease of the norm of wave functions due to the non-Hermitian part of the effective Hamiltonian which is due to dissipation. This decrease determines the probability of a quantum jump – a measurement of the trajectory by one of the dissipative operators. In the generalized case, an additional source of norm decrease is present, but it does not lead to quantum jumps. Still, it has to be addressed separately as it contains information about the full density matrix norm decay. Thus the primary goal is to allow for the systematic norm decay due to certain parts of the dissipative operators, and for standard norm decay associated with quantum jumps due to other parts. We follow the derivation structure from [172] with necessary modifications.

Let us consider an element of the projected subspace $\perp$, wave function $\psi$. We introduce an effective Hamiltonian that governs its evolution,

$$H_{eff} = H_{\perp} - \frac{i}{2}\sum_i A^{\dagger}_{\perp,i}A_{\perp,i} + A^{\dagger}_{d,i}A_{d,i}. \tag{3.19}$$

Let us consider an infinitesimal time-step $\delta t$. We calculate the wave-function norm loss due to only the jump-related dissipators $A_{\perp,i}$,

$$\begin{aligned}\delta p_{\perp} &= \left|\langle\psi(t)|1-e^{-i(-\frac{i}{2}\sum_i A^{\dagger}_{\perp,i}A_{\perp,i})\delta t}|\psi(t)\rangle\right| \\ &= \sum_i \delta p_i = \sum_i \langle\phi(t)|A^{\dagger}_{\perp,i}A_{\perp,i}|\phi(t)\rangle\delta t,\end{aligned} \tag{3.20}$$

where $\delta p_i$ is norm loss only due to the $i$-th channel and has the meaning of the probability of a jump in the corresponding channel. Based on this we define evolution at each moment of time to randomly follow one of the following transformations. Either to propagate with the effective Hamiltonian (3.19) (taking into account correct normalization),

$$|\psi(t+\delta t)\rangle = \frac{(1-iH_{eff})|\phi(t)\rangle}{(1-\delta p_{\perp}/p)^{1/2}} \text{ with probability } 1-\frac{\delta p_{\perp}}{p}, \tag{3.21}$$

where $p$ is a total norm, or to perform a jump in one of available, channels,

$$|\psi(t+\delta t)\rangle = p^{1/2}\frac{A_{\perp i}|\phi(t)\rangle}{(\delta p_m/\delta t)^{1/2}} \text{ with probability } \frac{\delta p_{\perp}}{p}\frac{\delta p_i}{\delta p_{\perp}}, \tag{3.22}$$

where $\delta p_{\perp}/p$ is the total probability of one of the jumps to happen, and $\delta p_i/\delta p_{\perp}$ defines probability of the channel $i$ to be activated. Averaging projections on the quantum trajectories $\sigma(t) = |\psi(t)\rangle\langle\psi(t)|$ over all the realizations gives

$$\begin{aligned}\overline{\sigma(t)} &= \left(1-\frac{\delta p_{\perp}}{p}\right)\frac{(1-iH_{eff})|\phi(t)\rangle\langle\phi(t)|}{(1-\delta p_{\perp}/p)^{1/2}} \\ &+\frac{\delta p_{\perp}}{p}\sum_i\frac{\delta p_i}{\delta p_{\perp}}p^{1/2}\frac{A_{\perp i}|\phi(t)\rangle\langle\phi(t)|}{(\delta p_m/\delta t)^{1/2}},\end{aligned} \tag{3.23}$$

which is equivalent to (3.15), and thus, this is the correct stochastic unraveling procedure. Thus, almost always the state is evolved by the effective Hamiltonian, while only occasionally one of the jump channels activates. To get a much more computationally preferred version of the QJ, we generate the jump-time itself as a random value, while propagating the state with the exponentiated Hamiltonian in between these events. This leads to the following algorithm,

1. Generate a random number $\eta$ uniformly distributed in $[0, 1]$.

2. Propagate the state using small time $\Delta t$ with the following operator $e^{-iH_{eff,\perp}\Delta t/2}\mathfrak{N}e^{-1/2\sum_i A^\dagger_{d,i}A_{d,i}\Delta t}e^{-iH_{eff,\perp}\Delta t/2}$ (utilizing Trotter decomposition), with $\mathfrak{N}$ being a non-linear operator that normalizes wave-function to the norm that was before the action of decaying (middle) part of the Trotter decomposition. This renormalized fraction of norm should be saved and collected. Note, that one may use higher-order Trotter decomposition without significant modifications.

3. When $p(t)$ reaches $\eta$, chose the jump channel using standard procedure only considering operators $A_{\perp,i}$. Normalize the wave function to unity. Go to step 1.

4. When the desired evolution time reached, average $|\phi(t)\rangle\langle\phi(t)|$ with the weight collected during steps 2.

This procedure produces a non-normalized density matrix of the orthogonal subspace with the trace indicating the probability of non-absorption. One may apply numerical techniques developed for conventional quantum jumps for accuracy and performance benefits.

### 3.4.3 Meta-population model

For the sake of continuity, in this section, we generalize the classical quasi-stationary model developed for biological meta-population problems [208]. Consider a system which state is a set of $n$ Boolean variables, which we will call "sites". The values indicate either the "extinct" or "populated" state of the site. If the sites are indistinguishable, all the states with the same number of populated sites are the same, so the probabilistic system state is given by an $n+1$-long vector $r = (p_0, p_1, \ldots, p_n)$, where $p_k$ is the probability of exactly $k$ sites being populated. We introduce two competing random processes: (a) "colonization" with the rate $c$ – each already populated site may populate any of the extinct sites; (b) "extinction" with the rate $e$ – each populated site has a probability to get extinct. The generator of the corresponding classical continuous-time Markovian

process is

$$L = \begin{pmatrix} 0 & e & 0 & \dots & 0 \\ 0 & -1e - c*1*(n-1) & 2e & \dots & 0 \\ 0 & c*1*(n-1) & -2e - c*2*(n-2) & \dots & 0 \\ \vdots & \vdots & \ddots & \dots & \vdots \\ 0 & \dots & \dots & c*(n-1)*1 & -n*e \end{pmatrix}, \tag{3.24}$$

so $\dot{r}(t) = Lr(t)$. The unique stationary solution of the equation is $r_{\rm st} = (1\ 0\ 0 \dots 0)$, as there are no processes that can re-populate a completely extinct state. Nevertheless, for adequately large system sizes, and if the colonization rate is large enough compared to the extinction rate, i.e. $c \gtrsim e$, the relaxation to the stationary state time becomes extremely large. One can understand this fact as the stationary state can only be reached from a moderately-populated state through a chain of low-probability extinctions, which is a rare fluctuation. The quasi-stationary state of this classical Markovian system can be calculated by constructing a reduced generator $\tilde{L}$ similarly to (3.15) [206]. It is done by simply removing the first raw and the column of (3.24). This procedure lifts the zero-total-columns condition (thus leading to the non-zero largest eigenvalue), but still is a Metzler matrix (all off-diagonal elements are non-negative), thus providing a proper probability vector as the eigenvector associated to the largest eigenvalue. This probability vector, quasi-stationary state, is a conditional probability vector, where the condition is that the state is not absorbed into the stationary state.

Quantum generalization of this problem is straightforward if one introduces dissipation-only Lindbladian dynamics on a hard-core boson system. Consider $n$ sites populated by a non-conserved number of hard-core bosons. The Hilbert space size is thus $N = 2^n$. The extinction and colonization processes are then generalized as dissipative channels with the rates $e$ and $c$ respectfully,

$$A_i^{(e)} = c_i \quad i = 1..n; \qquad A_{i,j}^{(c)} = c_i^\dagger c_j^\dagger c_j \quad i,j = 1..n, i \neq j, \tag{3.25}$$

where $c_i$ is the $i$-th site annihilation operator. An empty state $|00\dots\rangle$ is a dark state of each of the dissipation operators, and thus, this is again a stationary state in the absence of Hamiltonian coupling, and overall observable dynamics is exactly the same as of the classical counterpart (3.24). To introduce quantum generalization we add Hamiltonian coupling in a way that does not violate assumptions stated earlier. So consider Heisenberg XYZ Hamiltonian

$$H = \sum_{i}^{N} \sigma_i^{(1)} \sigma_{i+1}^{(1)} + a\, \sigma_i^{(2)} \sigma_{i+1}^{(2)} + b\, \sigma_i^{(3)} \sigma_{i+1}^{(3)} + h_i \sigma_i^{(3)}, \tag{3.26}$$

mapped into bosonic space by Jordan-Wigner transformation. In the absence of bosons it does not change the state, thus not alternating the stationary state. In the case of the XXX model ($a = b = 1$), this Hamiltonian still does not alter evolution as it is diagonal in the basis of classical states, hence quasi-stationary state is still the same diagonal state with zero coherence. But altering Hamiltonian parameters immediately leads to non-trivial quantum structure of the quasi-stationary state and evolution, see Fig.3.9. There the total number of particles is defined as $\# = \sum_i i\, \mathrm{tr}\{\rho(t) q_i\}$, where $q_i$ is the projector on the subspace with the total number of particles equals $i$.

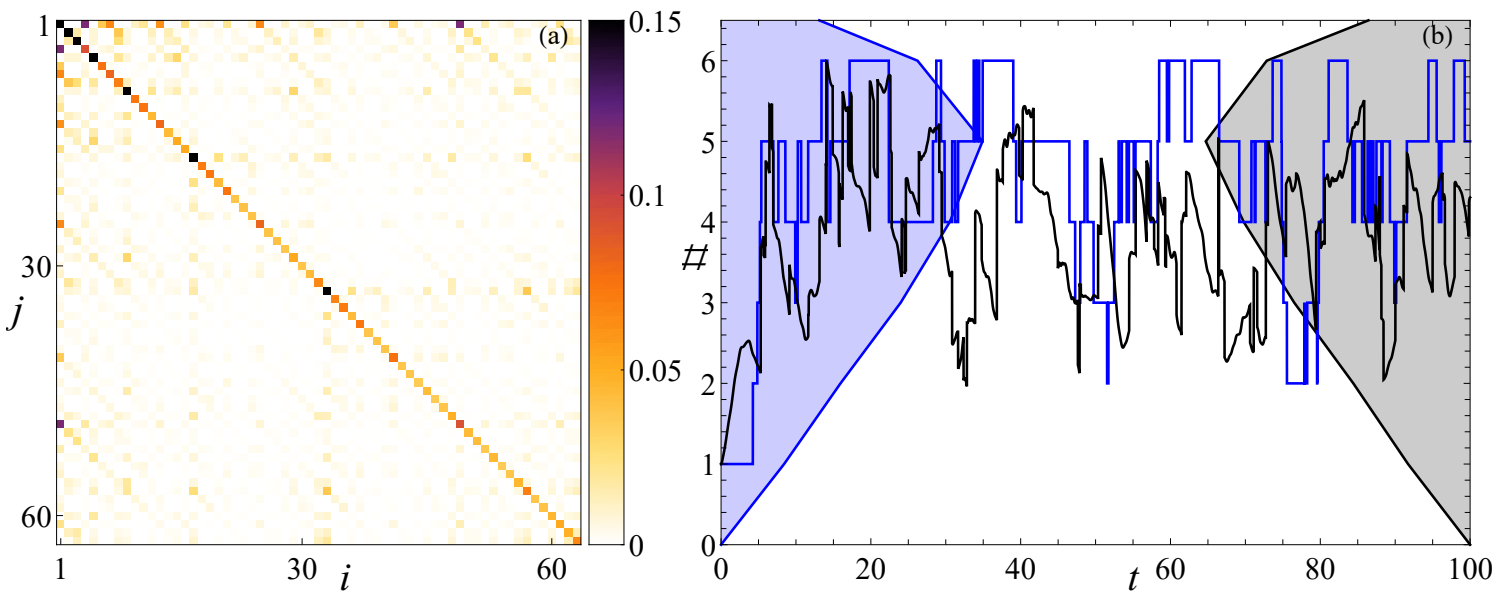

Figure 3.9: (a) Absolute values of the quasistatinary-state density matrix of metapopulation model (3.25) with $c = 0.1, e = 0.1$ and the XXZ Hamiltonian (3.4.3) with $a = 1.5, b = 2.3$, $n = 6$. (b) Dependence of the particle number on time for a single realization of a classical (blue line) and quantum generalization (black line). The shaded areas are corresponding quasi-stationary distributions of the total particles number: classical (left) and quantum generalization (right).

### 3.4.4 Discussion

By decomposing a general Hilbert-Schmidt space with Lindbladian dynamics into several components, and projecting dynamics on those, we derive a generalized master equation. This procedure constitutes a direct implementation of the classical QS concept for open quantum systems. Notably, but expectedly, unlike in the classical counterparts, quantum systems generally maintain components of evolution that correspond to the correlation between the stationary state and the rest of the system. We use this formalism to develop a generalized quantum trajectories algorithm and demonstrate the potential on a toy quantum meta-population model. This technique creates a new approach for studying out-of-equilibrium dynamics both numerically and analytically.

# Chapter 4

# Final remarks

In this thesis, we studied two classes of open quantum systems – discrete-time quantum walks and Lindblad master equation. We demonstrated a plethora of non-equilibrium properties and phenomena. In DTQW, we studied Anderson localization with several types of disorder. We showed that while many regimes qualitatively repeat the known results in the traditional lattice Anderson model, the unique Floquet nature leads to completely new features, most notably, the existence of regimes with the same unique Anderson localization length for every state of the system, that is also analytically calculable. Before this, uniform localization length regimes were only studied in very different setups of Aubry-Andre-like models. Using these results we studied nonlinear spreading of initially localized wave packets in random fields. Using numerical benefits of DTQW, we managed to verify previous conjectures on delocalization dynamics by several orders of magnitude. Developing on the nonlinear properties, we studied a zoo of various soliton-like excitations in flatband plus nonlinearity DTQW. We show the existence of both stationary and moving, stable and unstable structures and study them numerically and analytically. Our results demonstrate the power and flexibility of the DTQW approach. It is not only rich in potential applications, but provide a computationally and analytically promising toolbox. Future applications may include both many-body phenomena, such as MBL, thermalization, ergodicity, and transport, and nonlinear effects such as nonlinear spreading in many-dimensional disordered systems.

Using Lindblad formalism, we studied the effects of dissipation on MBL transition. Similar to Anderson localization, MBL is a purely quantum phenomenon heavily relying on coherence and interference. Naturally, any dissipation effects bring in decoherence and thus compete with MBL. Specifically, pure dephasing disorder would completely grind any (connected) Hamiltonian system into the maximally mixed state. We show that the addition of engineered dissipation allows for survival if not of the complete MBL transition, but at least of certain statistical properties, even in presence of additional dephasing. In order to be able to obtain these numerical results, we implemented and carefully benchmarked the time-evolving block decimation algorithm for one-dimensional dissipative many-body systems. We also studied quasi-stationary states of the Lindblad master equation and developed a theory that generalizes existing classical theory of quasi-stationary states in Markov systems. We developed the quantum trajectories method to deal with systems with quasi-stationary states and applied the methodology to a toy model. The future development in these directions may include application of engineered disorder to hybrid quantum computations, design of Hamiltonians supporting quasi-stationary dynamics, and thus long-lived long-range entanglement. Our computational results provide a basis for numerical studies for systems and regimes previously unattainable by conventional techniques.